\documentclass[twocolumn]{aa}
\usepackage{aas_macros}
\usepackage[varg]{txfonts}
\usepackage{natbib, twoopt}
\usepackage{scalerel,stackengine,amsmath}
\usepackage{siunitx} %[=v3]
\usepackage{amssymb}
\usepackage{grffile}
\usepackage{pdflscape}
\usepackage{afterpage}
\usepackage{comment}
\usepackage{datatool}
\usepackage{graphicx}   % need for figures
\usepackage{verbatim}   % useful for program listings
\usepackage{color}      % use if color is used in text
\usepackage{subcaption}  % use for side-by-side figures
\usepackage{rotating}
\usepackage[hyphenbreaks]{breakurl}
\usepackage[breaklinks]{hyperref}

\usepackage{tikz}
\usepackage[export]{adjustbox}
\usetikzlibrary{shapes,arrows}
\newcommand{\RM}[1]{% \RM{variable}
  \ensuremath{\mathrm{RM_{#1}}}}

\newcommand{\DM}[1]{% \DM{variable}
  \mathrm{DM_{#1}}}

\newcommand{\EM}[1]{% \EM{variable}
  \mathrm{EM_{#1}}}

\newcommand{\Len}[1]{% \Len{variable}
  \mathrm{L_{#1}}}

\newcommand{\Bp}[1]{% \Pb{variable}
  B_{\parallel\mathrm{#1}}}
\newcommand{\Bpw}[1]{% \Pb{variable}
  \overline{\Bp{}}_{\mathrm{#1}}}

\newcommand{\LoS}[2]{% \LoS{variable}{cond}
  \ensuremath{\left\langle {#1} \right\rangle_{\mathrm{L_{#2}}}}}

\tikzstyle{block} = [rectangle,thick, draw=black!100, %fill=blue!30,
    text width=1.2em, text centered, rounded corners, minimum height=1em,minimum width = 0.6em]
\tikzstyle{block2} = [rectangle,thick, draw=black!100, %fill=blue!30,
    text width=2.4em, text centered, rounded corners, minimum height=1em,minimum width = 1.2em]
\tikzstyle{block3} = [rectangle,thick, draw=black!100, %fill=blue!30,
    text width=3.6em, text centered, rounded corners, minimum height=1em,minimum width = 1.8em]
\tikzstyle{block4} = [rectangle,thick, draw=black!100, %fill=blue!30,
    text width=4.8em, text centered, rounded corners, minimum height=1em,minimum width = 1.8em]
\tikzstyle{line} = [draw,ultra thick, -latex']
\tikzstyle{curved line} = [draw, bend right=45,ultra thick, -latex']
\tikzstyle{dashed line} = [draw,ultra thick, -latex',dashed]
\tikzstyle{cloud} = [thick, ellipse, draw=black!100, %fill=green!30,
    node distance=0.6cm, minimum width=2em,
    minimum height=1.2em]

\bibpunct{(}{)}{;}{a}{}{,}             %% natbib format for A&A and ApJ
\definecolor{cobalt}{rgb}{0.06, 0.2, 0.65}
\hypersetup{
  colorlinks,
  citecolor=cobalt,
  linkcolor=[rgb]{0.8, 0.2, 1.0},
  urlcolor=cobalt,
}
\makeatletter
  \newcommandtwoopt{\citeads}[3][][]{\href{http://adsabs.harvard.edu/abs/#3}%
    {\def\hyper@linkstart##1##2{}%
     \let\hyper@linkend\@empty\citealp[#1][#2]{#3}}}
  \newcommandtwoopt{\citepads}[3][][]{\href{http://adsabs.harvard.edu/abs/#3}%
    {\def\hyper@linkstart##1##2{}%
     \let\hyper@linkend\@empty\citep[#1][#2]{#3}}}
  \newcommandtwoopt{\citetads}[3][][]{\href{http://adsabs.harvard.edu/abs/#3}%
    {\def\hyper@linkstart##1##2{}%
     \let\hyper@linkend\@empty\citet[#1][#2]{#3}}}
  \newcommandtwoopt{\citeyearads}[3][][]%
    {\href{http://adsabs.harvard.edu/abs/#3}
    {\def\hyper@linkstart##1##2{}%
     \let\hyper@linkend\@empty\citeyear[#1][#2]{#3}}}
\makeatother

\defcitealias{2019Hutschenreuter}{HE20}
\defcitealias{2022Hutschenreuter}{HE22}

\newcommand{\rmdm}{./images/RMDM_skies/}
\newcommand{\rmdmem}{./images/RMDMEM_skies/}
\newcommand{\dmem}{./images/EMDM_skies/}
\newcommand{\data}{./images/data/}
\newcommand{\illus}{./images/illustrations/}
\newcommand{\prior}{./images/prior/}

\newcommand{\mockdata}{./images/mock/data/}
\newcommand{\mocklow}{./images/mock/results_low/}
\newcommand{\mockhigh}{./images/mock/results_high/}

\newcommand{\skypath}[2]{#1#2}
\newcommand{\correlationpath}[1]{./images/correlations/#1_scatter.png}
\newcommand{\cutoutpath}[2]{./images/cutouts/#1/#1_#2_contour.png}
\newcommand{\powerpath}[1]{./images/power/#1_power.png}

\begin{document}

% \sisetup{exponent-mode=scientific}

\title{Disentangling the Faraday rotation sky}
\author{Sebastian Hutschenreuter \inst{\ref{inst:radboud}} \and
		    Marijke Haverkorn \inst{\ref{inst:radboud}} \and
        Philipp Frank \inst{\ref{inst:mpa}} \and
        Nergis C. Raycheva \inst{\ref{inst:radboud}} \and
		    Torsten A. En{\ss}lin \inst{\ref{inst:mpa},\ref{inst:lmu}}
        }
\institute{Department of Astrophysics/IMAPP, Radboud University, P.O. Box 9010,6500 GL Nijmegen, The Netherlands \label{inst:radboud}
\and
           Max Planck Institute for Astrophysics, Karl-Schwarzschildstr.1, 85741 Garching, Germany \label{inst:mpa}
\and
           Ludwig-Maximilians-Universit\"at M\"unchen, Geschwister-Scholl-Platz 1, 80539 Munich, Germany \label{inst:lmu}
}
\abstract {Magnetic fields permeate the diffuse interstellar medium (ISM) of the Milky Way, and are essential to explain the dynamical evolution and current shape of the Galaxy.
Magnetic fields reveal themselves via their influence on the surrounding matter, and as such are notoriously hard to measure independently of other tracers.
}{
In this work, we attempt to disentangle an all sky map of the line-of-sight parallel component of the Galactic magnetic field from the Faraday effect, utilizing several tracers of the Galactic electron density $n_\mathrm{e}$.
Additionally, we aim to produce a Galactic electron dispersion measure map and quantify several tracers of the structure of the ionized medium of the Milky Way.
}{
The method developed to reach these aims is based on information field theory, a Bayesian inference framework for fields, which performs well when handling noisy and incomplete data and constraining high dimensional parameter spaces.
We rely on compiled catalogs of extragalactic Faraday rotation measures and Galactic pulsar dispersion measures, a well as data on bremsstrahlung and the hydrogen $\alpha$ spectral line to trace the ionized medium of the Milky Way.
}{
We present the first full sky map of the line-of-sight averaged Galactic magnetic field.
Within this map, we find LoS parallel and LoS-averaged magnetic field strengths of up to 4 $\mu$G, with an all-sky root-mean-square of $1.1$ $\mu$G, which is consistent with previous local measurements and global magnetic field models.
Additionally, we produce a detailed electron dispersion measure map, which agrees with already existing parametric models at high latitudes, but suffers from systematic effects in the disk.
Further analysis of our results with regard to the 3D structure of $n_\mathrm{e}$ reveals that it follows a Kolmogorov-type turbulence for most of the sky.
From the reconstructed dispersion measure and emission measure maps we construct several tracers of variability of $n_\mathrm{e}$ along the LoS.
}{
This work demonstrates the power of consistent joint statistical analysis including multiple data sets and physical quantities and defines a roadmap towards a full three-dimensional joint reconstruction of the Galactic magnetic field and the ionized ISM.
}

\maketitle

\section{Introduction}
\label{sec:intro}

The structure of the Milky Way is best expressed in terms of density, velocity, and force fields, as these can be related to the dynamical laws that govern the formation of structure and the overall shape of the Galaxy today.
To our fortune, these fundamental quantities can be determined by a variety of observables, which, to our misfortune, in general do not yield constraints for those individually, but are mixed and entangled in nontrivial and sometimes nonlinear ways.

Prominent examples of such observables combining several Galactic constituents are e.g. synchrotron radiation coupling the relativistic electron density $n_\mathrm{rel}$ with the perpendicular component of the Galactic magnetic field $B_\perp$, or stellar polarization in the both the optical and infra-red regime, which gives information on both the magnetic field direction and dust properties.
The only route to disentangle these and similar observables and to map out the structure of the Galaxy is to cross-correlate them with each other and/or to compare with simulations.
For some observables, such a component separation has been successfully performed, e.g. \citet{2016Planck_X, 2020BeyondMain} separate the microwave sky into the CMB and astrophysical component maps or \citet{2015Selig, 2023Platz} separate the Fermi photon-count maps into point sources and diffuse emission of hadronic and leptonic origin.

In this paper, we aim to disentangle the Galactic Faraday sky into its physical components.
The Faraday effect entangles information on the line-of-sight (LoS) component of the magnetic field $B_\parallel$ with the plasma electron density $n_\mathrm{e}$ into the Faraday depth $\phi$.
It is the only observable that provides direct information on $B_\parallel$ for most regimes in the interstellar medium (ISM), apart from longitudinal Zeeman splitting, which may be observable in very dense regions such as molecular clouds \citep{2013Beck} and circular polarization of synchrotron radiation, which will only be observable with next generation radio telescopes \citep{2017Ensslin}.
The Faraday effect has long been used to constrain the Galactic magnetic field strength, see e.g. \citet{1972Manchester, 1974Manchester, 1994Rand, 2001Frick, 2006Han, 2019Sobey, 2022Pandhi}.
Some of these works exploit the information of pulsars, which have the advantage of potentially providing Faraday data along with an independent measure on the column density of the interjacent electrons, i.e. the dispersion measure (DM).
If one is interested to constrain large scale Galactic models, pulsars come with the disadvantage that only few with independent distance measurements are known (about 250 at the time of writing this paper) and hence one has to appeal to strong a priori assumptions on the magnetic field \citep{2010Jaffe, 2012Jansson} and/or the electron density \citep{2001Cordes, 2012Schnitzeler, 2016Maksim, 2017Yao} of the Milky Way.
Extra-Galactic Faraday data has also been used to constrain the Galactic magnetic field in conjuncture with electron data stemming from free-free or H-$\alpha$ emission \citep{2019Hutschenreuter, 2019Betti, 2022Nergis}.
In the first reference, henceforth abbreviated with \citetalias{2019Hutschenreuter}, the free-free sky was included as a phenomenological proxy for the Faraday depth amplitude.
This permitted qualitative statements on the local structure of the Galactic magnetic field such as the alignment of the magnetic field with the local Orion spiral arm.

In this work we will replace the phenomenological model used in \citetalias{2019Hutschenreuter} with a more physical one, with the aim to turn qualitative predictions into quantitative ones.
In particular, we are interested in reconstructing the averaged LoS component of magnetic field  and the integrated electron density, also known as the dispersion measure (DM) for the full Galaxy.
This will be attempted with the help of four different data sets, namely extra-Galactic Faraday data as compiled by \citet{2023van_Eck}, pulsar data \citep{2005ATNF}, the free-free map of the \textit{Planck} survey \citep{2016Planck_X} and a H-$\alpha$ map \citep{2003Finkbeiner}. \\
The statistical methodology of this work foots on the same grounds as previous inferences of the Galactic Faraday depth sky in \citet{2012Oppermann}, \citetalias{2019Hutschenreuter} and \citet{2022Hutschenreuter}, namely Information Field Theory (IFT).
IFT is information theory for fields and field-like quantities and can cope with large, incomplete, and noisy data sets.
For references to IFT see \citet{2019Ensslin} and for the accompanying python package \footnote{https://ift.pages.mpcdf.de/nifty/}{NIFTy}, in which the algorithm used in this work is implemented in, see \citet{2019NIFTY}.

We structure the paper as follows:
Sect. \ref{sec:observables} describes the physical observables relevant for this paper, putting special emphasis on effects that correlate the different relevant physical quantities.
Sect. \ref{sec:models} then explains the different models the data is interpreted in.
Sect. \ref{sec:results} discusses the results and finally Sect. \ref{sec:conclusions} gives a conclusive summary.

\begin{figure*}

\centering
\begin{subfigure}{0.45\textwidth}
\includegraphics[width=\textwidth, draft=false]{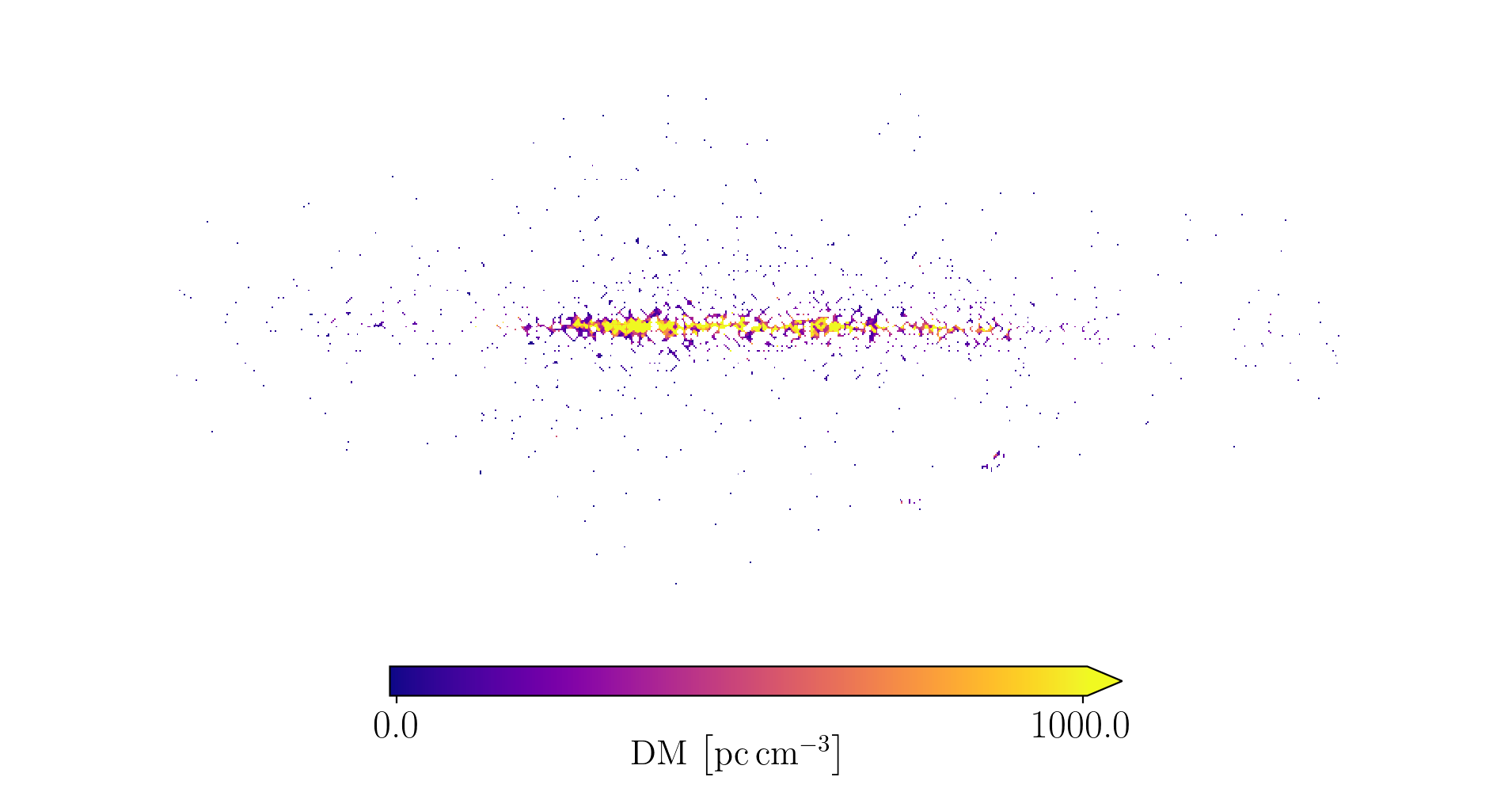}
\caption{\label{fig:DM_data}}
\end{subfigure}
\hfill
\begin{subfigure}{0.45\textwidth}
\includegraphics[width=\textwidth, draft=false]{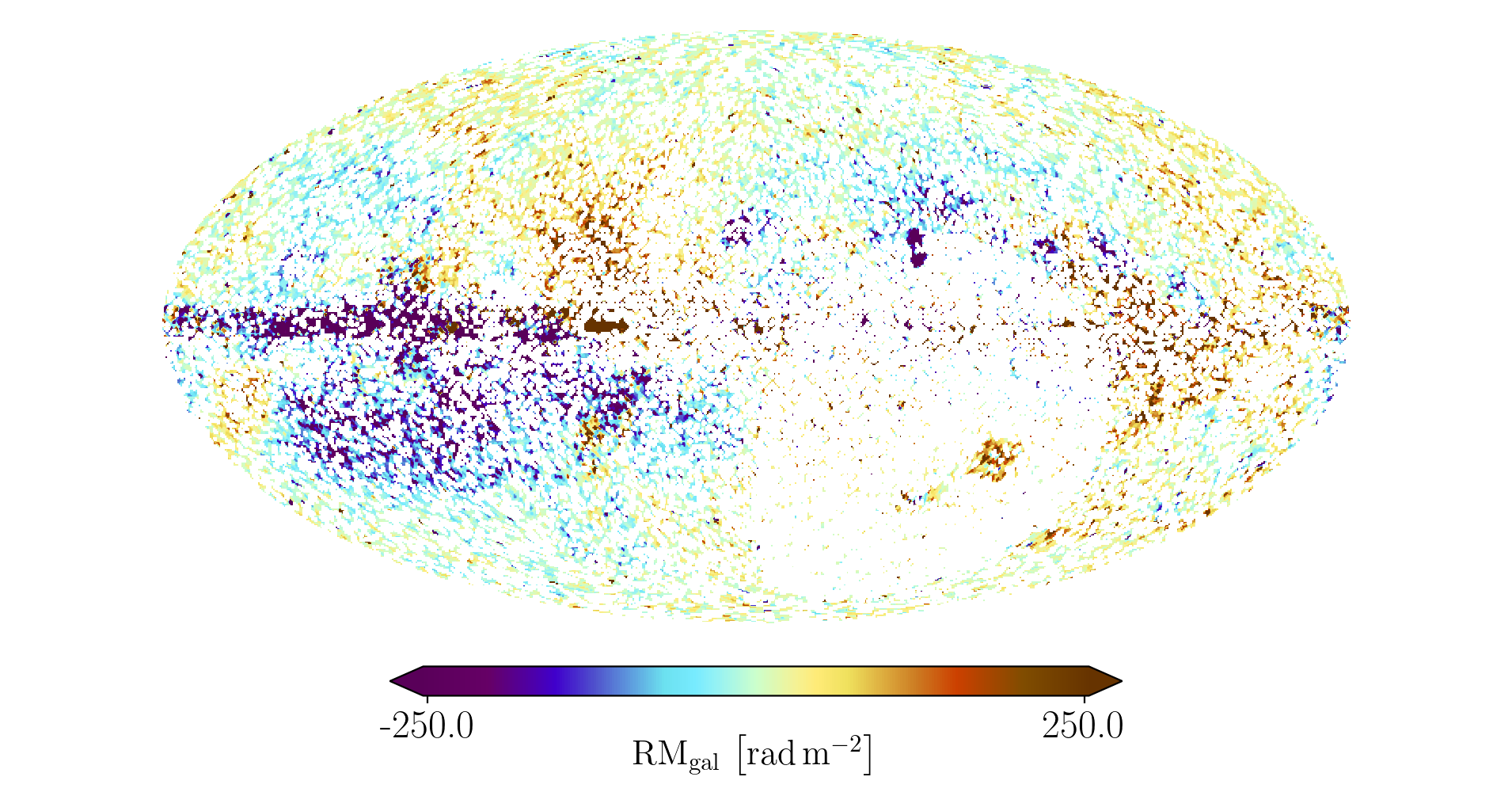}
\caption{\label{fig:RM_data}}
\end{subfigure}
\hfill
\begin{subfigure}{0.45\textwidth}
\includegraphics[width=\textwidth, draft=False]{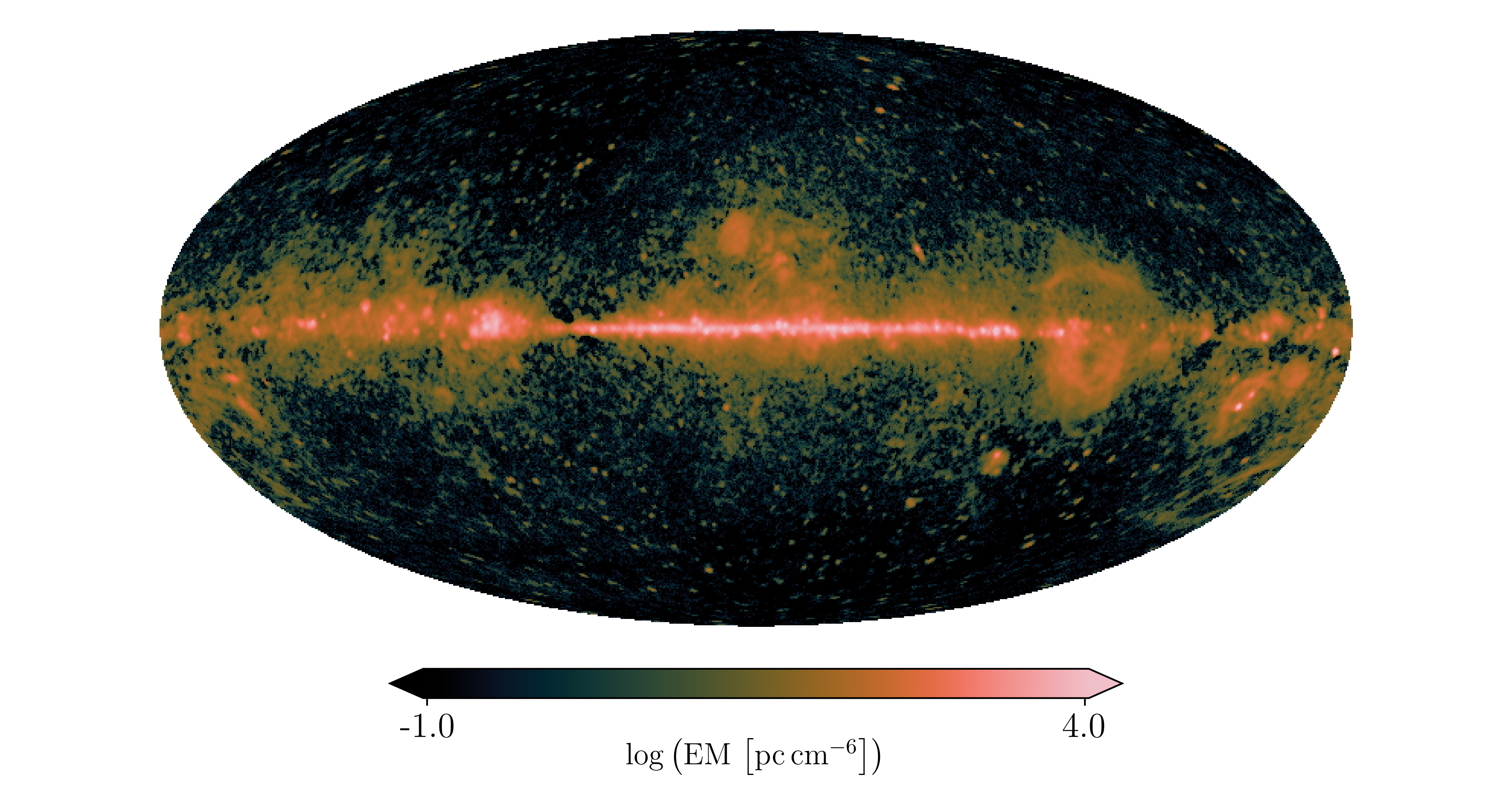}
\caption{\label{fig:log_planck_data}}
\end{subfigure}
\hfill
\begin{subfigure}{0.45\textwidth}
\includegraphics[width=\textwidth, draft=False]{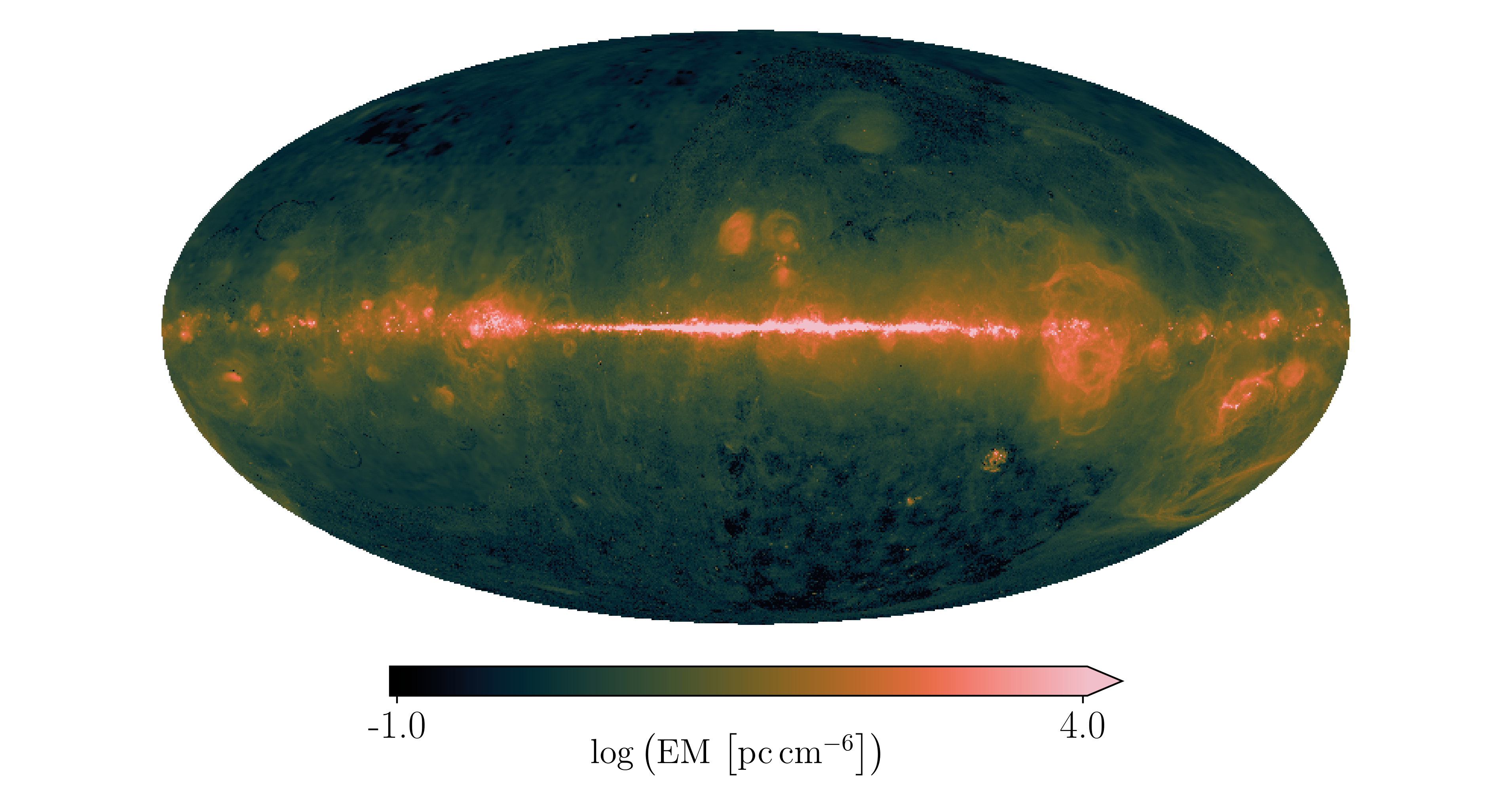}
\caption{\label{fig:log_halpha_data}}
\end{subfigure}
\hfill
\begin{subfigure}{0.45\textwidth}
\includegraphics[width=\textwidth, draft=False]{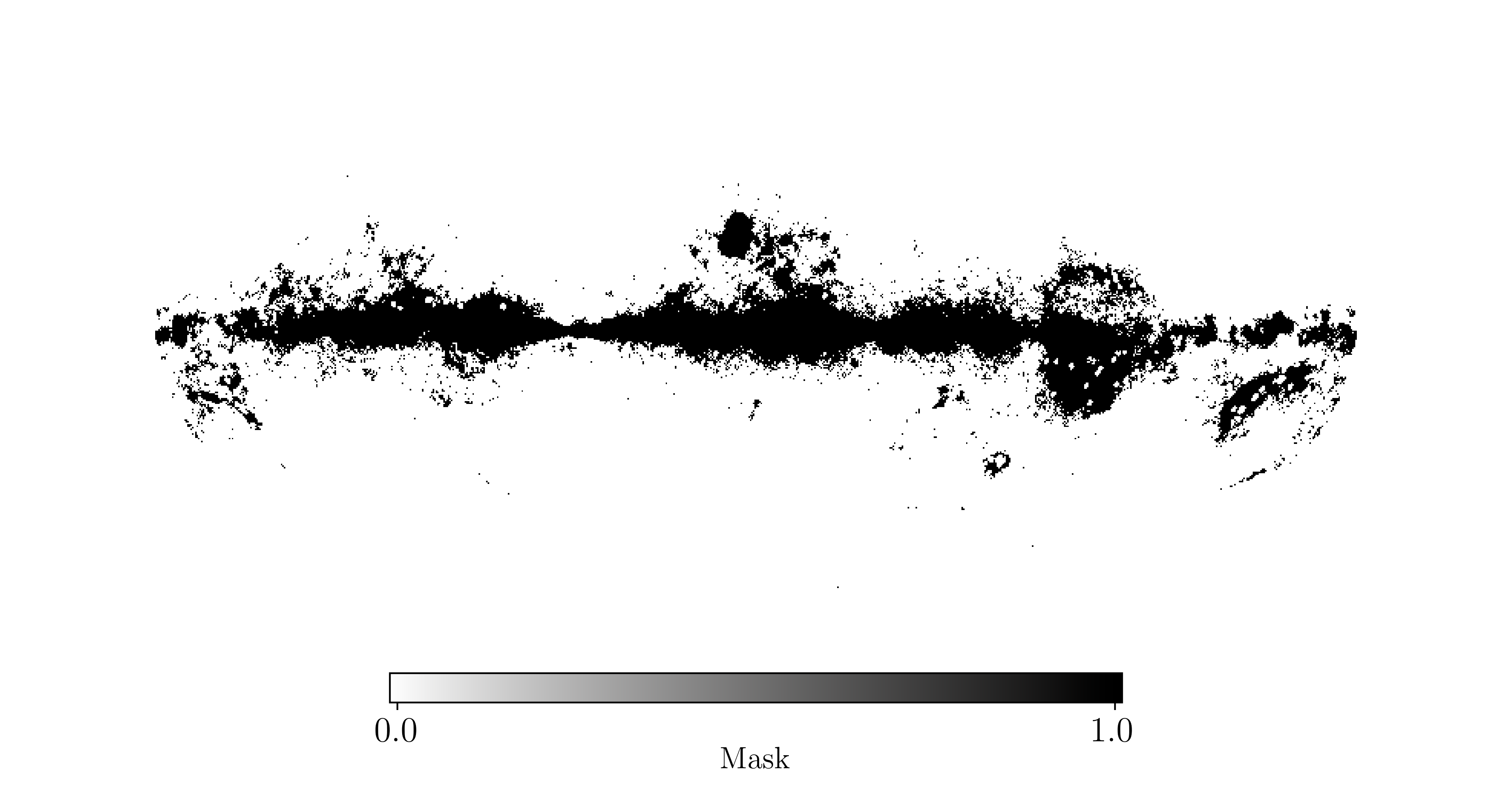}
\caption{\label{fig:planck_filter}}
\end{subfigure}
\hfill
\begin{subfigure}{0.45\textwidth}
\includegraphics[width=\textwidth, draft=False]{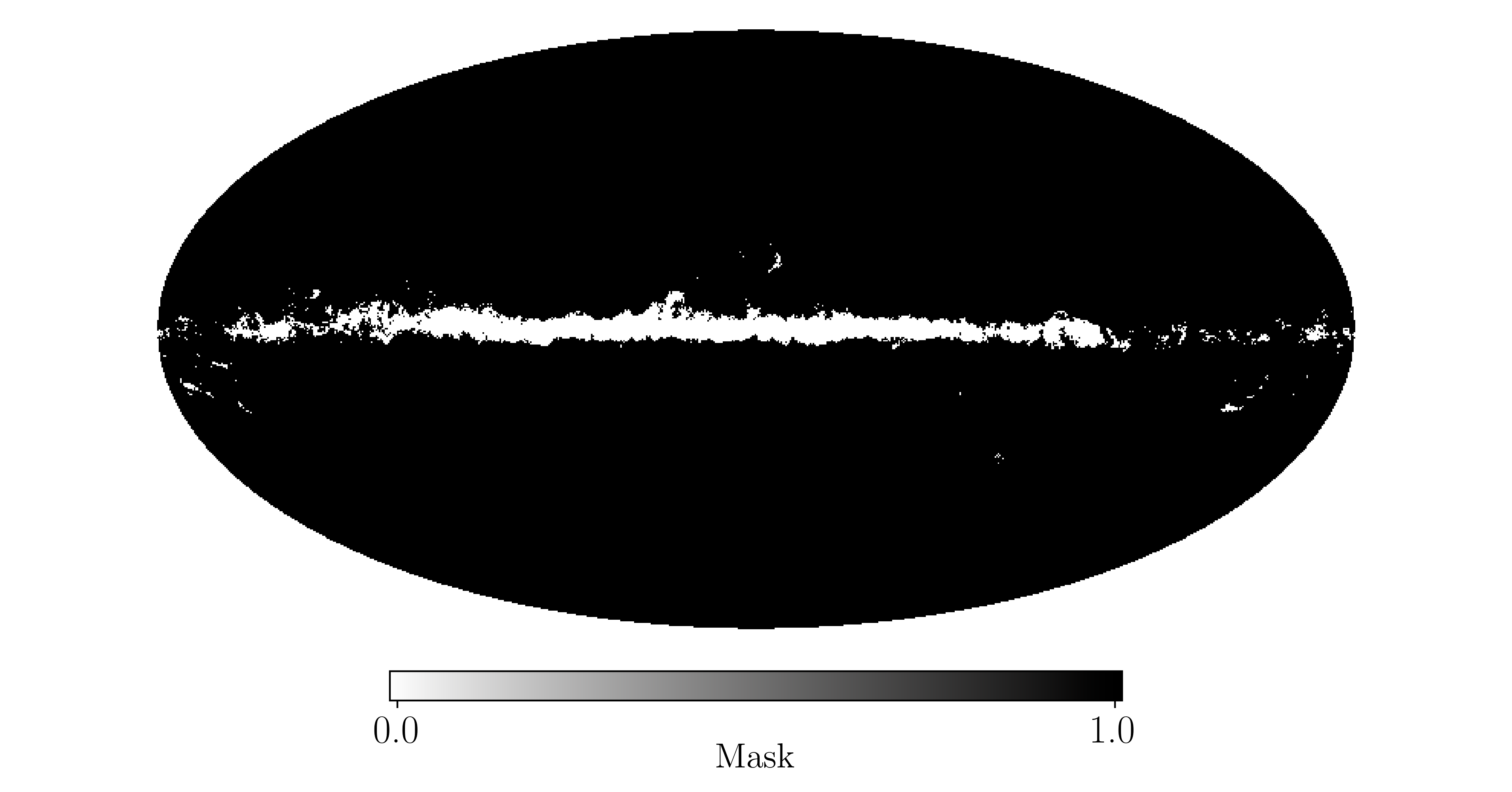}
\caption{\label{fig:halpha_filter}}
\end{subfigure}

\caption{\label{fig:data} Sky projections of all data sets used in this work. These are pulsar DMs from the ATNF catalog (Fig. (a), \citep{2005ATNF}), RMs stemming from the van Eck catalog (Fig. (b), \citep{2023van_Eck}) and the EMs calculated from free-free (Fig. (c), \citep{2016Planck_X}) and H-$\alpha$ (Fig. (d), \citep{2003Finkbeiner}) emission.
The latter two data sets are shown on a logarithmic scale.
We show the quality masks employed for the EM data (see discussion in Sect. \ref{subsubsec:emission measure - data}) in Figs. (e) and (f), with 0 indicating a masked pixel.
These and all sky maps following are presented in Galactic coordinates centered at $(l,b) = \left(\ang{0}, \ang{0}\right)$, with longitude increasing to the left.}
\end{figure*}

\section{Observables}
\label{sec:observables}

In order to simplify notation in the following, we introduce the LoS-average over a quantity $x$
\begin{equation}
\label{eq:notation}
\LoS{x}{i} \equiv \frac{1}{\mathrm{L_i}} \int_0^{\mathrm{L_i}} dl\, x,
 \end{equation}
with $\mathrm{L_i}$ indicating the length of the LoS $i$.
In a slight abuse of notation, we use $L$ for both the identification and the length of the LoS.
In this spirit, the subscript $i$ can be used to identify either a specific LoS or subclasses of LoS.
For example, we will use $\mathrm{L_{gal}}$ to refer to all LoS which trace the full Galaxy, where the boundary is implicitly defined via the physical processes that generate the data.
We assume that all processes used in this work trace the same LoS's, if not explicitly modelled otherwise.
We also would like to note that the electron density referred to in this work mostly describes the thermal component, although strictly speaking the RM, DM and free-free EM measurements are also slightly sensitive to the relativistic electrons, which are, however, only providing a very small contribution to all three tracers.
 In theory, suprathermal free electrons could also contribute to the electron population.
 However, as there is very little evidence of pervasive suprathermal electron populations in the ISM \citep{2021Gurnett}, we will not explicitly include this component in the current analysis.

\subsection{Pulsar dispersion measures}
\label{subsec:physics_pulsars}

\subsubsection{Physics}
Pulsars are magnetized and rapidly rotating neutron stars \citep{2004Lorimer}, emitting beamed electromagnetic radiation.
This results in periodic radio pulses.
As light travels slower within interstellar plasma at lower frequencies, the arrival time $t$ of the pulse varies with frequency $\nu$, which can be expressed as \citep{2011Draine}
\begin{equation}
\label{eq:pulsar_timing}
\frac{\partial \mathrm{t}}{\partial \nu} = -\frac{e^2}{\pi\nu^3m_ec}\DM{P}.
\end{equation}
The physical constants $e, m_e$ and $c$ describe the elementary charge, the electron mass and the speed of light, respectively.
We introduced the dispersion measure as the integral over the electron density $n_{\mathrm{e}}$,
\begin{equation}
\label{eq:dm_fundamental}
\DM{P} = \mathrm{L_p}\LoS{n_{\mathrm{e}}}{p}.
\end{equation}

The LoS $\mathrm{L_P}$ goes from the pulsar P to Earth.
Therefore, DMs obtained from pulsars provide a lower limit on the Galactic DM sky for the respective LoS they probe and, due to the fact that the DM monotonically increases along the LoS, can serve as a distance proxy.
This can be expressed in several ways, e.g. on the DM level by defining the residual DM `behind' the pulsar as
\begin{equation}
\label{eq:DM_as_average}
\DM{res} = \DM{gal} - \DM{P},
\end{equation}
where $\Len{Gal} \geq \Len{P}$ and $\DM{res} \geq 0$.
Some pulsars have distance measurements independent of $n_\mathrm{e}$, mostly determined via parallaxes, association with known structures or HI absorption.
The pulsar distances $\mathrm{L}_{\mathrm{P}}$ can be related to the $\DM{P}$ models via
\begin{align}
\label{eq:pulsar_distance}
\mathrm{L}_{\mathrm{P}} = f(\DM{P}).
\end{align}
Here, $f$ is a monotonically increasing function which encodes the electron distribution along the given LoS, and is hence different for each position on the sky.

\subsubsection{Data}
\label{subsubsec:DM_data}
We use the Australia Telescope National Facility (ATNF) \citep{2005ATNF} catalog of pulsars\footnote{\url{http://www.atnf.csiro.au/research/pulsar/psrcat}}.
The DM error bars given in the catalog are determined by propagating the arrival time uncertainties by taking Eq. \eqref{eq:pulsar_timing} at face value.
As the arrival timing is very precise, this leads to signal-to-noise ratios (SNR) in the DM of up to $10^8$.
Several systematic effects may slightly limit the interpretation of Eq. \eqref{eq:pulsar_timing} as a direct measure of the electron column density \citep{2020Kulkarni}.
We hence adapt a maximum signal-to-noise ratio (SNR) of $0.001$ for the DM data, i.e. sources with higher SNR get their error bars adapted accordingly.
The catalog also provides distance measurements, mostly from parallaxes, associations with known objects (e.g. globular clusters) or HI absorption.
If an object has several independent distance measurements available, we include all measurements in our analysis.
In some cases distances provided in the catalog are processed data (e.g. \citet{2012Verbiest}), which combine several distance measures.
In these cases, we only consider the processed data and disregard data points which have been used in the processing.
In general, we only consider measurements with well-defined error bars.
In total, the catalog includes 264 independent distance measurements usable in this work, 130 of which are parallaxes.
We show a projection of all pulsar DMs used in this work in Fig. \ref{fig:DM_data}.

\subsection{Faraday rotation measures}
\label{subsec:faraday}

\subsubsection{Physics}
\label{subsubsec:faraday_physics}

The differential angle of rotation $\Delta_\lambda$ of the polarization plane of linearly polarized light travelling through a magneto-ionic plasma, can be described by the following formula \citep{1966Burn}
\begin{equation}
\label{eq:faraday_angle}
\Delta_\lambda = \mathrm{RM}\,\lambda^2,
\end{equation}
where $\lambda$ is the observational wavelength and RM is the rotation measure, defined by this equation.
Determining RMs is usually done by observing $\Delta_\lambda$ at different wavelengths
and fitting the result in $\lambda^2$ space.
In the ideal case of a thin plasma screen being the only source for the rotation effect, the RM is equal to the Faraday depth $\phi$, which is defined via
\begin{align}
\phi  &= \frac{e^3}{2\pi m_e^2 c^4}\int_{\mathrm{LoS}} dl\, n_{\mathrm{e}} \Bp{} \nonumber \\
&= 0.812\,\int_{\mathrm{LoS}} dl\left[\mathrm{pc}\right] \, n_{\mathrm{e}}\left[\mathrm{cm}^{-3}\right] B_{\mathrm{\parallel}}\left[\mu\mathrm{G}\right],
\label{eq:faraday_fundamental}
\end{align}
where $\Bp{}$ is the LoS-parallel component of the magnetic field.
We can use Eq. \eqref{eq:faraday_fundamental} to define a $n_\mathrm{e}$-weighted average of $\Bp{}$ in the Milky Way,
\begin{align}
\label{eq:B_average}
\Bpw{} \equiv \frac{\phi_\mathrm{gal}}{\mathrm{DM}_\mathrm{gal}} = \frac{\LoS{n_\mathrm{e}\Bp{}}{gal}}{\LoS{n_\mathrm{e}}{gal}}.
\end{align}
This gives a direct and simple connection to extract information on the magnetic field from the $\phi_\mathrm{gal}$ and $\DM{gal}$ skies.
We note that in the case $\Bp{}$ and $n_\mathrm{e}$ are statistically independent, $\LoS{n_\mathrm{e}\Bp{}}{gal}$ simplifies to $\LoS{n_\mathrm{e}}{gal}\LoS{\Bp{}}{gal}$, which implies $\overline{\Bp{}}_{\mathrm{gal}} = \LoS{\Bp{}}{gal}$, i.e. we can extract the unbiased average of the LoS parallel component of the magnetic field from $\phi_\mathrm{gal}$.

Since this a very important case for the interpretation of $\overline{\Bp{}}_{\mathrm{gal}}$,  we investigate the correlation between $\Bp{}$ and $n_\mathrm{e}$ in more detail.
We start with the relationship between the absolute value of the full magnetic field vector $|\mathbf{B}|$ and $n_\mathrm{e}$, which is physically more tangible.
From a theoretical perspective, both correlation (due to compression in shock fronts \citep{1969Roberts} or in gravitationally collapsing structures) or anti-correlation (due to magnetic pressure compensating lacking gas pressure in conditions close to pressure equilibrium \citep{2003Beck}) are fathomable.
Observationally, the two quantities are generally decoupled in the warm inter-stellar medium (ISM) \citep{2010Crutcher}, but maybe correlated in denser regions such as molecular clouds \citep{2011Harvey-Smith, 2015Purcell} which may dominate the average for certain LoS (see also discussion in \citetalias{2019Hutschenreuter}).
Simulations have shown that for typical ISM conditions both quantities are uncorrelated on kpc scales in the subsonic regime  \citep{2021Seta}.

However, from a purely mathematical standpoint, statistical independence of $|\mathbf{B}|$ and $n_\mathrm{e}$ does of course not imply the same for  $\Bp{}$  and $n_\mathrm{e}$.
Since our position in the Milky Way is in no way special, only geometric effects and/or an alignment of the magnetic field with Galactic structure should introduce such correlations, which makes the scenario of a strong all sky correlation between the two quantities (in the case $\LoS{|\mathbf{B}|\,n_\mathrm{e}}{\mathrm{gal}}=0$ holds) unlikely.
But locally some correlation is expected and indeed, Faraday rotation and Planck dust polarization data have indicated that $\Bp{}$ might be correlated with specific Galactic structures such as the Local Arm \citepalias{2019Hutschenreuter}.
Furthermore, the THOR survey \citep{2019THOR} has found evidence for a correlation of a strong Faraday excess and the Sagittarius Galactic arm, and such a correlation between Galactic arms and Faraday excesses is expected by simulations  \citep{2020Reissl}.
All of these are correlations of projected quantities which in principle do not yet imply correlation along the LoS, but we do regard it as plausible in these cases.

To summarize, there is a solid body of evidence that the assumption of statistical independence between $\Bp{}$ and $n_\mathrm{e}$ holds for a wide range of scales in the Milky Way and on the sky, but also the indication that it might break down for specific structures on the sky.
Eq. \eqref{eq:B_average} has been used throughout the literature to estimate the average magnetic field strength, mostly for specific pulsars \citep{1972Manchester, 1974Manchester, 2009Han}.
It should be noted that the LoS averaged magnetic field strength $\Bpw{gal}$ can significantly underestimate the typical magnetic field strength if field reversals occur, i.e. in strongly turbulent environments.

\subsubsection{Data}
\label{subsubsec:faraday_data}
In this work, we rely on the same pre-compiled data catalog of extragalactic RMs \citep{2023van_Eck} as in \citet{2022Hutschenreuter}, including the same pre-processing routines.
We furthermore use the error estimate provided by \citet{2022Hutschenreuter} instead of the observational errors, as in those the potential extra-galactic components and observational systematic effects from e.g. n$\pi$-ambiguities \citep{2019Ma} are already factored in.
A projection of the data set is shown in Fig. \ref{fig:RM_data}.
We do not include RMs of pulsars in the inference model, as the small number of pulsars with associated RMs does not warrant the more complex modelling required for the residual Faraday depth behind the pulsars.
\footnote{This would require introducing additional fields that describe the dispersion and correlation of $\Bpw{gal}$ along a LoS.
As no independent measurement information is available in our approach on these quantities, they are unconstrained and thus not much information from pulsar RMs on other quantities can be expected.
For pulsar DMs the situation differs, as they always provide a strict lower limit on $\DM{gal}$ and auxiliary information on the electron density dispersion is provided via the EM.
}
We do, however, utilize these RMs to validate our model in Sect. \ref{subsec:correlations}.

\subsection{The electron emission measure}
\label{subsec:emission measure}

\begin{figure*}
\centering
\includegraphics[width=\textwidth, draft=False]{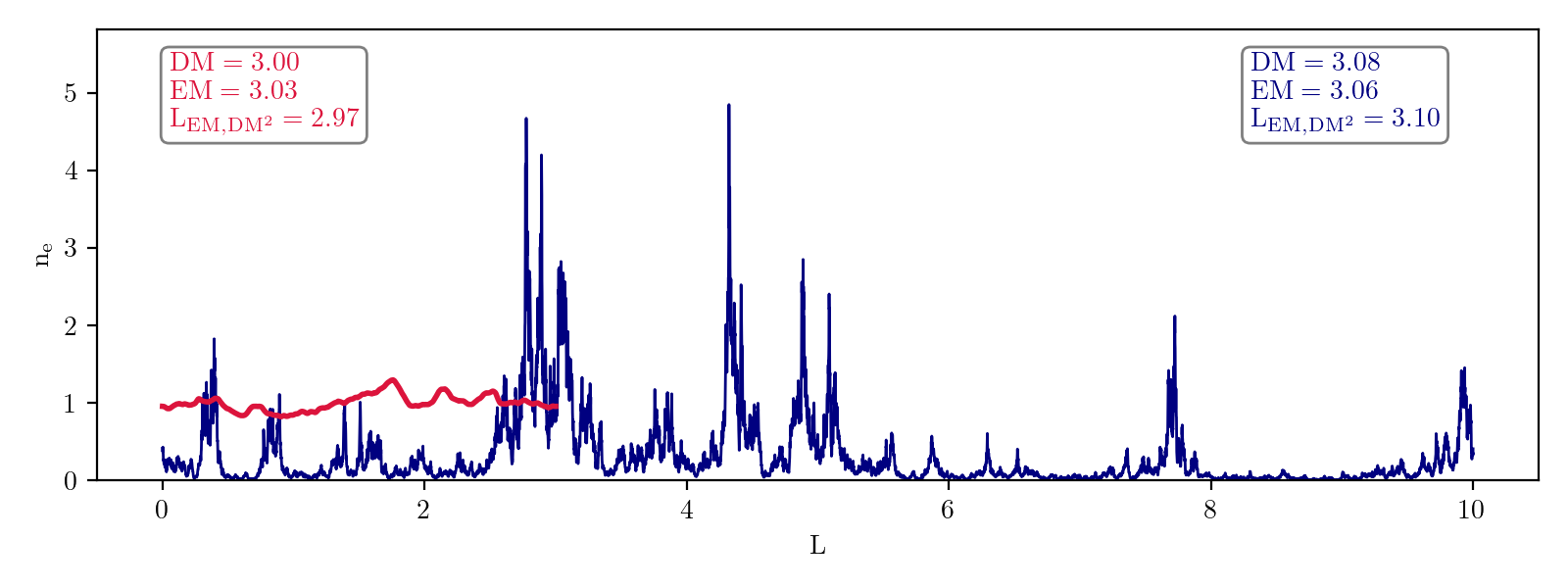}
\caption{\label{fig:em_dm_simple}
An illustration of the EM-DM relation for two simple models, as discussed in Sec. \ref{subsubsec:emission measure - physics}.
The red graph shows a short LoS with relatively weak variations in $n_\mathrm{e}$, while the blue one shows a long LoS with strong variations.
Albeit depicting two completely different environments, the calculated $\DM{}$, $\EM{}$ and $\Len{DM^2/EM}$ values for an observer sitting at $L=0$ approximately agree.
The units in this illustration are arbitrary, as the effect is scale invariant.
Its main intention is to demonstrate the interpretational limitations of Eq. \ref{eq:em_dm2_ratio}.
A more physical illustration for the Milky Way is provided later in Sect. \ref{subsec:results_sky_filling}.}
\end{figure*}

\subsubsection{Physics}
\label{subsubsec:emission measure - physics}

A further tracer of the electron column density is the Galactic emission measure, defined via
\begin{equation}
\label{eq:em_fundamental}
\EM{gal} = \mathrm{L_{gal}}\LoS{n_\mathrm{e}^2}{\mathrm{gal}}.
\end{equation}
As we are interested in the connection to the Galactic DM sky, it is worthwhile to note that one can interpret $n_\mathrm{e}$ as a statistical process along the LoS.
One can then write for the variance $\sigma^2_{n_\mathrm{e}, \mathrm{LoS}}$ of this process along a Galactic LoS:
\begin{align}
\label{eq:em_dm_variance}
\sigma^2_{n_\mathrm{e}} &=  \LoS{n_\mathrm{e}^2}{\mathrm{gal}} -  \LoS{n_\mathrm{e}}{\mathrm{gal}}^2 = \frac{\EM{gal}}{\mathrm{L_{gal}}} - \frac{\DM{gal}^2}{\mathrm{L_{gal}}^2}
\end{align}
There are several possible ways to connect the $\EM{gal}$ and $\DM{gal}$ skies.
Out of these, models which are easily interpretable and for which a priori constraints can be formulated are preferable.
Motivated by simple dimensional analysis to cancel the density units, we then calculate
\begin{align}
\label{eq:em_dm2_ratio}
\frac{\EM{gal}}{\DM{gal}^2} &= \frac{1}{\mathrm{L_{gal}}}\left(\frac{\sigma^2_{n_\mathrm{e}}}{\LoS{n_\mathrm{e}}{\mathrm{gal}}^2} + 1\right) = \frac{1}{\Len{gal}f_{\mathrm{EM}, \mathrm{DM}^2 }} = \frac{1}{\Len{DM^2/EM}}.
\end{align}
Here, we introduced the unitless conversion factor $f_{\mathrm{EM}, \mathrm{DM}}$, which is strictly positive and smaller than one, and the EM-DM path length $\Len{DM^2/EM}$.
We will construct a model for the $\Len{DM^2/EM}$ sky in Sec. \ref{subsec:results_sky_filling}, which allows us to connect the EM and DM data sets.

%% Interpretation
In order to be able to construct a prior on the $\Len{DM^2/EM}$ sky and to interpret the results of the inference, we discuss the possible limits and interpretations of Eq. \ref{eq:em_dm2_ratio}.
The $f_{\mathrm{EM}, \mathrm{DM}}$ factor approaches one if $n_\mathrm{e}$ is close to constant ($\sigma_{n_\mathrm{e}} \ll \LoS{n_\mathrm{e}}{\mathrm{gal}}$) and zero if $n_\mathrm{e}$ has strong variability along the LoS (i.e $\sigma_{n_\mathrm{e}} \gg \LoS{n_\mathrm{e}}{\mathrm{gal}}$).
The simplest conceivable case is hence that of a constant density along the LoS, implying $\sigma_{n_\mathrm{e}} = 0$ and therefore $\Len{gal}\EM{gal} = \DM{gal}^2$, i.e. the EM-$\DM{}^2$ ratio simply gives the inverse length of the LoS.
$\Len{DM^2/EM}$ thus provides a lower limit to the length of the LoS.
A slightly more complicated model already often used in the literature (i.e. \citep{1993Pynzar, 2008Berkhuijsen, 2011Harvey-Smith}) assumes the ionized ISM to be composed of internally uniform clouds with little to no ionized matter in between.
In this particular case, $f_{\mathrm{EM}, \mathrm{DM}}$ can be viewed as fraction of the LoS that is ionized, and is equal to the LoS filling factor $f$.
Allowing for internal density fluctuations in these clouds and cloud-to-cloud variations leads to the ionized-cloudlet model \citep{1991Cordes, 2001Cordes, 2020Ocker}, in which case $f_{\mathrm{EM}, \mathrm{DM}}$ can be parametrized in terms of these variations and the overall filling factor $f$.
In the general case, $\Len{DM^2/EM}$ approximates the size of the part of a LoS that exhibits the largest $n_\mathrm{e}$ values.

Since $\LoS{n_\mathrm{e}}{\mathrm{gal}}$, $\sigma_{n_\mathrm{e}}$ and $\Len{gal}$ can vary strongly and do not depend on each other, the interpretation of the $\Len{DM^2/EM}$ calculated from EM and DM measurements must rely on additional assumptions.
We illustrate this point by presenting two examples of possible $n_\mathrm{e}$ configurations along a LoS in Fig. \ref{fig:em_dm_simple}, which give very similar observational results, albeit depicting completely different scenarios for the $n_\mathrm{e}$ distribution.
It should further be noted that $\Len{DM^2/EM}$ can be directly calculated from EM and DM values, while $\LoS{n_\mathrm{e}}{\mathrm{gal}}$, $\sigma_{n_\mathrm{e}}$ and $\Len{gal}$ are 3D-model dependent quantities, in particular all three depend on the definition of the border region of the Milky Way.
All our sky models are independent of this choice, but their interpretation with regard to the 3D structure of the Galaxy is not.
We give a simple example of such a 3D model in Sect. \ref{subsec:results_sky_filling}, as it is more connected to the concrete interpretation of our results.

At this point we would also like to note that the dimensional analysis argument used to motivate Eq. \ref{eq:em_dm2_ratio} could also be used to cancel the units of length instead of the density units, i.e. to calculate:
\begin{align}
  \label{eq:em_dm_ratio}
  \frac{\EM{gal}}{\DM{gal}} &=  \frac{\sigma^2_{n_\mathrm{e}}}{\LoS{n_\mathrm{e}}{\mathrm{gal}}} + \LoS{n_\mathrm{e}}{\mathrm{gal}} = \frac{ \LoS{n_\mathrm{e}}{\mathrm{gal}}}{f_{\mathrm{EM}, \mathrm{DM}}} \equiv \varrho_{\mathrm{EM}/\mathrm{DM}}
\end{align}
This equation is not as useful as Eq. \ref{eq:em_dm2_ratio} for our sky model, as we found it more difficult to put physical a priori assumptions on this fraction.
Nonetheless, the quantity $\varrho_{\mathrm{EM}/\mathrm{DM}}$ is still interesting for the interpretation of our results, especially in the limit $\sigma^2_{n_\mathrm{e}} \ll \LoS{n_\mathrm{e}}{\mathrm{gal}}$, in which it is simply equal to the average electron density along the LoS.
In general, due to the limits of $f_{\mathrm{EM}, \mathrm{DM}}$, $\varrho_{\mathrm{EM}/\mathrm{DM}}$ provides an upper limit to $\LoS{n_\mathrm{e}}{\mathrm{gal}}$, and can be interpreted as the LoS averaged electron density weighted by itself.

\subsubsection{Data}
\label{subsubsec:emission measure - data}

There are several tracers which give information on $\EM{gal}$.
In this work, we use the EM map provided by the \textit{Planck} mission, derived from microwave free-free emission (the bremsstrahlung resulting from the interaction of free protons and electrons),
and a H-$\alpha$ emission map compiled by \citet{2003Finkbeiner} based on several surveys \citep{1998Dennison, 2001Gaustad, 2003Wisconsin} tracing the optical hydrogen Balmer-$\alpha$ line.
Both effects are not clean tracers of $\EM{gal}$, but require a careful consideration of systematic biases.
In this work, we utilize data from both sources, but seek to minimize possible biases by either correcting for them or, if this is not possible, mask the respective sky areas where the data sets become unreliable.

\paragraph{Free-free EM}

Free-free emission is an important Galactic foreground in the microwave sky, especially in the regime around $30$ GHz, and was therefore accurately determined by Cosmic Microwave Background (CMB) missions such as \textit{Planck}  \citep{2016Planck_X} or WMAP \citep{2013WMAP}.
The observed free-free emission depends on two Galactic environmental variables, namely the (squared) electron density $n_\mathrm{e}$ and the electron temperature $T_\mathrm{e}$.
The \textit{Planck} team has produced both $T_e$ and EM maps  \citep{2016Planck_X} via an elaborate component separation technique \citep{2008Eriksen}.
At high latitudes, the \textit{Planck} EM map contains many bright point-like objects, which correspond to extragalactic objects \citep{2016Planck_X}.
Since this is only partially reflected in the statistical uncertainties, these sources contaminate the data set, especially at high latitudes.
The \textit{Planck} has provided a mask of point sources for all frequency channels.
We have used the 30 GHz mask to mask all point sources with  $|l| > \deg{10}$.
At lower latitudes, the \textit{Planck} mask also excludes structures clearly belonging to the disk.
Additionally, we have hence decided to mask all points in the \textit{Planck} EM data set with an SNR $< 10$, which effectively removes most remaining point sources.
We note that this SNR cut results in a rather conservative mask, as this discards many more areas of the sky than just the point sources and effectively only leaves the disk and some very bright regions of the sky (about 12.3 \% of the full sky remain).
We are ready to accept this downside, as H-$\alpha$ emission provides a good complementary data set.

\subsubsection{\texorpdfstring{H-$\alpha$}{} EM}

With knowledge on the electron temperature, the H-$\alpha$ intensity $I_\mathrm{H-\alpha}$ can be related to the EM via
\begin{equation}
  \EM{} = 2.75\,T_4^{0.9} I_\mathrm{H-\alpha},
  \label{eq:ha_to_em}
\end{equation}
with $T_4$ being the temperature in units of 10000 K  and $I_\mathrm{H-\alpha}$ being measured in Rayleigh \citep{1998Haffner}.
We use the \textit{Planck} temperature map for the $T_4$ factor, noting that for many parts of the sky this map is basically unconstrained and shows the 7000K prior adopted by \textit{Planck}.
While Eq. \eqref{eq:ha_to_em} is a very well established relation, light stemming from H-$\alpha$ emission is obscured by dust, which limits its use as a tracer of the full $\EM{gal}$, especially in the inner parts of the Galaxy.
\citet{2003Dickinson} have developed a method to correct for this effect by utilizing the dust extinction $(E-B)_V$ map provided by \citet{1998Schlegel}.
In that, they calculate the optical depth $\tau$ for H-$\alpha$ via
\begin{equation}
  \label{eq:dust_tau}
  \tau = 2.1 f_D  E\left(B-V\right),
\end{equation}
where $E\left(B-V\right)$ is the extinction between $B$ and $V$ band in magnitudes, the 2.1 factor encompasses the conversion from magnitudes to the natural logarithm and a correction factor for converting $B-V$ to H-$\alpha$ extinctions \citep{2003Finkbeiner}.
The factor $f_D$ is an additional fudge factor which encodes the proportion of dust being in front of behind H-$\alpha$ emitting regions, with 0 indicating no absorption and 1 full dust absorption (i.e. all dust is in front of the emitting gas).
In principle, this factor most likely has a strong all-sky dependence, but \citet{2003Dickinson} have determined $f_D =0.33 \pm 0.1 $ to be a good approximation for most of the sky.
We adopt this value, but note that this a simplification and may bias our results locally (see also Sect. \ref{subsec:model_summary}).
We further mask all sky pixels with $\tau > 1$, in  accordance with \citet{2003Dickinson, 2008Berkhuijsen}, which disregards most of the inner Galactic disk.
This leaves about 96\% of the EM sky constrained by H-$\alpha$ data.

In summary, about 8.9\% of the sky are constrained by both H-$\alpha$
and free-free data and  0.15\% are covered by none of the data sets.
In the former regions, we note a Pearson correlation coefficient of $\approx 0.7$ between both EM data sets.
Calculating the error weighted average of the difference of the two data sets $\left(d_{\EM{H\alpha}}- d_{\EM{ff}}\right)/\left(\sqrt{\sigma^2_{\EM{H\alpha}} + \sigma^2_{\EM{ff}}}\right)$ of about $-1.7$, with a scatter of about $9.5$.
While this demonstrates acceptable correspondence, we note that several pixels show discrepancies over an order of magnitude or more, indicating that some residual biases are still present in the data sets.
We show both the free-free and H-$\alpha$ EMs which we use as input for our model in Fig. \ref{fig:data}, as well as the respective masks.

\section{The models}
\label{sec:models}

In the following, we use Eqs. \eqref{eq:dm_fundamental} \eqref{eq:faraday_fundamental}, \eqref{eq:em_fundamental}, to construct a joint inference model connecting all observables discussed in Sect. \ref{sec:observables} via full sky maps.
We will refer to the Faraday rotation, DM, EM sky maps as observables, as they are directly related to the data sets.
We will develop a model for each sky map, which we annotate with $s_{x}\left(l, b\right)$, where $x \in \left(\DM{gal}, \Bpw{gal}, \phi_{\mathrm{gal}}, \EM{gal}, \Len{DM^2/EM}\right)$.
This notation is used to clearly distinguish between the data sets (annotated similarly with $d_{x}$) and our sky models.
We constrain ourselves to models where all components have a physical interpretation, with potential degeneracies minimized as far as possible.
The fundamental building blocks of these sky models are Gaussian fields (indicated with Greek letters), each of which has an a priori independent unknown homogenous correlation structure that is inferred simultaneously.
These correlation structures are each dependent on a set of hyperparameters, which encode our a priori expectations on the structure, e.g. their expected correlation lengths or the expected range of fluctuations.
These parameters are usually set conservatively, i.e. they will allow for a somewhat larger range of possible field realizations then what is physically plausible.
A more detailed discussion on the correlation model can be found in \citet{2022Arras}.
We discuss and illustrate the prior of each sky map in the Appendix \ref{app:prior}.

The sky models for the observables $\mathcal{O} \in \left(\DM{gal}, \phi_{\mathrm{gal}}, \EM{gal}\right)$ are connected to the data-sets via
\begin{equation}
\label{eq:data}
d_{O} = \mathcal{R}s_{\mathcal{O}}\left(l, b\right) + m_{\mathcal{O}} + n_{\mathcal{O}},
\end{equation}
where $\mathcal{R}$ is a response operator projecting the sky model to data space, $n_{\mathcal{O}}$ is the random observational noise term and $m_{\mathcal{O}}$ contains systematic data biases which are modeled explicitly as far as possible.

\subsection{Galactic Sky models}

We begin with the development of the sky model $s_{\DM{gal}}\left(l, b\right)$, as the  $\mathrm{DM}_\mathrm{gal}$ sky is a component that affects all other observable sky quantities considered here.
We note that it is a strictly positive quantity, with expected all-sky variations of the Galactic DM over several orders of magnitude.
Such a pattern can be naturally modeled via a log-normal model, i.e. we set
\begin{equation}
\label{eq:model_DM}
s_{\DM{gal}}\left(l, b\right) = e^{\rho},
\end{equation}
and assume $\rho$ to be a field with homogeneous and isotropic Gaussian statistics.
We adjust the hyperparameters of $\rho$ such that the value range of $1\,\mathrm{pc}\, \mathrm{cm}^{-3}$ to $10000\, \mathrm{pc}\, \mathrm{cm}^{-3}$ is covered conveniently (see Appendix \ref{app:prior}).

We can then use Eq. \eqref{eq:B_average} to write the sky model for the Galactic Faraday sky as
\begin{align}
\label{eq:model_phi}
s_\phi\left(l, b\right) = 0.812 s_\mathrm{DM}s_{\Bpw{gal}} = 0.812 e^{\rho} \chi
\end{align}
i.e. the model for the magnetic field sky $\chi$ obeys Gaussian prior statistics.
Here, the hyperparameters on $\chi$ are set to easily encompass the typical amplitude of the magnetic field in the Milky Way (i.e. in the order of a few $\mu G$, we again illustrate the prior in the Appendix \ref{app:prior}).
This model is analogous to the model introduced in \citetalias{2019Hutschenreuter} and \citet{2022Hutschenreuter}, with the difference that the degeneracy between amplitude field $e^\rho$ and sign determining field $\chi$ is now broken by the DM data, which constrains Eq. \eqref{eq:model_DM}.

The last observable to connect is the Galactic EM.
Observing Eq. \eqref{eq:em_dm2_ratio}, we propose the following model
\begin{equation}
\label{eq:em_model}
s_{\EM{gal}}\left(l, b\right) =  \frac{s_{\DM{gal}}^2}{s_{\Len{DM^2/EM}}} = e^{2\rho - \psi}
\end{equation}
The newly introduced model $s_{\Len{DM^2/EM}}\left(l, b\right) = e^\psi$ captures the $\Len{DM^2/EM}$ factor used to translate the EM and DM skies.
We choose again a log-normal model, as this factor is strictly positive and most likely varies over orders of magnitude.
The $\Len{DM^2/EM}$ sky is expected to vary between the order of $1$ pc and several $10^5$ pc (i.e. the size of the Milky Way) a priori.
We discuss and illustrate the prior on $\psi$ in more detail in the Appendix \ref{app:prior}.

There are good physical reasons to believe that the true logarithmic DM, $\Bpw{gal}$ and logarithmic $\Len{DM^2/EM}$ maps (which are the sky maps with Gaussian priors in our model) also contain structures which are a priori unlikely in our model, such as e.g. filaments which are seen in Galactic HI \citep{2014Clark} or dust maps \citep{2016Planck_XXXII}.
We emphasize that the assumption of statistically homogenous and isotropic Gaussian fields only constitute an a priori constraint, and the posterior maps may contain such structures if driven by the data.
To our knowledge, there exists no physically informed, efficient and unbiased model for the correlation structure of the full sky, which we could use to improve our prior.

\subsection{Data space models}
The integrals in Eqs. \eqref{eq:dm_fundamental}, \eqref{eq:faraday_fundamental}, and \eqref{eq:em_fundamental} will typically receive their most significant contributions from the Milky Way for most LoS and the observables in question.
Nonetheless, all three data sets are subject to systematic effects, which are in part caused by additional or neglected physical contributions, but also by systematic effects in the data processing.
In the case of Faraday RM and free-free EM data, we use the updated error bars inferred by \citepalias{2019Hutschenreuter} and \citet{2022Hutschenreuter}, which implicitly correct for such effects.
For the pulsar DM and distance data sets used in our work, we develop new models as detailed below.

\subsubsection{Pulsar DM}
As discussed in Sect. \ref{sec:observables}, observational errors of pulsar DMs are generally small and well understood.
The biggest systematic effect of $\DM{P}$ as a tracer of $\DM{gal}$ comes from the fact that most pulsars lie within the Milky Way, i.e. we trace only a limited part of the respective $\Len{gal}$.
In order to model this effect, we introduce an explicit model for the pulsar DM.
This is necessary, as we have no good statistical measure to estimate the Galactic contribution similarly as for the extragalactic Faraday data, which relied on all-sky correlations.

A very simple model for the pulsar DM can be constructed by introducing a factor $w_\mathrm{P}$ for each pulsar,
\begin{equation}
\label{eq:pulsar_dm}
m_\DM{P} = w_\mathrm{P} \mathcal{R}_\mathrm{\DM{P}} s_\DM{gal}\left(l, b\right),
\end{equation}
where $w_\mathrm{P} \leftarrow \mathcal{U}\left(0, 1\right)$, i.e. assuming a uniform prior on the relative position of the pulsar along the LoS, if measured in DM units.
This model does not make any a priori assumptions on the geometry of the Milky Way, which implies that we put equal likelihood on a pulsar tracing any percentage of the LoS, irrespective of their angular position on the sky.
This model has the advantage of being almost entirely data-driven, i.e. it does not require an explicit three-dimensional model of the Milky Way.
One should note at this point, that an implicit assumption of Eq. \eqref{eq:pulsar_dm} is that the pulsars are tracing a significant portion of the LoS, as the uniform weighting disfavors a $\DM{P}$ orders of magnitudes smaller than the respective $\DM{gal}$ on the sky.
Nonetheless, this model still leaves room for some degeneracy, as it might still be possible to consistently decrease $w_\mathrm{P}$ for the pulsars and increase the $\DM{gal}$ and still explain the data with the same explanatory power.
Hence, if constrained with pulsar data alone, the model can only provide a lower limit on $\DM{gal}$ and offset inconsistencies when explaining pulsars with different DMs along the same LoS.
We furthermore do not cover any selection effects, i.e. if certain parts of the Milky Way are not sampled by pulsars, this will lead to systematic effects in our results.
We note that a viable alternative might be to use an existing 3D electron model to construct this prior, but this would lead to the problem to quantify the uncertainty on the DM predicted by these models.
In the specific case of the most recent \citet{2017Yao} electron model (henceforth abbreviated YMW), \citet{2021Ocker} have indicated that the model uncertainties are widely not understood, we have hence opted for our simpler uniform model.
Constructing a better informed prior on the relative pulsar DM and distances would increase the accuracy of our result, but is left to future work.

\subsubsection{Pulsar distances}
\label{subsubsec:model_pulsar_distance}

Due to the relative scarcity of pulsars, we additionally use independently measured distance data that is available for some pulsars (see Sect \ref{subsubsec:DM_data}) to introduce an additional likelihood term.
To this end, we build a model for the path length $L_P$ by translating the predicted DMs for each pulsar to distances by translating Eq. \ref{eq:pulsar_distance} into a model equation:
\begin{equation}
m_{\mathrm{L_P}} = f_{\mathrm{DM}, L}\left(m_\DM{P}\right)
\end{equation}
In this work, we use the YMW model to construct the conversion function $f_{\mathrm{DM}, L}$.
The desired effect of this additional likelihood term is additional constraints on the large scale structure on the sky.
Given the small number of pulsars with independent distance measurements and the typically large error bars of these distances, we do not expect a strong impact for smaller structures.
We have tested the impact of this additional term by running inferences with and without the additional distance data.
We find that including the pulsar distances significantly stabilizes our results, and we hence accept the small downside of including a weak dependence on an explicit electron density model into our inference, especially as the same model is also implicitly present in the error bars of the Faraday data set (see Sect. \ref{subsubsec:faraday_data}).
We refrain from putting additionally a priori constraints on our model.
Any systematic effects resulting from this are discussed in the next section.

\begin{figure*}
\centering
\begin{tikzpicture}[auto]
    % \small
    % node placement with matrix library
  \matrix[ampersand replacement=\&, row sep=0.5cm, column sep=0.5cm] {
    % Place nodes
    \node [block4] (psi) {$s_{\log(\Len{DM^2/EM})}$};
    \&
    \&
    \&
    \&
    \&
    \node [block3] (rho) {$s_{\log(\DM{gal})}$};
    \&
    \&
    \&
    \&
    \&
    \node [block3] (chi) {$s_{\Bpw{gal}}$};
    \\  % newline
    \node [block2] (em) {$s_{\EM{gal}}$};
    \&
    \&
    \&
    \&
    \&
    \node [block3] (dm) {$w_\mathrm{P}, s_{\DM{gal}}$};
    \&
    \&
    \&
    \&
    \&
    \node [block2] (phi) {$s_{\phi_{\mathrm{gal}}}$};
    \\ %newline
    \node [block3] (data_emff) {$d_{\mathrm{EM}, ff}$};
    \&
    \&
    \node [block3] (data_emha) {$d_{\mathrm{EM}, H-\alpha}$};
    \&
    \&
    \node [block3] (data_dmp) {$d_{\mathrm{DM}, P}$};
    \&
    \&
    \node [block3] (data_distp) {$d_{L, P}$};
    \&
    \&
    \&
    \&
    \node [block3] (data_rmegal) {$d_{\mathrm{RM}, egal}$};
    \\
    };
    % Draw edges
    \path [line] (em) -- node{}(data_emff);
    \path [line] (em) -- node{}(data_emha);
    \path [line] (phi) -- node{}(data_rmegal);
    \path [line] (dm) -- node{}(data_dmp);
    \path [line] (dm) -- node{}(data_distp);

    \path [line] (psi) -- node{}(em);
    \path [line] (rho) -- node{}(em);
    \path [line] (rho) -- node{}(dm);
    \path [line] (rho) -- node{}(phi);
    \path [line] (chi) -- node{}(phi);

    % \path [curved line] (alpha_beta) to node{}(eta);

\end{tikzpicture}
\caption{\label{fig:model} Graph illustrating the hierarchical Bayesian model employed in this work.
The sky models in the upper layer correspond to the Gaussian fields $\psi, \rho$ and $\chi$ in that order.
This graph omits several layers to the top, which are similar for each Gaussian field and contain the correlation structure modelling.
For details on those, see the Appendix \ref{app:prior} and \citet{2022Arras}.
}
\end{figure*}
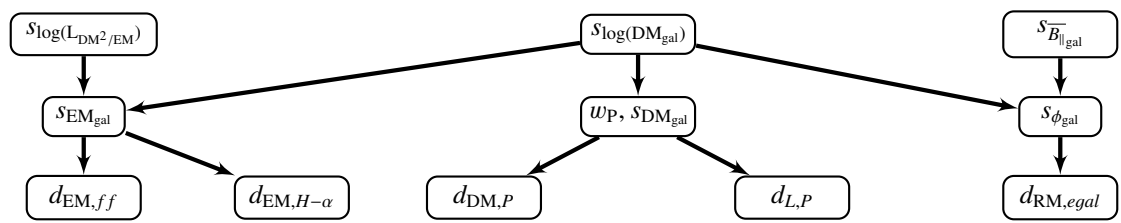

\subsection{Model summary and evaluation}
\label{subsec:model_summary}

The full model is summarized and illustrated in Fig. \ref{fig:model}.
This figure shows that our model consists of three branches stemming from the three data sets.
We hence refer to these as the \textit{RM-}, \textit{DM-}, and \textit{EM-branch} for the rest of the paper.
The construction of the model as well as the interpretation of the results is conditional to several assumptions and unresolved systematic effects, which we summarize below:

\begin{itemize}
\item \textbf{Insufficient volume sampling of $n_\mathrm{e}$ via pulsars} \newline
The volume density of observed pulsars in the Milky Way is unfortunately still very low and far from uniform.
Specifically the inner Galactic disk ($l \approx 0 $) is most likely not sampled at all beyond a certain distance.
If we take the YMW model as a reference, almost all pulsars lie in the nearest half of the Milky Way.
We hence expect our Galactic disk results to underestimate the true $\DM{gal}$ by at least a factor of two.
Following from this, our magnetic field estimates in the disk are most likely overestimated by the same factor, as the extragalactic RM data probes the full disk.
We have refrained from fixing this via e.g. a volume prior on the Milky Way, as this would only obfuscate the issue, while bringing little to no new insights.
Future pulsar surveys are projected to have a very deep luminosity limit in the Milky Way \citep{2008vanLeeuwen}, which will alleviate the issue.
At higher latitudes, the sampling in depth is more uniform, but unfortunately relatively sparse on angular scales (see next point).\newline
\item \textbf{Small scale DM structure} \newline
We have used the YMW electron model to translate the distances of pulsars to DMs, which has helped to stabilize our algorithm.
The same model was also used in \citet{2022Hutschenreuter} as an input to their error correction routine, which we used in this work.
The YMW model is a parametric electron density model, which only considers the largest scales apart from some local structures such as the local bubble.
Combined with the sparsity of pulsars at higher latitudes, we therefore expect that smaller structures at higher latitudes will be insufficiently constrained by pulsar data.
A detailed discussion of the YMW model and its possible shortcomings are given in \citet{2021Ocker}.
We have tested the effects of the sparse spatial and angular sampling of the Milky Way by pulsars in Sect. \ref{app:synthetic_tests}.
These tests demonstrate the insufficient sampling of the disk has precisely the effect discussed above, while in the high latitude regions, the underestimation only appears locally.
\newline
\item \textbf{Calculating EMs from H-$\alpha$}\newline
The correction of the  H-$\alpha$ emission for dust absorption is using a uniform mixing model \citep{2003Dickinson}, which includes a fudge factor $f_D$ to account for the proportion of dust that lies in front of the H-$\alpha$ emitting gas.
We have used $f_D = 0.33 \pm 0.1$ in this work for the full sky, again according to \citet{2003Dickinson}.
It is likely that this estimate is wrong for some LoS.
Additionally, the electron temperature necessary for the conversion is not strongly constrained, which may introduce additional biases.\newline
\item \textbf{Correlation of $n_\mathrm{e}$ and $\Bp{}$}\newline
The interpretation of $\Bpw{gal}$ depends on this relation, as it can both over- or underestimate the more interesting quantity $\LoS{\Bp{}}{gal}$ depending on whether $\Bp{}$ is correlated or anti-correlated with $n_\mathrm{e}$.
As discussed in Sect. \ref{subsubsec:faraday_physics}, there is considerable evidence that this relation holds for large portions of the sky, but it may break down e.g. along spiral arms.
\end{itemize}
To test the impact of these assumptions on our results, we have run two secondary inferences with either the \textit{RM-} or the \textit{EM-branch} left out.
We will refer to these inferences as \textit{EM-DM-} and \textit{RM-DM-run}, respectively.
These inferences have demonstrated that, while the overall scales of the respective inferred sky maps remains similar, the data sets have considerable impact on the small scale structure.

All models (the main one and the two secondary ones) are evaluated using the Geometric Variational Inference algorithm (geoVI) \citep{2021Frank}.
In variational inference (VI), the posterior distribution (which in our case is a highly non-gaussian distribution with several million dimensions) is approximated via a parameterized and analytically traceable distribution.
The idea of VI is then to minimize the forward Kullback-Leibler divergence, a measure for similarity between probability distributions, w.r.t. the parameters of this simpler distribution which yields the optimal VI approximation.
In the particular case of geoVI, this distribution is implicitly defined via a parameterized invertible function that maps from the space where the model parameters are defined into a space where the approximate distribution is a
standard normal distribution.
This function is constructed using the Fisher information metric as a proxy for the curvature of the logarithm of the posterior.
By ensuring that the metric and its first derivatives vanish at a given expansion point, the function is defined to map into a space where the posterior is locally close to a normal distribution.
The coordinates of this expansion point constitute the parameters of the function and its optimal location is found by minimizing the KL.
We note that this algorithm is an update to the Metric Gaussian Variational Inference algorithm \citep{2019Knollmuller} used in \citet{2022Hutschenreuter}.
For a detailed discussion and comparison between the methods as well as comparison to direct posterior evaluation, we refer the reader to \citet{2021Frank}.
Both MGVI and geoVI have been used successfully in the past for reconstructions of the Faraday sky \citep{2019Hutschenreuter,2022Hutschenreuter} and more general astrophysical \citep{2019Leike, 2022Arras, 2023Mertsch, 2023Platz, 2023Westerkamp, 2023Edenhofer} and non-astrophysical \citep{2022Guardiani, 2023Reeb} contexts.
We evaluate the resulting approximated posterior via drawing samples from it and report the corresponding empirical mean and standard deviations.

\begin{figure*}
  \includegraphics[width=\textwidth, draft=false]{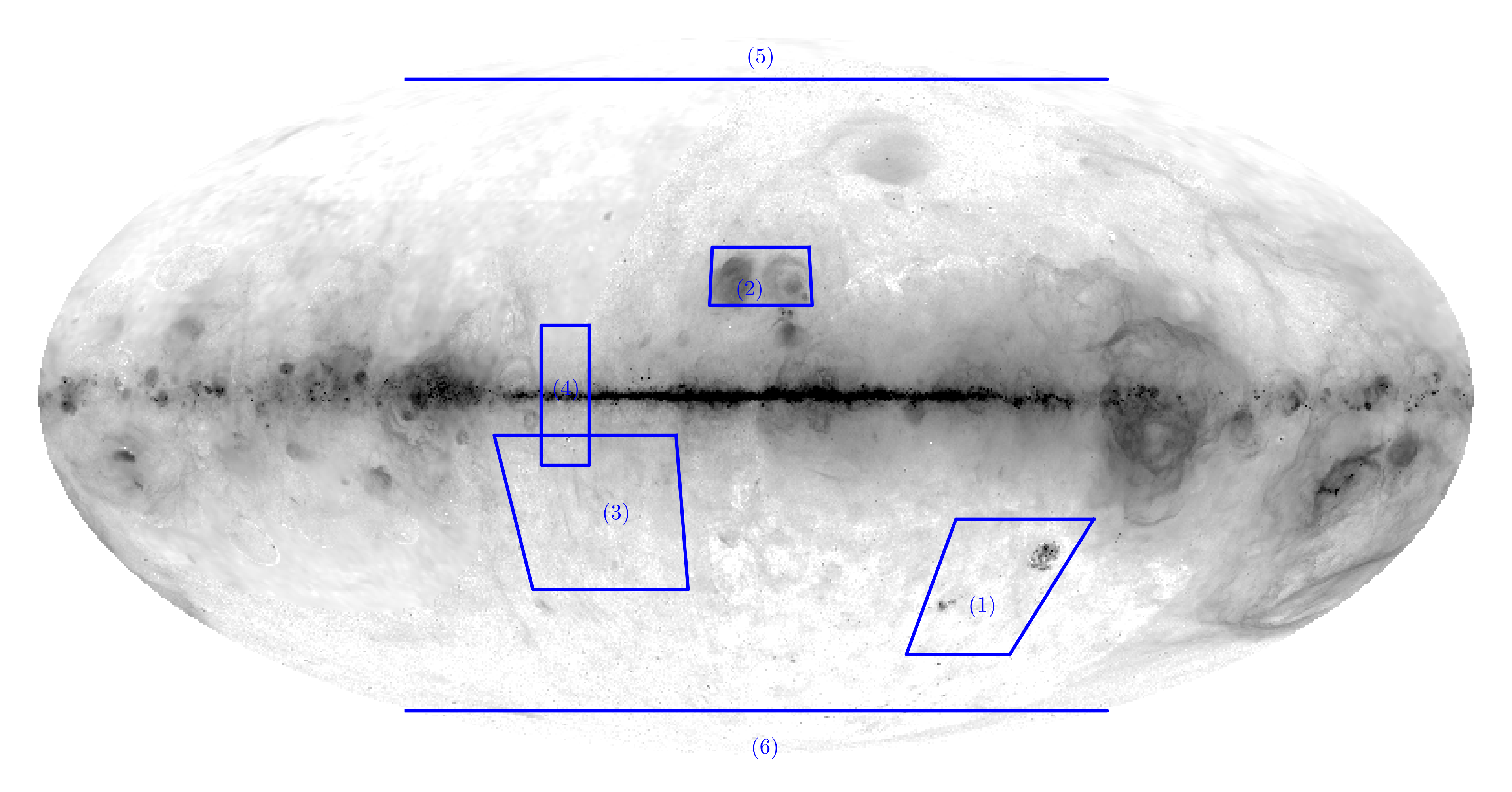}
  \caption{\label{fig:regions} Specific sky regions which are discussed in more detail in the text. They correspond to (1) the Magellanic clouds, (2) the HII regions S2-27 and S2-7, (3) the Smith high velocity cloud, (4) the Sagittarius arm region, and the North (5) and South (6) Galactic poles.
  We have used the logarithmic H-$\alpha$ EM data as the background.}
  \end{figure*}
\section{The results}
\label{sec:results}

In the following, we illustrate our results and compare them to existing work.
All results have been inferred using the full model illustrated in Fig. \ref{fig:model}, unless stated otherwise.
All sky regions explicitly mentioned in the discussion are marked in Fig. \ref{fig:regions}.
We show the full sky posterior mean maps of the $\Bpw{gal}$ (Fig. \ref{fig:Bpar}), $\DM{gal}$ (Fig. \ref{fig:dm}), $\Len{DM^2/EM}$ and $\varrho_{\mathrm{EM/DM}}$ (Fig. \ref{fig:distance}) skies.
We discuss and illustrate our results on the $\Len{DM^2/EM}$ sky using a simplified model of $n_\mathrm{e, gal}$ in Fig. \ref{fig:em_dm_illustration}.
The power spectra of the three main sky maps are shown in Fig. \ref{fig:power}.
We discuss two smaller sky regions (see Figs. \ref{fig:sh2_27} and \ref{fig:smith}) in Sect. \ref{subsec:results_cutout} in more detail to compare our results to existing works.
We further correlate the $w_\mathrm{P}$ factor with different external data sets in Sect. \ref{subsec:correlations}.
The Faraday and EM sky maps (\ref{fig:em_and_phi}) are discussed in the Appendix \ref{app:results_rm_em}.
We summarize the results of the secondary models in the Appendix \ref{app:results_sky_secondary}.
The logarithmic $\DM{gal}$ skies for the two secondary models are shown in Fig. \ref{fig:dm_compare}, while the $\Bpw{gal}$ map from the \textit{RM-DM-run} and the $\Len{DM^2/EM}$ map from the \textit{EM-DM-run} are shown in Fig. \ref{fig:bpar_filling_secondary}.
The error bars quoted in the text are derived from posterior samples.
Given the approximations made in the variational inference, these errors have to be considered a lower limit.
The results (i.e. all sky maps and power spectra from all models, including corresponding uncertainties) are accessible via \href{https://zenodo.org/records/10523170}{Zenodo}.

\subsection{Sky maps}
\label{subsec:results_sky}
\subsubsection{Magnetic field sky}
\label{subsec:results_sky_magnetic}

\begin{figure*}
\includegraphics[width=\textwidth, draft=false]{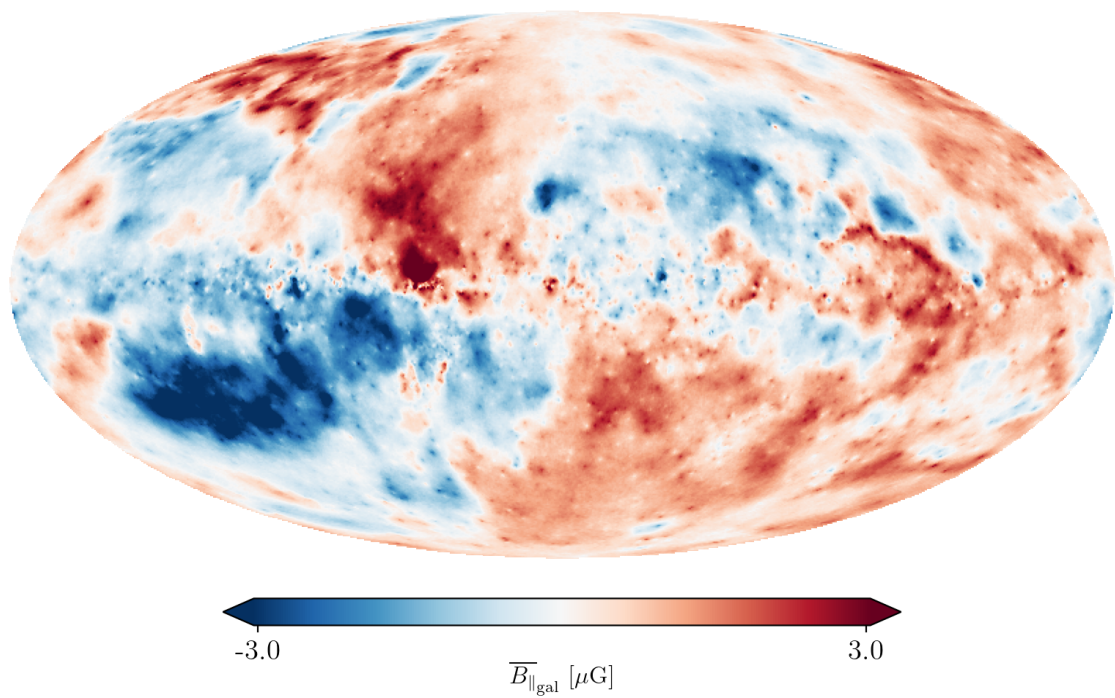}
\caption{\label{fig:Bpar}  The posterior mean for the Los averaged parallel magnetic field of the Milky Way.
The color scale is saturated at $\pm 3.\, \mu\mathrm{G}$.}
\end{figure*}

The posterior mean of the LoS averaged LoS-parallel and electron density weighted magnetic field sky is shown in Fig. \ref{fig:Bpar}.
We record a root-mean-square (RMS) magnetic field value of $\LoS{\Bp{}}{gal, rms} = 1.1 \pm 0.0039 \, \mu\mathrm{G}$ averaged over the full sky.
The map shows several distinct regions at higher latitudes with coherent LoS magnetic field strengths in the order of $2\, \mu\mathrm{G}$.
The sign pattern at high latitudes has long been studied (albeit in the RM sky) to find evidence for an axi-symmetric spiral or bi-symmetric spiral, which would point to evidence for the existence of a Galactic dynamo, see \citet{2022Dickey} for recent results in this regard and \citet{2022Brandenburg} for a review.
Our results allow for a much clearer fit of 3D magnetic field models, since they also allow a fit of the amplitude.
We will attempt such a fit in future work.
These reported magnetic field strengths are consistent with measurements of the large scale magnetic field \citep{2015Haverkorn}.
We further validate our results for specific regions in the sky in Sect. \ref{subsec:results_cutout}.

The inner disk (i.e. the area within $b=\pm\ang{5}$ over the full longitude range) shows a similar RMS of $1.0 \pm 0.0053\,\mu\mathrm{G}$, but shows rather small scaled structures, which is consistent with a picture of many field reversals, that also lead to a suppression of the LoS averaged disk field LoS component compared to the in-situ values of the LoS component at typical disk locations.
We note that the averaged LoS field strengths reported in the disk are most likely an over-estimate of more than a factor of 2, due to the fact that we do not probe the full Galaxy in DM with pulsars and hence most likely under-estimate $\DM{gal}$ (see discussions in Sect. \ref{subsubsec:DM_data} and \ref{subsec:results_sky_dm} and also Fig. \ref{fig:dm}).
There are prominent outliers in the disk, which contain the highest magnetic field strengths, found at $(l, b) \approx (\ang{48.2}, \ang{4.3})$ with  $5.4 \pm 0.13 \,\mu\mathrm{G}$, and also  at $(l, b) \approx (\ang{87.0},  \ang{-0.3})$ with $ -4.5 \pm 0.13 \, \mu\mathrm{G}$.
The former excess is a consequence of the extremely strong RMs reported in this area \citep{2019THOR}, which lead to a similar excess in the Faraday skies of \citet{2022Hutschenreuter}.
The LoS associated with the former region corresponds roughly to the tangent of the Sagittarius arm \citep{2018Vallee}.
The reported $\Bpw{gal}$ excess might then be a result of the large-scale, uniform magnetic field aligned with the spiral arm in this region, as already indicated in \citet{2019THOR}, and supported by simulations \citep{2020Reissl}.
We note that the $\Len{DM^2/EM}$ sky in Fig. \ref{fig:distance} also indicates a high amount of ionized material in this region, see discussions in Sect. \ref{subsec:results_sky_filling} and Sect. \ref{subsubsec:results_cutout_other}.

\subsubsection{DM sky}
\label{subsec:results_sky_dm}

\begin{figure*}
\centering
\begin{subfigure}{0.95\textwidth}
\includegraphics[width=\textwidth, draft=False]{\skypath{\rmdmem}{DM_mean.png}}
\caption{\label{fig:dm_linear}}
\end{subfigure}
\begin{subfigure}{0.95\textwidth}
\includegraphics[width=\textwidth, draft=False]{\skypath{\rmdmem}{log10_DM_mean.png}}
\caption{\label{fig:dm_log}}
\end{subfigure}
\caption{\label{fig:dm}
Inference results for the Galactic DM sky, constrained by pulsar DM, free-free and H$\alpha$ EM and extra-galactic Faraday data.
Fig. (a) shows the posterior mean,  Fig. (b) the same result in logarithmic scaling.
The color maps are saturated at $500\, \mathrm{pc}\,\mathrm{cm}^{-3}$ and between $10\, \mathrm{pc} \,\mathrm{cm}^{-3}$ and $1000\, \mathrm{pc} \,\mathrm{cm}^{-3}$, respectively.}
\end{figure*}

In Figs. \ref{fig:dm} we show the results for the Galactic DM sky on both linear and logarithmic scaling.
We find maximum DM values along lines of sight at $b = 0$ and near the GC at $l = 0$, reaching  $1.7 \times 10^3\pm 0.26 \, \mathrm{pc}\,\mathrm{cm}^{-3}$.
We find the minimum at higher latitudes, about $4.0 \pm 0.26\, \mathrm{pc}\,\mathrm{cm}^{-3}$ at (l $\approx \ang{64.2}, b \approx \ang{68.5}$).
Most of the inner galactic disk exhibits values above $1000\, \mathrm{pc}\,\mathrm{cm}^{-3}$.
The values in the disk are relatively low compared to e.g. the disk predicted by \citet{2017Yao} or \citet{2001Cordes}, which each predict DMs above $3000 \,\mathrm{pc}\,\mathrm{cm}^{-3}$ in the inner disk, with the Galactic center reaching up to $4000\, \mathrm{pc}\,\mathrm{cm}^{-3}$.
In \citet{2021Price}, the typical fractional relative error on the distances predicted by the YMW electron model is estimated to lie between $0.35-2.76$.
Assuming that the DM error for the full Galaxy is in the same order of magnitude, our results would be narrowly compatible.
But since pulsars with independent distances indicate that we do not probe the full Galaxy, it is highly likely that a significant tension to our results remains.
We attribute this discrepancy to the insufficient volume sampling of pulsars in the Galactic disk and consider our results to be an under-estimate of about a factor of 2 in the inner disk region, in line with the discussion in Sect. \ref{subsec:model_summary} and our tests in the Appendix \ref{app:synthetic_tests}.

Towards the Galactic poles, we can compare our results not only to the full Galaxy electron models of \citet{2017Yao} or \citet{2001Cordes}, but also to local models of the Galactic disk \citep{2008Gaensler, 2020Ocker}.
DMs calculated from these models in the Galactic pole regions are often reported as $\DM{\perp} = \DM{}|\sin(b)|$, i.e. the DM perpendicular to the Galactic plane, as in this projection the DM only depends on the Galactic scale height of the models.
These models predict $\DM{gal, \perp}$ values between $18.9 \pm 0.9\, \mathrm{pc}\,\mathrm{cm}^{-3}$ \citep{2017Yao} and  $25.6 \pm 2.6\, \mathrm{pc}\,\mathrm{cm}^{-3}$ \citep{2008Gaensler}.
We calculate mean values of $\DM{\perp}$ from our maps in the Galactic North and South Pole regions above and below $\pm \ang{70}$ to compare to these results.
We report $12 \pm 0.15 \, \mathrm{pc}\,\mathrm{cm}^{-3}$ (north) and  $13 \pm 0.29 \,  \mathrm{pc}\,\mathrm{cm}^{-3}$ (south) and note maximum values of $26 \pm 0.28 \mathrm{pc}\,\mathrm{cm}^{-3}$  and  $25 \pm 0.82 \mathrm{pc}\,\mathrm{cm}^{-3}$ in the same respective regions.
While the maximum values agree well with the other models, the mean values are considerably lower.
Under the presumption that these numbers are straightforwardly comparable with the aforementioned model predictions, this would again indicate a slight underestimation of DM in our results, which might be explainable by the sparse sampling by pulsars at high latitudes, as indicated by our tests with synthetic data, in Appendix \ref{app:synthetic_tests}.
Additionally, since the other models do not incorporate the small scaled structure as we do, it is possible that some subregions in the polar regions have a smaller DM as predicted by the parametric models, which might be dominated by the pulsars with larger DM.

Another explanation can be motivated by comparing with the logarithmic DM maps from the \textit{RM-DM} and \textit{EM-DM run} in
Fig. \ref{fig:dm_compare}.
In the \textit{EM-DM run}, we calculate mean values for $\DM{\perp}$ of $\num{1.83E+01}\, \mathrm{pc}\,\mathrm{cm}^{-3}$ (north) and $\num{1.87E+01}\, \mathrm{pc}\,\mathrm{cm}^{-3}$ (south), which are much closer to the predictions of the parametric electron models discussed before.
Visual comparison reveals that there are regions of very low DM in both polar regions, which are not present in the \textit{EM-DM run}, indicating that these structures are driven by the RM data.
These regions correspond to areas of increased magnetic field strength in Fig. \ref{fig:Bpar} and \ref{fig:bpar_phidm}.
The RM data seems to be better explained by an increase in $\Bpw{gal}$ and the EM data does not show such a correlation, hence the DM maps are not affected in the \textit{EM-DM run}.
If our DM result in the polar regions is indeed underestimated, this may point to a significant  $\Bp{} - n_\mathrm{e}$ correlation, as the RM amplitude seems to be clearly driven by $\Bpw{gal}$.

\subsubsection{The EM-DM conversion factors}
\label{subsec:results_sky_filling}

\begin{figure*}
\centering
\begin{subfigure}{0.99\textwidth}
\includegraphics[width=\textwidth, draft=False]{\skypath{\rmdmem} log10_EMoverDM_mean.png}
\caption{\label{fig:log_EMoverDM}}
\end{subfigure}
\begin{subfigure}{0.99\textwidth}
\includegraphics[width=\textwidth, draft=False]{\skypath{\rmdmem} log10_distance_mean.png}
\caption{\label{fig:log_distance}}
\end{subfigure}
\caption{\label{fig:distance}
Inference results for the logarithmic $\varrho_{\mathrm{EM}/\mathrm{DM}}$ (Fig. a)) and $\mathrm{L}_\mathrm{\mathrm{DM^2}/\mathrm{EM}}$  (Fig. b)) skies.
Both quantities highlight different aspects of the ionized ISM, see Eqs. \eqref{eq:em_dm2_ratio} and \eqref{eq:em_dm_ratio} for their definitions and Sect. \ref{subsubsec:emission measure - physics} for the accompanying discussion.
The $\mathrm{L}_\mathrm{\mathrm{DM^2}/\mathrm{EM}}$ sky map is directly modelled in our work, while the $\varrho_{\mathrm{EM}/\mathrm{DM}}$ map is calculated in a post-processing step.}
\end{figure*}

In Fig. \ref{fig:distance}, we show the posterior means of $\varrho_{\mathrm{EM}/\mathrm{DM}}$ and $\Len{DM^2/EM}$ over the sky, as defined in Eqs. \eqref{eq:em_dm2_ratio} and \eqref{eq:em_dm_ratio}.
While the latter map is a direct result of the inference, the former was calculated in a post-processing step from posterior samples.
As detailed in Sect. \ref{subsubsec:emission measure - physics}, both quantities illustrate different aspects of the structure of the Galactic electron density, with $\varrho_{\mathrm{EM}/\mathrm{DM}}$ being related to the average electron density within the emitting plasma and $\Len{DM^2/EM}$  to its volume.

The $\varrho_{\mathrm{EM}/\mathrm{DM}}$ sky shows variations between $\num{9.6e-3} \pm \num{1.3e-3} \, \mathrm{cm}^{-3}$ to $17 \pm \num{1.3e-3} \, \mathrm{cm}^{-3}$.
The predominantly blue areas in this map show values around  $0.1 \mathrm{cm}^{-3}$.
This is consistent with the typical value assumed for the warm ionized ISM \citep{2020Ferriere}, which indicates that the electron density is rather uniform (i.e. implying $\sigma_{n_\mathrm{e}} \lesssim \Len{n_\mathrm{e}}$ in Eq. \eqref{eq:em_dm_ratio}) along these LoS.
Regions with $\varrho_{\mathrm{EM}/\mathrm{DM}} \gg 0.1 \mathrm{cm}^{-3}$ are most likely dominated by strong fluctuations in electron density.
Given that the aforementioned consistency of $\varrho_{\mathrm{EM}/\mathrm{DM}}$ with the literature value for $n_\mathrm{e}$, we regard the former option as more likely in most cases.

Regarding the  $\Len{DM^2/EM}$ map, we note that the map varies over orders of magnitude, from $21 \pm 1.1$ pc to $\num{2500} \pm 1.1$ pc.
Qualitatively, it appears to be a reliable tracer of turbulent structures on the sky, as e.g. the inner Galactic disk, shock structures and known HII regions all stand out with a small  $\Len{DM^2/EM}$.
We note that $\Len{DM^2/EM}$ in the thin inner disk is most likely overestimated by the same factor as the magnetic field estimate, as it is similarly indirectly affected by under-sampling of pulsars in the Galactic plane, as discussed previously.
On the other hand, several regions stand out with a large $\Len{DM^2/EM}$.
These are most notably the Sagittarius arms region, the halos of the Magellanic clouds, and the tail of the Smith high velocity cloud (HVC), all marked in Fig. \ref{fig:regions}.
It is notable that these regions are less discernible in the $\varrho_{\mathrm{EM}/\mathrm{DM}}$ map, indicating that the high $\Len{DM^2/EM}$ values trace regions with similar density but trace a larger volume than other regions.

\begin{figure*}
\includegraphics[width=\textwidth, draft=False]{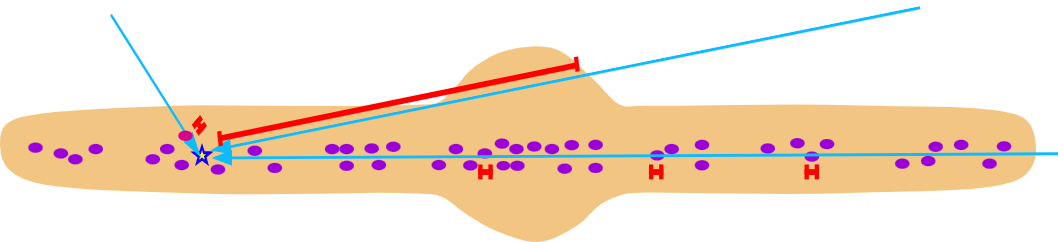}
  \caption{\label{fig:em_dm_illustration} Sketch of the geometry proposed to explain the appearance of the $\Len{DM^2/EM}$ sky in Fig. ~\ref{fig:distance}.
  In a low electron density hot ISM (light orange) small sized but dense electron clouds are embedded (deep purple ellipses).
  The observer's position is marked by the blue star symbol.
  A LoS at high galactic latitudes (blue arrow from top right) usually misses the dense clouds and therefore probes the geometric length of the dilute ISM (long red bar) resulting in a $\Len{DM^2/EM} \sim \mathcal{O}(\mathrm{kpc})$.
  A LoS in the Galactic plane (horizontal blue arrow) usually passes through at least one of such clouds and therefore probes their much smaller size (short red bars), leading to $\Len{DM^2/EM} \sim \mathcal{O}(10 \mathrm{pc})$.
  Some high latitude LoS pass through such clouds as well (blue arrow from top left) and exhibit such very short $\Len{DM^2/EM}$ as well (see short red bar next to that LOS).
  As high latitude clouds need to be nearby to be within the thin Galactic disk, the set of LoS's passing through them span a larger area of the high latitude sky.
  We note that the scales or shape of this illustration are not intended to realistically match the Milky Way.}
\end{figure*}

In order to illustrate these interpretations, we show a simplified model of the ionized ISM in Fig. \ref{fig:em_dm_illustration}.
In there, we assume the ISM to mainly consist of a rather uniform low density plasma, in which small ionized regions of high $n_\mathrm{e}$ are embedded (e.g. HII clouds).
All LoS going through these small regions will have a small $\Len{DM^2/EM}$, as the density along these LoS shows large fluctuations.
Assuming that the clouds have similar density, $\Len{DM^2/EM}$  will be equal to the portion of the LoS within them.
Since most of these embedded regions are located in the thin Galactic disk,
Some regions in the vicinity of Earth with a notable vertical offset will appear as extended regions at higher latitudes, again with a small $\Len{DM^2/EM}$.
Most high latitude LoS, however, will not go through such regions and hence trace the more uniform low density plasma.
The $\Len{DM^2/EM}$ values will then agree with the physical length of the LoS.
The sketch also illustrates that $\Len{DM^2/EM}$ values of LoS not hitting high density regions may still vary strongly by geometric effects, i.e. by probing large structures outside (and not necessarily connected to) the thin disk. This may explain the large $\Len{DM^2/EM}$ values towards the regions mentioned earlier.
We devote a separate section (Sect. \ref{subsec:results_cutout}) to discuss effects related to such specific structures in more detail.

\subsection{Power spectra}
\label{subsec:results_power}

\begin{figure*}
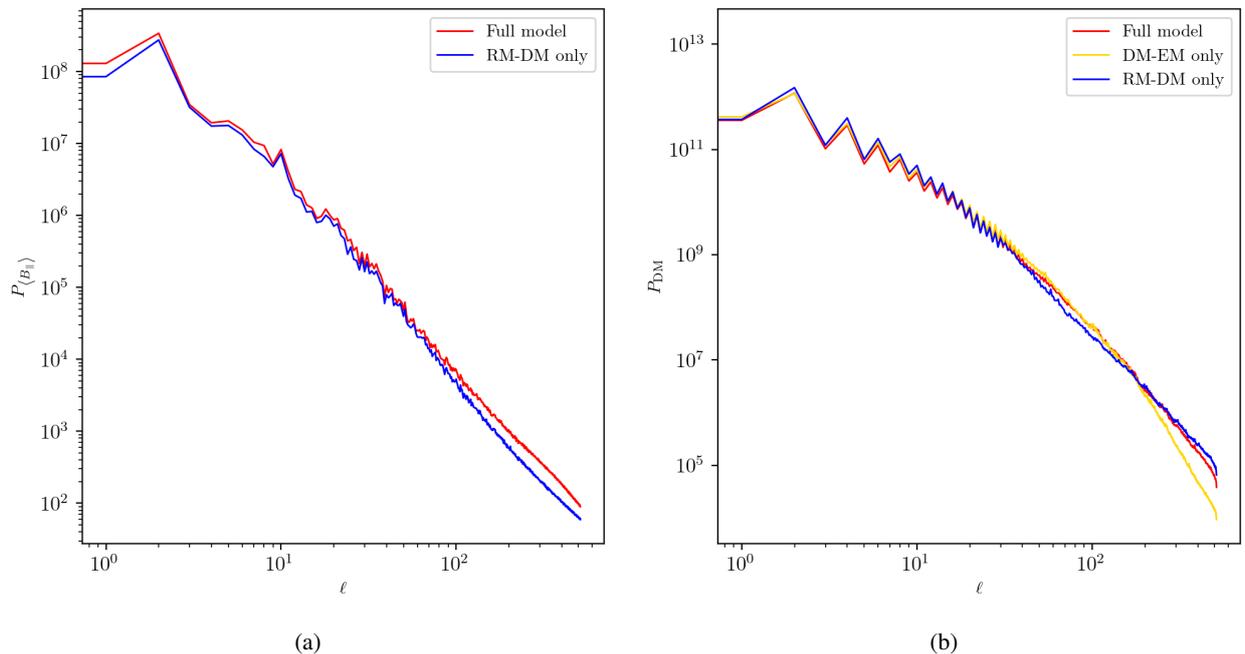

\centering
\begin{subfigure}{0.45\textwidth}
\includegraphics[width=\textwidth, draft=False]{\powerpath{Bpar}}
\caption{\label{fig:power_bpar}}
\end{subfigure}
\begin{subfigure}{0.45\textwidth}
\includegraphics[width=\textwidth, draft=False]{\powerpath{DM}}
\caption{\label{fig:power_dm}}
\end{subfigure}

\caption{\label{fig:power} Inferred angular power spectra of the $\Bpw{gal}$ (Fig. (a)) and  $\DM{gal}$ (Fig. (b))) sky maps.
The colors indicate the different runs from which the spectra where calculated, for details see Sect. \ref{subsec:model_summary}.}
\end{figure*}

We calculate the angular power spectra of the $\DM{gal}$ and $\Bpw{gal}$ skies in order to describe the statistical properties of our results and to compare between the full model and the two secondary models.
The results are shown in Fig. \ref{fig:power}.
We note that the spectra of the $\DM{gal}$ skies do not capture their full statistics, due to the non-linear nature of the $\DM{gal}$ sky models.
We have nonetheless opted to show the $\DM{gal}$ power spectra instead of the Gaussian logarithmic $\DM{gal}$, as $\DM{gal}$ is much more likely to be reported by both simulations and observations.
In order to represent the power spectra concisely, we additionally fit the following parametric model as a function of multipole $\ell$

\begin{equation}
\label{eq:power}
C_{\ell, fit} = \frac{A}{\left(\ell_0 + \ell\right)^{-s}},
\end{equation}
to the posterior power spectra with free parameters $A$, $s$ and $\ell_0$, using the maximum a posteriori (MAP) method.
We report on the fitted values in Table \ref{tab:power}.
Both the figures and the fits demonstrate that the power spectra show little variance between the different models.
The most interesting parameter is the spectral slope parameter $s$, as it gives information on the turbulent scaling of the underlying 3D quantities, i.e. the magnetic field and the electron density.
\citet{2010Chepurnov} give an analytic formula to relate the power spectrum of an 3D isotropic and homogenous field to an integrated 2D angular spectrum, which essentially demonstrates that the spectral slope of a simple power law spectrum remains the same when projected on the sky under these idealized conditions and for $\ell > 15$.

The Galactic disk has most likely different statistical properties than the higher latitude regions, which breaks these assumptions.
Due to the relatively small area occupied by the disk, the spectral slope inferred here will be dominated by these higher latitude regions.
The values found for $s$ in the DM case are all somewhat close to $11/3 = 3.67$, indicating Kolmogorov turbulence in the fluctuations of the electron density distribution.
This is consistent with the `Big power-law in the sky', i.e. other tracers of the spectral slope of the nearby $n_\mathrm{e}$ which have shown its Kolmogorov-like behavior over many orders of magnitude \citep{1995Armstrong, 2020Ferriere}.
The spectral slope of the $\Bpw{gal}$ sky is generally flatter, which might be a result of magnetic field reversals, which introduce sharp edges in the map.
Another possible explanation might be a spectral flattening of the 3d spectrum of the magnetic field, marking the injection scale of the Kolmogorov spectrum.
Such an effect has been observed by \citet{1996Minter} for a high latitude region using point source RM data and H$\alpha$ derived EM data.
The flatter-than-Kolmogorov angular spectrum observed in our case may then stem from an averaging effect over different spatial scales of this `broken' power law.
While this is a plausible explanation, it should be noted that the electron density power spectrum is also conjectured to flatten above a certain scale, albeit with some controversy about the actual value (most likely between 3 and 100 pc \citep{2020Ferriere}).
The fact that we do observe a Kolmogorov spectrum for the $\DM{gal}$ but not for the $\Bpw{gal}$ might then point to a larger injection scale for turbulence in the electrons, which decreases the impact of the flatter parts on the angular spectrum.
We note that we do not directly observe a spectral break in any of the angular power spectra, which can be explained by the fact that different 3D scales are averaged to the same angular scales, thereby smoothing out any breaking scale.

\begin{table*}[htbp]
  % \small
  \centering
\begin{tabular}{l | c c c |  c c c  }
       &\multicolumn{3}{c|}{$\Bpw{gal}$}  & \multicolumn{3}{c}{$\DM{gal}$}   \\
       \hline
       & $A$ & $s$ & $\ell_0$ & $A$ & $s$ & $\ell_0$  \\
       \hline
    Full Model &  $\num{5.00e9}$ & 2.87 & 1.34 & $\num{2.20e14}$ & 3.41 & 2.37 \\
    RM-DM & $\num{4.56e9}$ & 2.95 & 1.65 & $\num{1.82e14}$ & 3.39 & 2.40 \\
    DM-EM & \multicolumn{3}{c|}{NA} & $\num{1.39e15}$ & 3.81 & 3.92 \\
\end{tabular}
\caption{Results of the MAP fit of Eq. \eqref{eq:power} to the power spectra of the $\Bpw{gal}$ and $\DM{gal}$ sky maps.}
\label{tab:power}
\end{table*}

\subsection{Specific objects on the sky}
\label{subsec:results_cutout}

The physical assumptions of our model laid out in Sect. \ref{subsec:model_summary} are based on our general understanding of the diffuse ionized ISM.
To compare our results, we have selected regions which have already been subject to similar analyses to infer $\Bp{}$ from RM and DM or EM measurements, and hence provide an ideal testing ground.
In the following, we exemplarily analyze two regions on the sky, corresponding to two different physical environments at intermediate latitudes, namely the HII region Sh2-27 and the Smith high velocity cloud (HVC).
An additional incentive to choose these regions comes from their apparent extreme values in $\Len{DM^2/EM}$ visible in Fig. \ref{fig:distance}, indicating very different physical environments.
In the end of the section, we also comment on the Magellanic clouds and the Sagittarius region in the Milky Way.

\subsubsection{Sh 2-27}
\label{sububsec:results_cutout_sh2_27}

Sh 2–27 is a prominent HII region centered approximately at $(l, b) = (\ang{6}, \ang{23})$ and is easily discernible as a distinct region of negative RM above the Galactic center in the Faraday map (Fig. \ref{fig:phi}).
We show cutouts of the $\Len{DM^2/EM}$, $\Bpw{gal}$ and $\varrho_{\mathrm{EM/DM}}$ skies corresponding to this region in Fig. \ref{fig:sh2_27}.
Fig. \ref{fig:sh2_27_distance} seems to reliably trace the region as a compact object of low $\Len{DM^2/EM}$, indicating a high level of turbulence.
We find a minimum value of $22 \pm 1$ pc for $\Len{DM^2/EM}$, which is much lower than the immediate surroundings, which show values well above 100 pc, but is consistent with the size of the HII cloud of 35 pc at distance of 160 pc \citep{2019Thomson}.

Within the area of the cloud, we find minimum magnetic field strengths in Sh 2-27 of $ -4. \pm 0.16 \,\mu G$ (Fig. \ref{fig:sh2_27_bpar}) and a rms of 2.12 $\pm$ 0.048 $\,\mu G$.
For $\varrho_{\mathrm{EM/DM}}$, we find values up to  3.5 $ \mathrm{cm}^{-3}$ (Fig. \ref{fig:sh2_27_rho}).
The mean DM is $88.4 \pm  1.61\, \mathrm{pc}\,\mathrm{cm}^{-3}$, with a maximum of $120 \pm 4.7 \, \mathrm{pc}\,\mathrm{cm}^{-3}$.
We find 6 pulsars within or very close by to the boundary of Sh 2-27.
Comparison of their DM with the full DM sky reveals that four of them seem to have similar or higher DM as Sh 2-27, indicating that they lie most likely behind the region.
For one of them (PSR J1643-1224), this is confirmed by independent distance measurement, as already found by \citet{2011Harvey-Smith}.
In order to calculate the average magnetic field strength within the HII cloud $\LoS{\Bpw{}}{Sh2-27}$, one has to subtract any potential fore- and background $\LoS{\Bp{}}{fg}$ which impacts the LoS average.
Indeed, \citet{2019Thomson} have found considerable magnetized foreground structure in diffuse RM in the LoS towards Sh 2-27, which they attribute to the Local Bubble and several ionized dust clouds.
In order to provide a rough estimate for $\LoS{\Bp{}}{Sh2-27}$, we assume that $\LoS{\Bp{}}{fg}$ is similar in the close environment to the HII clouds, which is dominated by a large diffuse region with about $0.5\, \mu$G.
If we additionally assume no correlation between $n_\mathrm{e}$ and $\Bp{}$, the aforementioned size of the HII cloud of 35 pc at distance of 160 pc \citep{2019Thomson}, and the mean value for $\LoS{\Bp{}}{gal} = 2\, \mu$G in the region, a back-of-the-envelope calculation gives
\begin{equation}
  \label{eq:sh227_mag_estimate}
  \LoS{\Bp{}}{Sh2-27} = \frac{\Len{gal}}{\Len{Sh2-27}}\left(\LoS{\Bp{}}{gal} - \LoS{\Bp{}}{fg} \right) + \LoS{\Bp{}}{fg} \approx -11 \mu G
\end{equation}

Repeating the same exercise as above for $\varrho_{\mathrm{EM/DM}}$ gives $\LoS{n_\mathrm{e}}{Sh2-27} \approx 7\, \mathrm{cm^{-3}}$,

We can compare these value to previous analyses of \citet{2011Harvey-Smith} and \citet{2022Nergis}.
The former reference uses 57 extragalactic RM's to constrain the magnetic field strength and find a median value of $-6.1 \pm 2.8\, \mu G$.
Similarly, \citet{2022Nergis}, have analyzed the region using diffuse polarized radio emission and find a median magnetic field strength of $-4.1 \pm 0.1 \mu$G, with a maximum near -9 $\mu$G.
Both works use H-$\alpha$ EMs to constrain the electron density and assume Sh 2-27 to have close to spherical geometry, with a LoS path length of about $35$ pc, and a filling factor $f_{\EM{}, \DM{}^2}$ of 0.2.
We show both the magnetic field estimate and projected geometry of the HII region from \citet{2022Nergis} in Fig. \ref{fig:sh2_27_raycheva}.
Regarding the electron density, \citet{2011Harvey-Smith} and \citet{2022Nergis} find $10.6\, \mathrm{cm^{-3}}$ and $8.2\, \mathrm{cm^{-3}}$, respectively, which is higher than our estimate.
It should be mentioned that in the latter reference, if the elliptical model used in this work is aligned with the LoS, the predicted average electron density lowers to $7.3\, \mathrm{cm^{-3}}$,

It is possible that the difference in the magnetic field and electron density estimates can be attributed to our very simple foreground estimation.
There are also some notable differences in modelling, as both \citet{2011Harvey-Smith} and \citet{2022Nergis}  assume all dust to lie in front of the HII region (i.e. assuming $f_D=1$ in Eq. \eqref{eq:dust_tau} in contrast to 0.33 used in this work), which increases their estimate of the EM of Sh 2-27 by a factor 2-5 compared to our work, depending on the amount of observed dust.
This implies an increase in their estimate of the electron density in the cloud, and in turn decreases their magnetic field estimate compared to our work.
We note that the Planck mission also predicts in the order of 300 pc $\mathrm{cm}^{-6}$ for the EM, in line with the $360 \pm 29 \,\mathrm{pc}\,\mathrm{cm}^{-6}$ found in this work.
However, we note that the low Planck EM may suffer from the same bias stemming from the treatment of dust emission in the astrophysical component separation of the microwave sky.

\begin{figure*}
\centering
\begin{subfigure}{0.45\textwidth}
\includegraphics[width=\textwidth, draft=False]{\cutoutpath{Sh227}{EMDM2}}
\caption{\label{fig:sh2_27_distance}}
\end{subfigure}
\begin{subfigure}{0.45\textwidth}
\includegraphics[width=\textwidth, draft=False]{\cutoutpath{Sh227}{B}}
\caption{\label{fig:sh2_27_bpar}}
\end{subfigure}
\begin{subfigure}{0.45\textwidth}
\includegraphics[width=\textwidth, draft=False]{\cutoutpath{Sh227}{rho}}
\caption{\label{fig:sh2_27_rho}}
\end{subfigure}
\begin{subfigure}{0.45\textwidth}
\includegraphics[width=\textwidth, draft=False]{\cutoutpath{Sh227}{Bnergis}}
\caption{\label{fig:sh2_27_raycheva}}
\end{subfigure}

\caption{\label{fig:sh2_27} Cutouts of several sky maps towards the HII region Sh 2-27.
The images show a) $\Len{DM^2/EM}$, b) $\Bpw{gal}$, c) $\varrho_{EM/DM}$, and d) the foreground reduced $\Bp{gal}$ image from \citet{2022Nergis}, which is derived from diffuse polarized radio emission.
The contours of the latter image are plotted over the first three images, with contour lines set at -2 $\mu$G (light-blue) and -6 $\mu$G (dark-blue).
The elliptic boundary in d) indicates the projected boundaries of Sh2-27 assumed in \citet{2022Nergis}.}
\end{figure*}

\subsubsection{Smith cloud}
\label{sububsec:results_cutout_smith}

\begin{figure*}
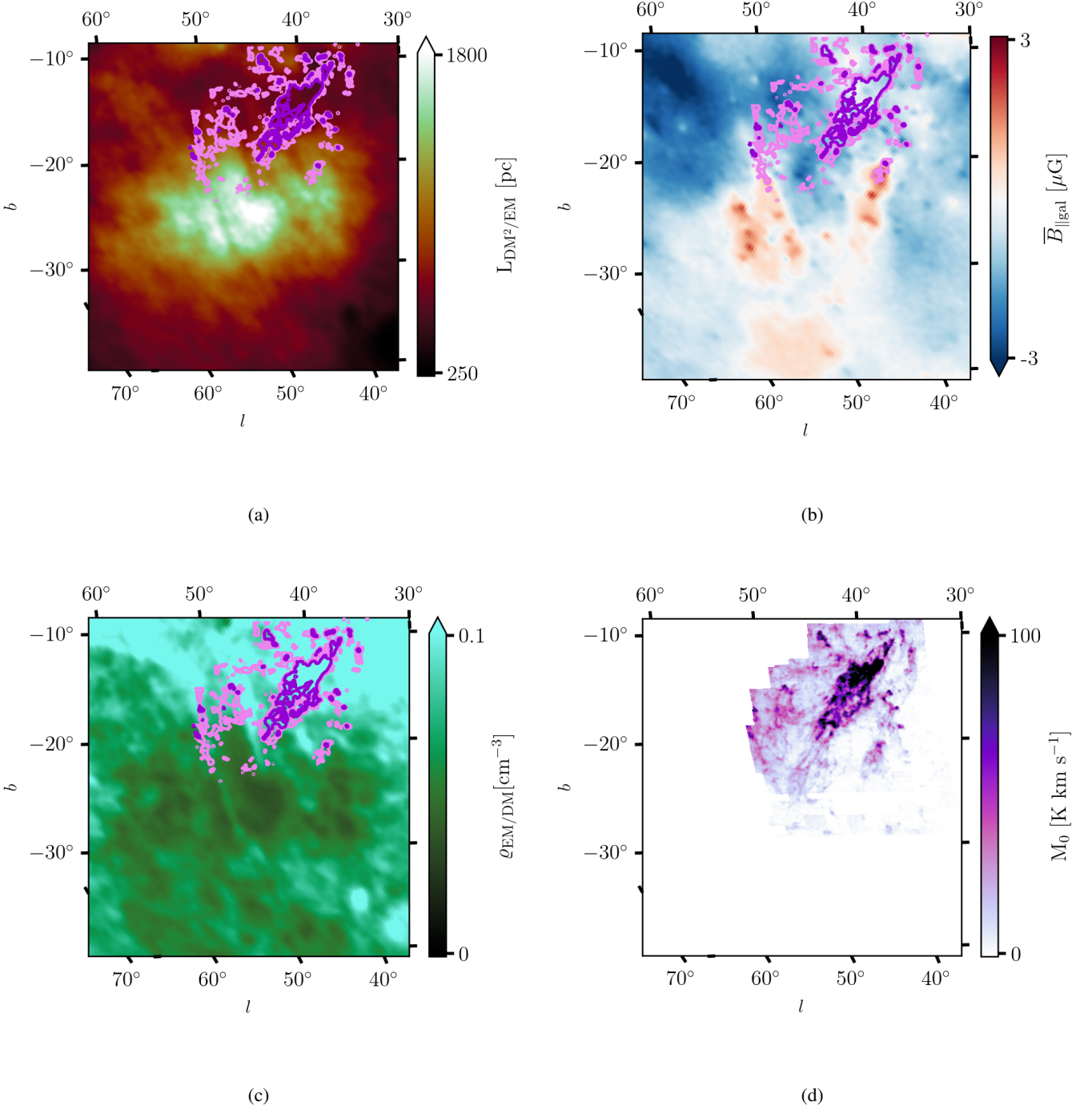

\centering
\begin{subfigure}{0.49\textwidth}
\includegraphics[width=\textwidth, draft=False]{\cutoutpath{smith}{EMDM2}}
\caption{\label{fig:smith_distance}}
\end{subfigure}
\begin{subfigure}{0.49\textwidth}
\includegraphics[width=\textwidth, draft=False]{\cutoutpath{smith}{B}}
\caption{\label{fig:smith_bpar}}
\end{subfigure}
\begin{subfigure}{0.49\textwidth}
  \includegraphics[width=\textwidth, draft=False]{\cutoutpath{smith}{rho}}
\caption{\label{fig:smith_dm}}
\end{subfigure}
\begin{subfigure}{0.49\textwidth}
\includegraphics[width=\textwidth, draft=False]{\cutoutpath{smith}{H1}}
\caption{\label{fig:smith_em}}
\end{subfigure}
\caption{\label{fig:smith}
Cutouts of several sky maps towards the Smith HVC.
The images show a) $\Len{DM^2/EM}$, b) $\Bpw{gal}$, c) $\rho_{EM/DM}$, and d) velocity integrated HI as observed by the Green Bank Telescope \citep{2008Lockman}.
Contours of the HI data are plotted over the first three images, with contour lines set at 30 K km s$^{-1}$ (light-violet) and 60 K km/s(dark-violet). }
\end{figure*}

The Smith cloud is a high velocity cloud (HVC) on the Southern sky, centered on $(l, b) = (\ang{42}, \ang{-16})$, at a distance of $12.4 \pm 1.3$ kpc \citep{2008Lockman}.
The cloud covers a large portion of the sky, with a $(l, b) = (\ang{39}, \ang{-13})$ nose pointing towards the Galactic disk and a tail extending well past $(l, b) = (\ang{45}, \ang{-20})$  \citep{2013Hill}.
We show cutouts of $\Len{DM^2/EM}$, $\Bpw{}$ and $\varrho_{\mathrm{EM/DM}}$ in this sky region in Fig. \ref{fig:smith}.
In there, we illustrate the shape of the cloud using HI data from the Green Bank Telescope \citep{2008Lockman} (\ref{fig:smith}).
While the nose seems to be barely magnetized, the tail shows relatively strong excesses in extragalactic RMs \citep{2013Hill, 2019Betti}, indicating significant magnetization.
Within the tail region, we derive magnetic field strengths between $5.5 \pm 0.18 \, \mu$G and $-2 \pm 0.13\, \mu$G and identify three compact subregions with alternating sign in the magnetic field map (Fig. \ref{fig:smith_bpar}), roughly separated by longitude lines $\ang{49}$ and $\ang{41}$.
These regions align very well with ridges seen in the HI data, which is consistent with the picture of the cloud falling into the Galaxy, leaving behind an ionized tail.
For a more detailed discussion on geometry and physical effects, we refer the reader to \citet{2019Betti}.
We estimate LoS averaged magnetic field strengths of approximately 2, -1.5 and 2.5 $\mu$G in these subregions, ordered with decreasing longitude.
These compact regions have been studied as well \citet{2019Betti} (there called regions 3, 4 and 5), who estimated lower limits of $2.86 \pm 0.3\, \mu$G, $-1.35 \pm 0.2 \, \mu$G  and $5.33 \pm 0.3\, \mu$G within these regions, respectively.
\citet{2013Hill} have estimated a maximum magnetic field strength of 8 $\mu$G in the Smith cloud.
Given the large angular size and distance of the cloud, we have refrained from performing a foreground estimation as we did for the Sh2-27 HII region, but only note the good correspondence of the sign patterns between our work and \citet{2019Betti}, as well as the very precise alignment with the HI data for the ridges.
In our work, the tail region is also noticeable as an area of strongly increased $\Len{DM^2/EM}$ Figs. \ref{fig:distance} and  \ref{fig:smith}, reaching values of $\num{2.5e3} \pm 150$  pc.
We note that the Smith cloud is naturally not constrained directly by pulsars, which makes our DM (and therefore $\Len{DM^2/EM}$ and $\varrho_{\mathrm{EM/DM}}$) estimate rather uncertain.

The examples of Sh 2-27 and the Smith HVC demonstrate that our method is in agreement with known results even in such extraordinary environments, but also that our results cannot be interpreted without a careful consideration of the particularities of the region in question, e.g. the Galactic fore- and background or the pulsar coverage.

\subsubsection{Other regions}
\label{subsubsec:results_cutout_other}

The halo of the Magellanic clouds is also noticeable in the $\Len{DM^2/EM}$ sky map.
Their angular sizes are visually consistent with the denser parts of the Magellanic corona found by \citet{2022Krishnarao}.
Regarding a quantitative comparison, \citet{2019Smart} give values of several kpc for the physical size of the tail region of the SMC (i.e. the region pointing towards the LMC), while we report a maximum of $\num{700}$ pc in $\Len{DM^2/EM}$ in the same area.
Given that $\Len{DM^2/EM}$ is a lower limit on the physical size, this indicates consistency.

Regarding the Sagittarius arm region (defined by the tangent point of the Sagittarius arm at $l=\ang{49}$ and $b=\ang{0}$, also marked in Fig. \ref{fig:regions}), which also displays very large values in $\Len{DM^2/EM}$ just a few degrees above and below the disk in latitude.
As mentioned before, this region also exhibits the strongest RM's and LoS averaged magnetic field, also pointing to a coherent magnetic field configuration with minimal averaging along the LoS.
However, the picture is obfuscated somewhat by the fact that it also contains the most significant tension between the H-$\alpha$ and free-free EMs, with the H-$\alpha$ estimate two orders of magnitude larger than the \textit{Planck} data, which points to strong systematic effects in one or both of the EM estimation techniques, which might have affected our results in $\Len{DM^2/EM}$.

\subsection{Correlations to other pulsar observables}
\label{subsec:correlations}

\begin{figure*}
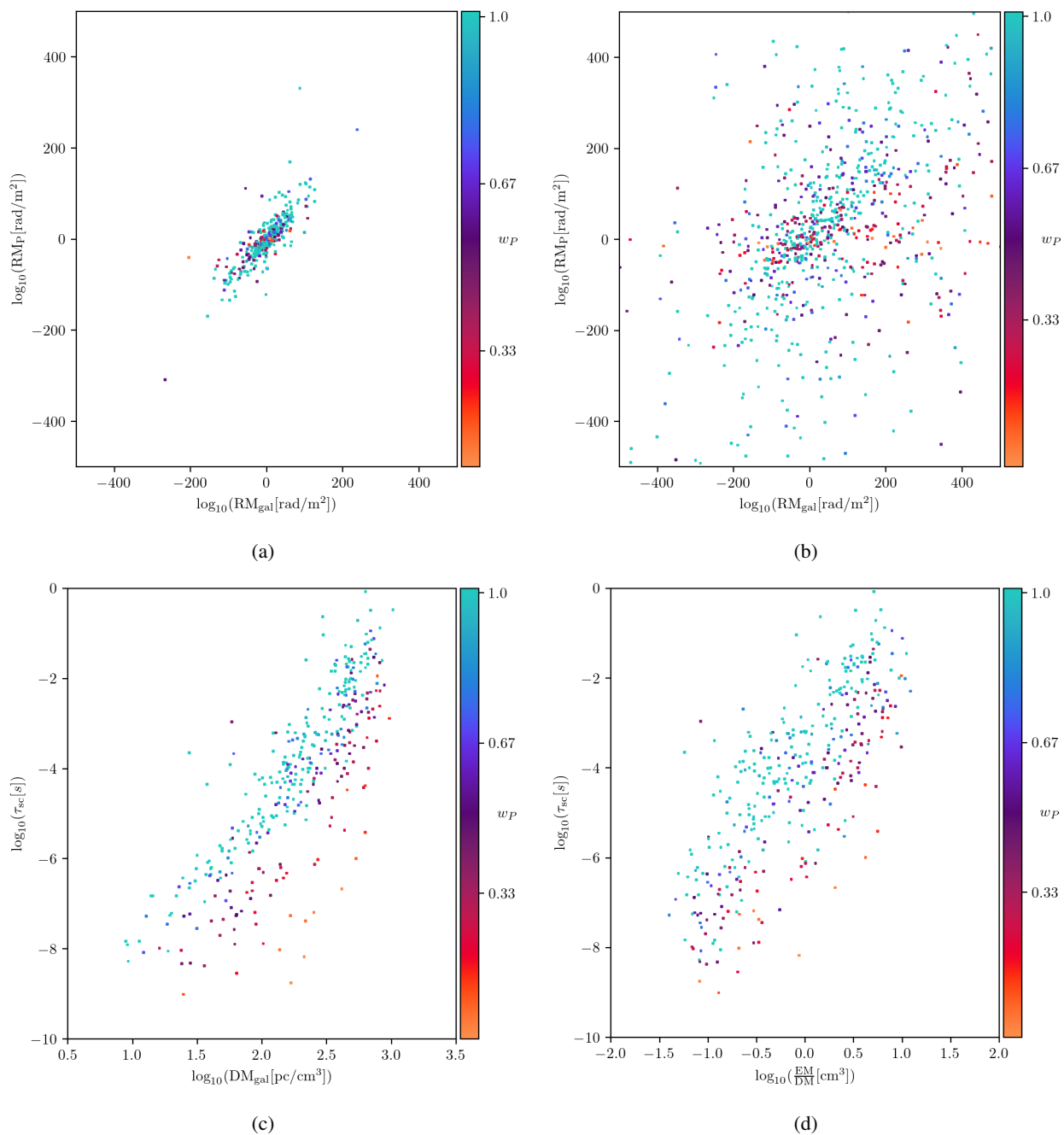

\centering
\begin{subfigure}{0.45\textwidth}
\includegraphics[width=\textwidth, draft=False]{\correlationpath{RM_filtered}}
\caption{\label{fig:pulsar_rm_scatter_high_lat}}
\end{subfigure}
\begin{subfigure}{0.45\textwidth}
\includegraphics[width=\textwidth, draft=False]{\correlationpath{RM_complement_filtered}}
\caption{\label{fig:pulsar_rm_scatter_low_lat}}
\end{subfigure}
\begin{subfigure}{0.45\textwidth}
\includegraphics[width=\textwidth, draft=False]{\correlationpath{tau_sc_dm}}
\caption{\label{fig:pulsar_tau_sc_scatter_dm}}
\end{subfigure}
\begin{subfigure}{0.45\textwidth}
\includegraphics[width=\textwidth, draft=False]{\correlationpath{tau_sc_emdm}}
\caption{\label{fig:pulsar_tau_sc_scatter_dist}}
\end{subfigure}
\caption{\label{fig:scatter} Scatter plots of pulsar observables vs. the respective inferred Galactic values. The coloring indicates the posterior mean of $w_\mathrm{P}$ parameter introduced in Eq. \eqref{eq:pulsar_dm}.
Figs. a) and b) illustrate the $\RM{p}$ -  $\RM{gal}$ correlation at high and low latitudes, respectively.
Figs. c) and d) illustrate the correlation of the pulsar temporal broadening parameter $\tau_\mathrm{sc}$ with $\DM{gal}$ and $\varrho_{\mathrm{EM/DM}}$, respectively.
}
\end{figure*}

Pulsars not only provide DM measurements, but a significant fraction also come with RM data (about $12 \%$) and data on interstellar turbulence via e.g. measuring the temporal broadening $\tau_\mathrm{sc}$ of pulses (again about $12 \%$ of all pulsars).
The pulsar RM data is not immediately suitable for our algorithm, as it in contrast to the pulsar DM, the modelling of the residual Galactic component is not possible with a simplistic effective model as for the pulsar DM in Eq. \eqref{eq:pulsar_dm}.
Nonetheless, this data is of course still informative on the Galactic ISM.
The pulsar RMs are compared to the respective RM values of the full Galaxy at the same LoS in Figs. \ref{fig:pulsar_rm_scatter_high_lat} and Figs. \ref{fig:pulsar_rm_scatter_low_lat}, where we separate the data into high and low latitude pulsars.
The plot reveals that high latitude pulsar RMs show a good correlation (Pearson-r: $0.82$) to the Galactic RM, indicating that they probe the same magneto-ionic medium.
This implies that our DM predictions in this region should not suffer strongly from the insufficient sampling of the ionized volume of the Milky Way.
At lower latitudes, the correlation vanishes (Pearson-r: $0.004$), consistent with only a small volume probed by pulsar RM.
In both cases, the posterior $w_\mathrm{P}$ factor defined as the relative DM occupied by the pulsar in Eq. \eqref{eq:pulsar_dm}, is in better correlation with the full respective population of pulsar RM.
This indicates, that the $w_\mathrm{P}$ factor is at least a good qualitative indicator for the fraction of the LoS probed by a pulsar.

In Figs. \ref{fig:pulsar_tau_sc_scatter_dm} and Figs. \ref{fig:pulsar_tau_sc_scatter_dist}, we correlate $\tau_\mathrm{sc}$ with the $\DM{gal}$ and $\varrho_{\mathrm{EM/DM}}$ skies.
This quantity is directly proportional to the scattering measure (SM), which is defined as the LoS integral of the square $n_\mathrm{e}$ fluctuation amplitude, and hence a good tracer of variability along the LoS, see. e.g. \citet{2020Ferriere} for a summary.
We expect it to be positively correlated with $\sigma:{n_\mathrm{e}}$ (i.e. more variations lead to more scattering), the DM (more material, more scattering) and with the length of the LoS due to the integral.
The DM is correlated with $\tau_\mathrm{sc}$, which is consistent with previous findings \citep{1977Rickett, 2019Krishnakumar}.
This effect can in simple terms be explained as more material leading to more scattering in the interstellar plasma, the details depend on the correlation structure of the electron density field.
The plot furthermore reveals that for a certain $\DM{gal}$ found at the position of a pulsar, a smaller $w_\mathrm{P}$ tends to imply a smaller $\tau_\mathrm{sc}$.
Since $w_\mathrm{P}\DM{gal}$ gives the pulsar DM which produces the scattering, this simply confirms that smaller $w_\mathrm{P}$ indeed probe smaller portions of the respective LoS.
Considering only the pulsars with $w_\mathrm{P}\approx 1$,  $\tau_\mathrm{sc}$ increases with a slope of about 2  below $\approx 100\, \mathrm{pc\, cm^{-3}}$, which is in general agreement with the literature \citep{1977Rickett, 2019Krishnakumar}.
For larger DM values the slope steepens.
As pulsars with such high DM are mostly placed in the disk, their $\DM{gal}$ is most likely underestimated in line with the discussions in \ref{subsec:model_summary} and \ref{subsec:results_sky_dm}, which would explain the steepening.

The correlations of both $\tau_\mathrm{sc}$ and $w_p$ with $\Len{DM^2/EM}$ are approximately zero, and are hence not shown here.
The $\Len{DM^2/EM}$ factor is inversely proportional to  $\sigma_{n_\mathrm{e}}$ but linearly proportional to the length of the LoS, which explains this lack of correlation.
The $\varrho_{\mathrm{EM/DM}}$ factor, however, shows an almost linear relation with $\tau_\mathrm{sc}$ on log-log scale, which is fit well by $\tau_\mathrm{sc} \propto \varrho_{\mathrm{EM/DM}}^{3.5}$.
According to Eq. \eqref{eq:em_dm_ratio} this fraction is proportional to $\sigma_{n_\mathrm{e}}$ in case of strong variability along the LoS.
Since $\sigma_{n_\mathrm{e}}$ is also directly related to the SM, this plot demonstrates the strong $n_\mathrm{e}$ variations along the LoS probed by pulsars.

We would like to emphasize that the Figs. \ref{fig:pulsar_tau_sc_scatter_dm} and \ref{fig:pulsar_tau_sc_scatter_dist} correlate the observed pulsar data with inferred Galactic quantities (which trace the full LoS).  
This explains the much higher scatter of $\tau_\mathrm{sc}$ per DM observed in these plots, in contrast to e.g. $\tau_\mathrm{sc} - \DM{P}$ analyses, see e.g. \citep{2019Krishnakumar}.
All plots in Fig. \ref{fig:scatter} also show that the $w_\mathrm{P}$ factors are approximately 1 for a large portion of pulsars, which, at least in the disk where most pulsars are found, is most likely not true as most pulsars are known not to probe the full Galaxy.
This again indicates that the inferred $\DM{gal}$ sky at low absolute latitudes is underestimated.
We hence deem our results on $w_\mathrm{P}$ in this area as an upper limit, and postulate that many values are most likely overestimated by at least a factor of two at low latitudes, again in line with the discussions in \ref{subsec:model_summary} and \ref{subsec:results_sky_dm}.

\section{Summary and Outlook}
\label{sec:conclusions}

In this work, we use several data sets in order to disentangle the Galactic Faraday sky into physical components, namely the averaged LoS parallel and electron density weighted magnetic field component and the DM, both for the full sky.
To constrain the DM, we rely on pulsars and additionally on EM-data stemming from collisional processes, which also opens a path to constrain the structure of ionized plasma of the Milky Way.

We have chosen to model all sky maps non-parametrically, i.e. each pixel is only constrained by data and a-priori assumptions on the all-sky correlation structure.
This allows us to represent Galactic structures in unprecedented detail, especially in comparison to existing parametric models for the Galactic magnetic field or the electron density.

At mid to high latitudes we find average magnetic field values of around $2\,\mu$G, consistent with local measurements.
The morphology of the map in Fig. \ref{fig:Bpar} indicated the strongest LoS-average magnetic field in mid-latitude regions outside the turbulent Galactic disk.
In the same regions, the DM values gradually decrease with absolute latitude towards a minimum between 10 and 20 $\mathrm{pc\, cm^{-3}}$ at the poles.
Comparison with previously published models and independent data sets reveal that our results agree well with previous findings in these regions.
In contrast, the DM results in the disk (with a maximum of $ \num{1.7e3} \mathrm{pc\, cm^{-3}}$) appear to be too small by at least a factor of two, if compared to other models.
This is easily explained by the insufficient volume sampling of the Galactic disk by pulsars, which we have not corrected for.
This is a deliberate modelling choice, which accepts this systematic error in order to not have to rely on strong assumptions on the structure of the Milky Way.
It should be noted that this error translates onto all other sky maps inferred in this work, in particular the magnetic field estimate in the disk is too large by the same factor.

In addition to our results on the Galactic magnetic field and the DM sky, we have derived several results on the structure of the electron density of the Milky Way.
We inferred angular power spectra for all our sky maps, which in case of the DM sky reveals a Kolmogorov like behavior, in accordance with the `Big power law in the sky' \citep{1995Armstrong}.
Additionally, we have defined two conversion factors to link the DM and EM skies, namely $\Len{DM^2/EM}$ (in Eq. \eqref{eq:em_dm2_ratio}, as the ratio between squared DM and EM) and $\varrho_{\mathrm{EM/DM}}$ (in Eq. \eqref{eq:em_dm_ratio}, as the ratio between squared DM and EM).
The former factor gives approximately the portion of the LoS effectively contributing to EM and DM via high densities, implying that it is equal to the length of the LoS in the special case of a uniform medium and in general a lower limit to it.
In a similar vein, $\varrho_{\mathrm{EM/DM}}$ gives the average electron density along the full LoS in the approximately uniform case.
Both maps agree well with literature values for the Milky Way.

We have devised a simple illustrative mock model of the Galactic electron density, which allowed us to explain this phenomenology in terms of simple assumptions on the 3D distribution of electrons.

We have also shown detailed analysis of two cutouts of the sky, centered on the HII cloud Sh 2-27 and the Smith high velocity cloud.
A comparison to previous works on these objects reveals good correspondence, but also demonstrates the need for a detailed foreground analysis, if the properties of peculiar objects were to be analyzed with our method.

Our results still suffer from several unaccounted potential biases in our model and the data sets we use, which we have partly verified.
These are the mainly large systematic uncertainties in the EM data derived from H$\alpha$, which mostly stem from the unknown relation to the Galactic dust distribution, as well as the sparse volume (at low latitudes) and angular (at high latitudes) sampling of pulsars.

The road ahead to improve our results must clearly be focussed to alleviate these issues, which can be achieved by two main paths.
At first, the inclusion of new data sets will be helpful.
Potential additions are mostly dust data, which would increase the fidelity in the H$\alpha$ derived EM data, as well as new pulsar data coming from next generation radio telescopes such as the Square Kilometer Array, which are projected to be able to observe almost all Galactic pulsars \citep{2008vanLeeuwen, 2015Keane}.
A second possible data source that may reduce the systematic biases stemming from the pulsar distribution are Fast Radio Bursts, which already have been demonstrated to allow for such an analysis \citep{2022Pandhi}.
See also \citet{2023Cook} and \citet{2023Ravi} for illustrations of the potential value of FRBs for the inference of the Galactic electron density.
FRB DMs provide a nice complementary data set to the pulsar DMs, as they yield an upper limit to the DM, in contrast to the lower limit provided by pulsars.

The second path to improve our results is concerned with modelling.
A common denominator for many of the shortcomings of our results is that they originate in some systematic effect along the LoS, e.g. via some selection effect in probing the Galactic volume or an unclear correlation of Galactic components before the LoS integration (e.g. $\rho_{\mathrm{Dust}}- n_\mathrm{e}$ correlations).
Additionally, the interpretation of our results often requires the assumption on properties of the three-dimensional structure of the Milky Way, e.g. with regard to the $\Bp{} - n_\mathrm{e}$ correlation or the turbulent properties of the electron density to explain the $\Len{DM^2/EM}$ factor.
It seems hence clear to us that a full resolution of these effects can only be achieved via inferring the full three-dimensional volume for both the interstellar plasma and the associated magnetic field, as envisioned by e.g. \citet{2018Boulanger}.

\begin{acknowledgements}
We would like to thank the referee for a constructive report.
Furthermore, we would like to thank Valentina Vacca for helpful discussions and Jay Lockman and Sarah Betti for providing the GBT HI data.

The following python packages were used in this work: \texttt{Numpy} \citep{2020Harris}, \texttt{Scipy} \citep{2020Virtanen}, \texttt{Nifty} \citep{2019NIFTY}, \href{https://mtr.pages.mpcdf.de/ducc/}{ducc}, \texttt{cmasher} \citep{2020VanDerVelden}, \texttt{astropy} \citep{2022astropy} and \texttt{h5py} \citep{2014Collette}.

SH and MH acknowledge funding from the European Research Council (ERC) under the European Union's Horizon 2002 research and innovation programme (grant agreement No. 772663).
PF acknowledges funding through the German Federal Ministry of Education and Research for the project ErUM-IFT: Informationsfeldtheorie für Experimente an Gro{\ss}forschungsanlagen (F\"orderkennzeichen: 05D23EO1).
NR acknowledges funding from the joint NWO-CAS research programme (project number 629.001.022) in the field of radio astronomy, which is (partly) financed by the Dutch Research Council (NWO).
\end{acknowledgements}

\bibliographystyle{aa}
\bibliography{lib}

\begin{appendix}

  \begin{table*}
    \begin{center}
      \begin{tabular}{|c|c|c|c|l|}
       Parameter & $\LoS{\Bp{}}{gal}$ & $\log(\DM{gal})$ & $\log(\Len{DM^2/EM})$ & interpretation  \\
       \hline
       $\left( m_0, \sigma_0, \sigma_{\sigma, 0} \right)$ & (0, 5, 4) & (5, 1, 0.001)  & (-3, 1, 0.001)  & mean of the field   \\
       $\left(m_s, \sigma_s\right)$ & (-3, 1) & (-3, 1) & (-3, 1)  &  power law slope  \\
       $\left(m_a, \sigma_a\right)$ & (6, 4) & (.7, .1)  & (.7, .1)  &  per-pixel variance \\
       $\left(m_\eta, \sigma_\eta\right)$ & (.1, .1) & (.1, .1)  & (.1, .1)  & amplitude of power law deviations \\
      \end{tabular}
      \caption{\label{tab:params} Hyperparameters of the correlation structure model used in this work.
      See Eqs. \eqref{eq:field_model} and \eqref{eq:kernel_model} for the definitions, \citep{2022Arras} for further details of the model and Fig. \ref{fig:priors} for illustrations of the resulting priors.}
      \end{center}
  \end{table*}

  \section{Sky Priors}
  \label{app:prior}

  \begin{figure*}
    \centering
    \begin{subfigure}{0.4\textwidth}
      \includegraphics[width=\textwidth, draft=False]{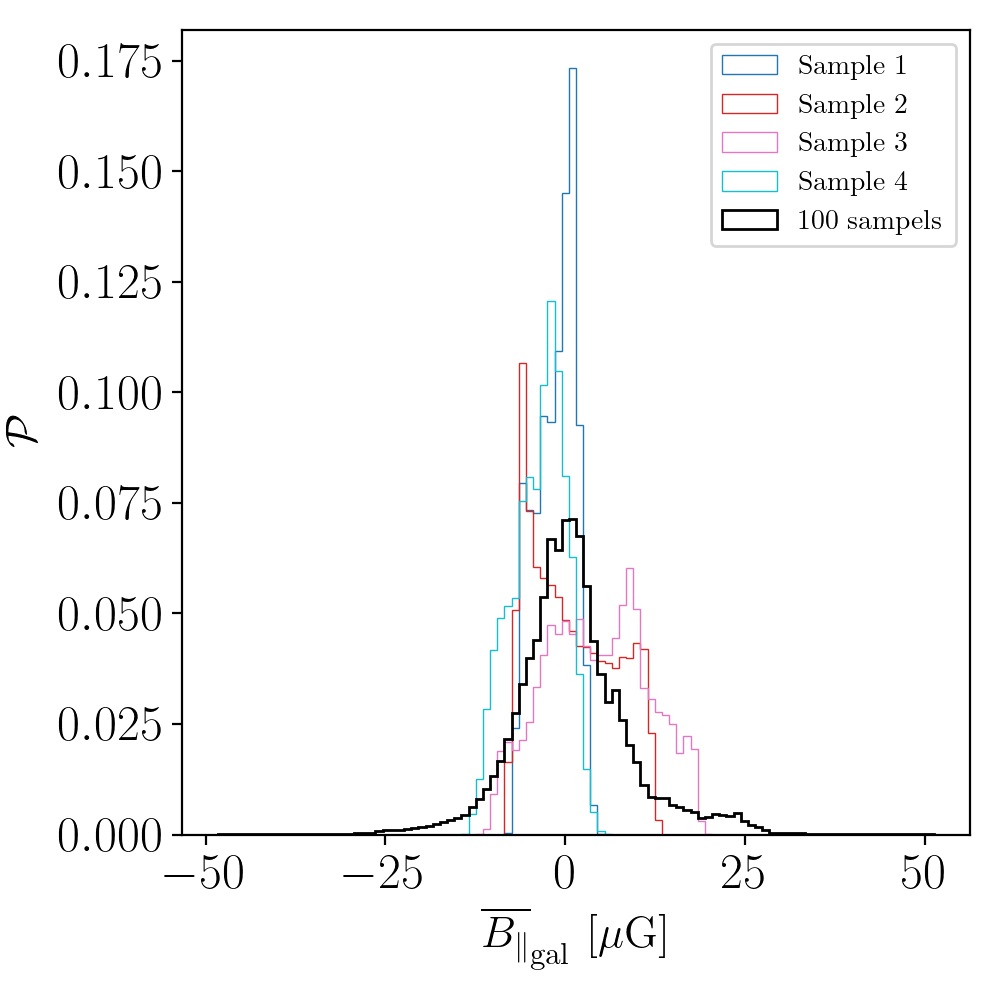}
      \caption{\label{fig:Bpar_prior_hist}}
    \end{subfigure}
    \hspace{0.1em}
    \begin{subfigure}{0.58\textwidth}
      \includegraphics[width=\textwidth, draft=False]{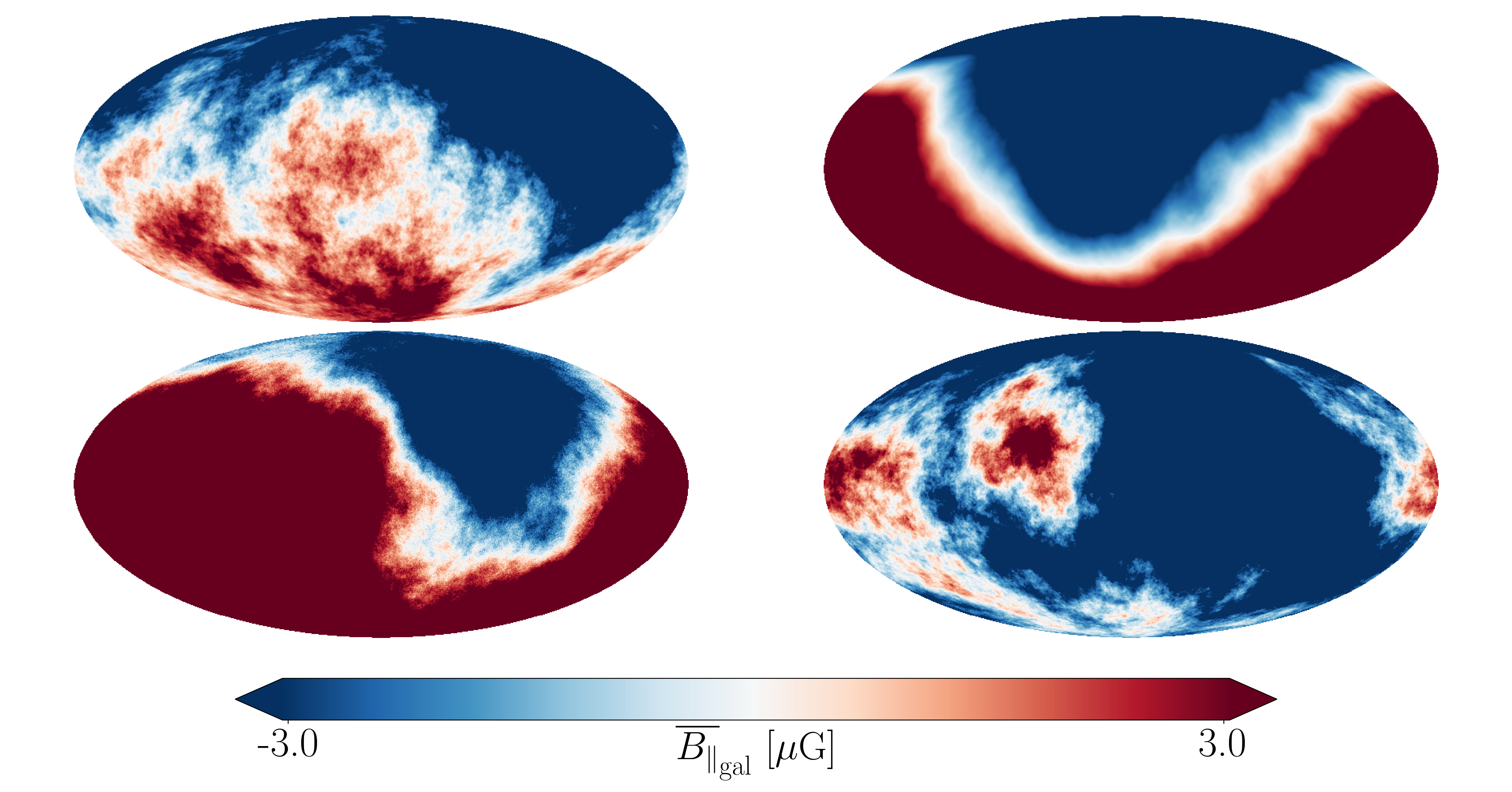}
      \caption{\label{fig:Bpar_prior_sky}}
    \end{subfigure}

    \begin{subfigure}{0.4\textwidth}
      \includegraphics[width=\textwidth, draft=False]{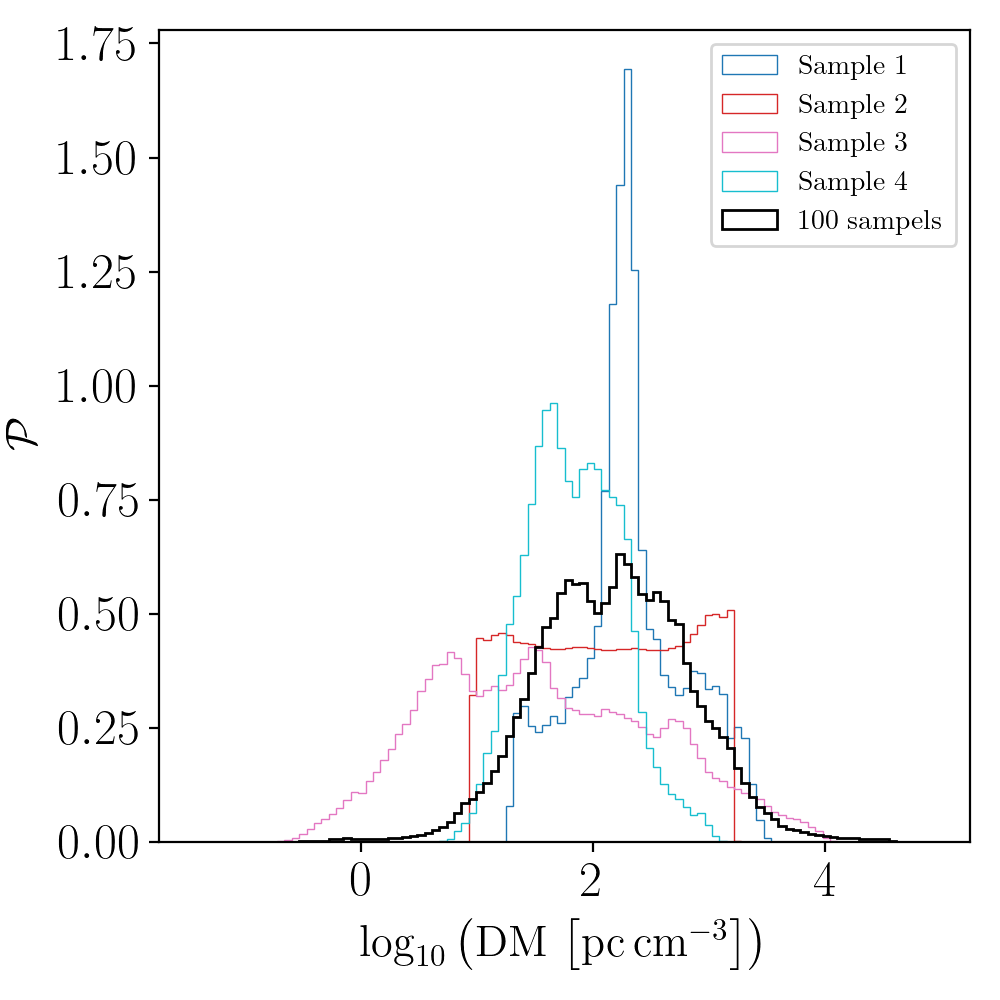}
      \caption{\label{fig:log_dm_prior_hist}}
    \end{subfigure}
    \hspace{0.1em}
    \begin{subfigure}{0.58\textwidth}
      \includegraphics[width=\textwidth, draft=False]{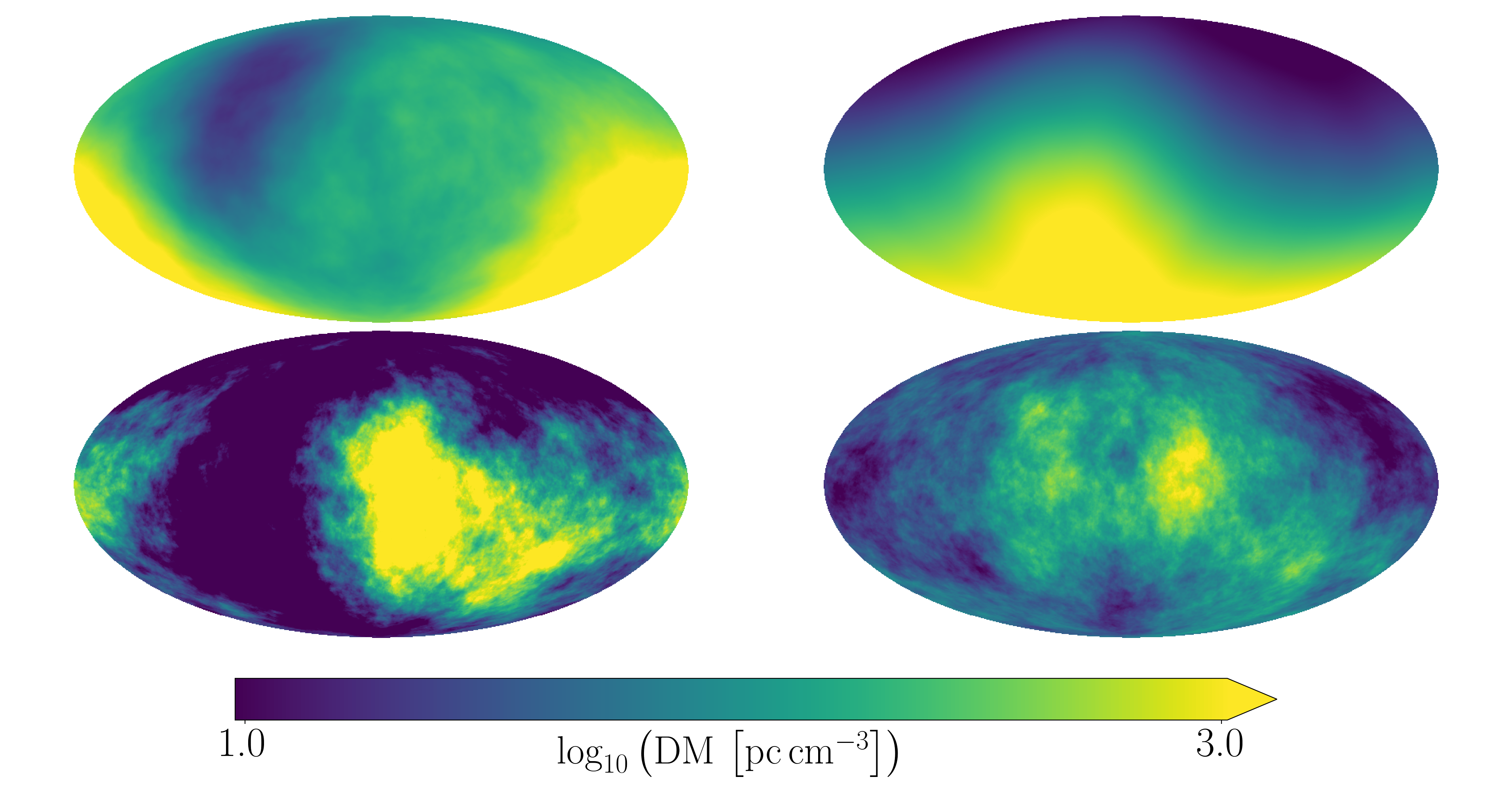}
      \caption{\label{fig:log_dm_prior_sky}}
    \end{subfigure}

    \begin{subfigure}{0.4\textwidth}
      \includegraphics[width=\textwidth, draft=False]{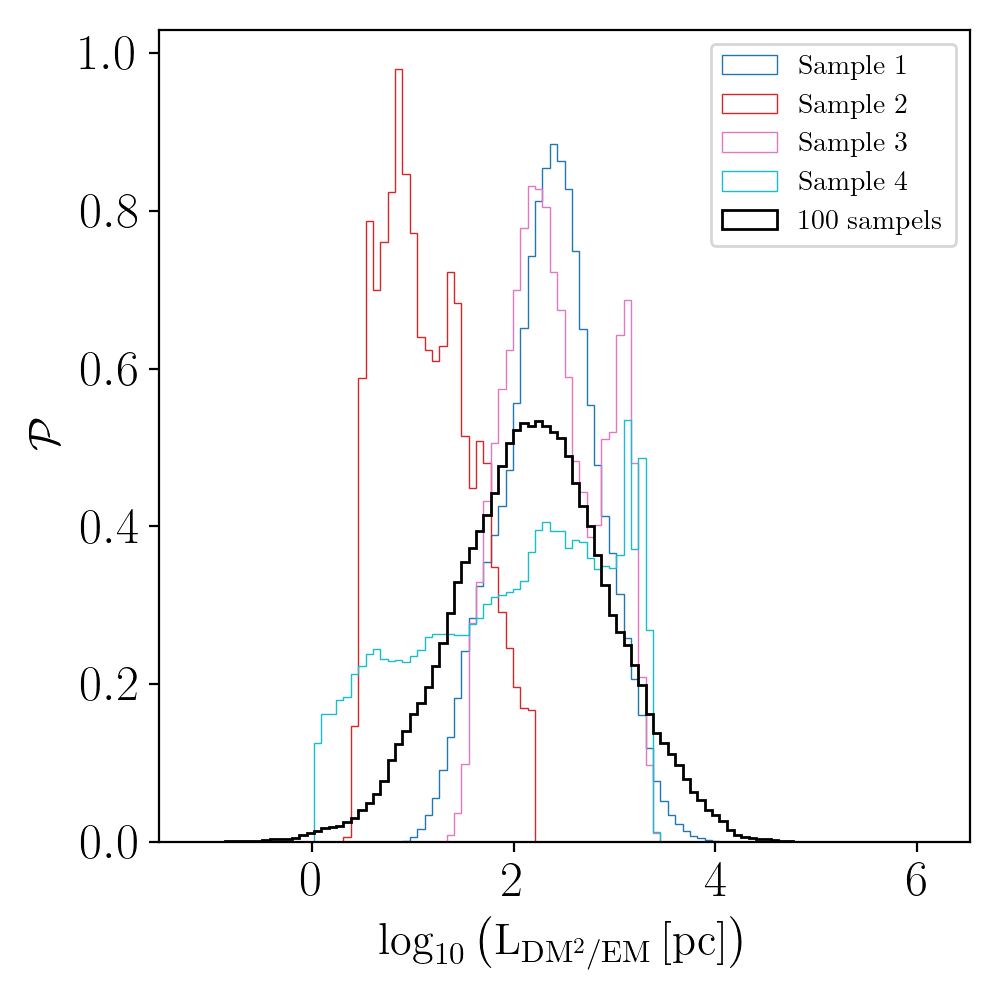}
      \caption{\label{fig:log_distance_prior_hist}}
    \end{subfigure}
    \hspace{0.1em}
    \begin{subfigure}{0.58\textwidth}
      \includegraphics[width=\textwidth, draft=False]{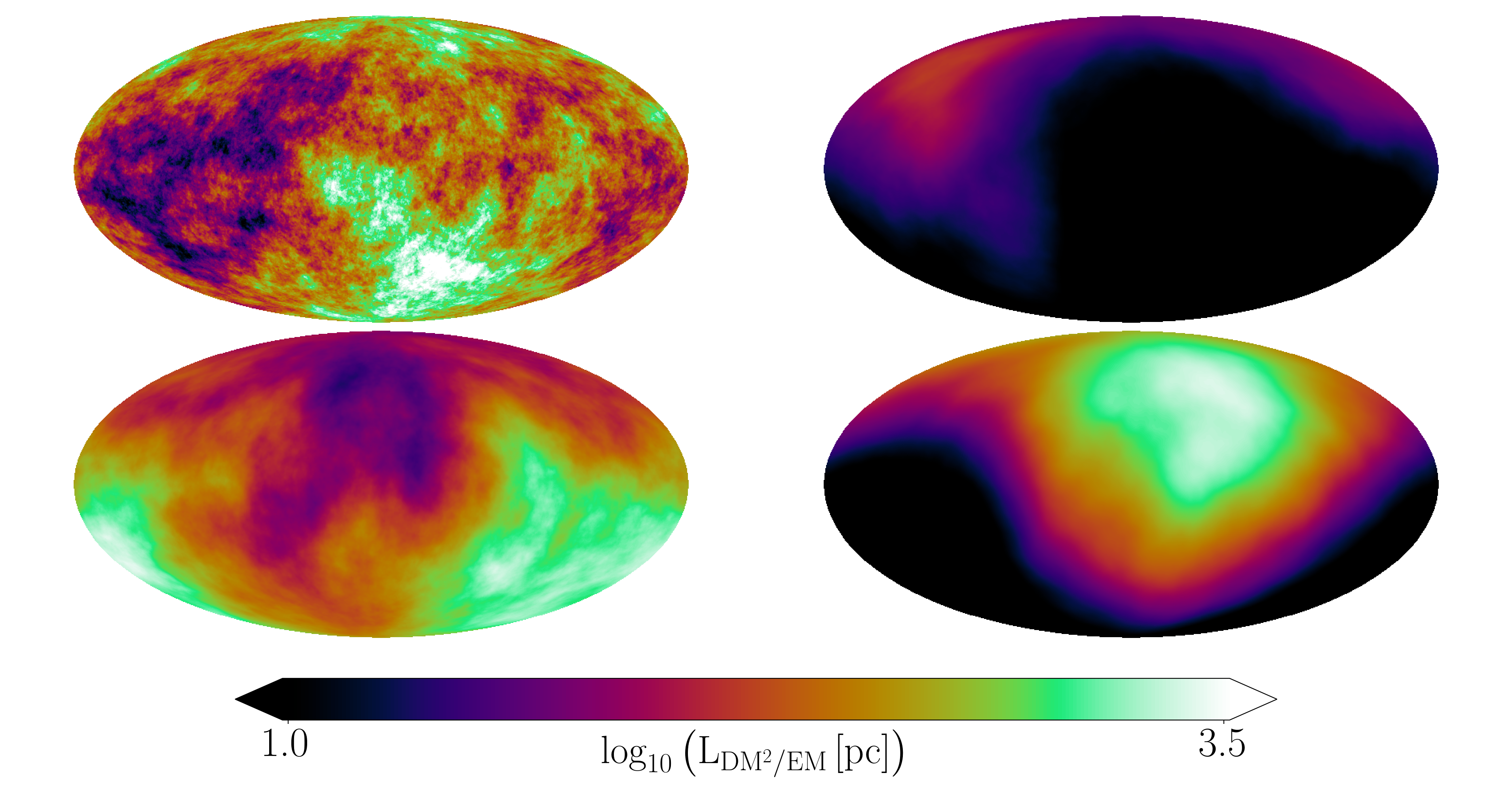}
      \caption{\label{fig:log_distance_prior_sky}}
    \end{subfigure}
    \caption{\label{fig:priors}}
    Illustrations of the priors used for the $\LoS{\Bp{}}{gal}$, $\DM{gal}$ and $\Len{DM^2/EM}$ sky models.
    The right side shows sky maps corresponding to four different samples drawn from the respective prior distributions.
    The left side shows histograms of the same prior samples, and additionally a line summarizing the distribution of values over 100 samples.
    The histograms were normalized to unity to ensure comparable scales.
  \end{figure*}

  Here, we summarize the priors on the $\LoS{\Bp{}}{gal}$, $\DM{gal}$ and $\Len{EM, DM}$ sky models.
  As discussed in Sect. \ref{sec:models}, these are modelled as (log-)normal processes with an unknown correlation function, for which we set a priori constraints, following the generic model developed in \citep{2022Arras}.
  In the specific instance of correlations on the unit sphere, this model aims to represent a field $s$ via
  \begin{equation}
    \label{eq:field_model}
    s = m_0 + (\sigma_0 + \sigma_{\sigma, 0} \xi_\sigma)\xi_0 + \mathcal{F}\left(\sqrt{T_\ell(\xi_T)}\xi_\ell\right).
  \end{equation}
  where the global mean of the field is modelled via the first three terms with hyperparameters $m_0, \sigma_0 $ and $\sigma_{\sigma, 0}$.
  The second part encodes the correlation structure of the field, with $\mathcal{F}$ denoting the spherical harmonic transform and $T_\ell$ an angular correlation kernel.
  The excitation terms $\xi_x$ form the parameter space of the model that is constrained in the inference and are a-prior standard normal distributed, while the hyperparameters encode a priori assumptions on the correlation function.
  The kernel is modelled via
\begin{equation}
  \label{eq:kernel_model}
  \sqrt{T_\ell} = \frac{e^{ m_a + \sigma_a\xi_a}}{N} \, \mathrm{Exp}* e^{\left[\left(m_s + \sigma_s\xi_s \right)\ln(\ell) + f(m_\eta, \sigma_\eta, \xi_\Phi)  \right]},
\end{equation}
  where have borrowed the notation of \citet{2020Leike}, in that $\mathrm{Exp}*$ is denoting exponentiation of the coordinates, i.e. going from $\log(\ell)$ to $\ell$.
  This equation parametrizes a power law in log-log space with the slope modelled by $m_s$ and $\sigma_s$ terms, amplitude modelled by the $m_a$ and $\sigma_a$ terms, the normalization of the power spectrum $N$ and an additional term $f$ responsible for deviations from the power law.
  We do not write out $f$ and $N$ here to keep our notation simple, but refer to \citet{2022Arras} for details.
  This field model depends on hyperparameters $\left(m_0, \sigma_0, \sigma_{\sigma, 0}, m_s, \sigma_s, m_a, \sigma_a, m_\eta, \sigma_\eta  \right)$.
  We summarize our choice of hyperparameters in Table \ref{tab:params}.

  The hyperparameters of the model are chosen such that mean and variance of the overall resulting field models easily cover the plausible scales for each of the three physical quantities, as discussed Sect. \ref{sec:observables}, in order to not bias the inference.
  The width of the priors is also often limited by numerical issues, as the exponentiation and multiplication of fields easily produces values beyond what is numerically representable.
  This has limited the possible variance of the $\log(\DM{gal})$ and $\log(\Len{DM^2/EM})$ priors somewhat, but as illustrated in Fig. \ref{fig:priors}, the relevant physical scales are still well represented.
  The figure shows four samples for each model.
  The morphology of the sky maps illustrate the strong variability in smoothness the priors allow.
  We also show histograms of the sky values of the four samples and a combined set of 100 sampled sky maps for each model, illustrating the wide range of a priori allowed values, easily encompassing the ranges discussed in Sect. \ref{sec:models}.

  \section{Tests with synthetic data}
  \label{app:synthetic_tests}

  In this section, we demonstrate the impact of selection effects in the observed spatial pulsar distribution, as well as our assumption of a-priori Gaussianity for the $\Bp{gal}$, log-$\DM{gal}$ and log-$\Len{DM^2/EM}$ skies.
  To this end, we generate synthetic data using the following recipe:
  \begin{enumerate}
    \item We draw random sky maps for the $\Bp{gal}$, $\DM{gal}$ and $\Len{DM^2/EM}$ skies using the \citet{2022Arras} correlation model assuming a Kolmogorov spectrum for all three random maps.
    \item We generate a profile template of the large scale features of the Galaxy using the YMW electron model.
    We multiply the synthetic DM map with this template, and accordingly divide the synthetic $\Len{DM^2/EM}$ map with it.
    This imprints the shape of the large scale Milky Way features on the data, making the synthetic date more realistic.
    Furthermore, features such as the thin disk are rather unlikely in our isotropic prior models, making their reconstruction an additional challenge for this test.
    \item We adapt the hyperparameters of the \citet{2022Arras} correlation model, such that the synthetic RM, DM and EM roughly show values expected for the Milky Way.
    \item We draw random angular positions for the extra-galactic RMs and the Galactic pulsar DMs and distribute the former uniformly over the sky, while we scale the density of the latter with $|l|^{1.5}$ and $|b|^{2.5}$, reflecting the Galactic origin of pulsars.
    We note that this results in a relative overdensity of pulsars at $l=0$ at higher latitudes, which is clearly not mirroring the real distribution, but serves as a proxy for a possible observational selection bias.
     We choose to draw positions for 50000 RMs, roughly the same as we have available in the \citet{2023van_Eck} catalog.
     In case of the pulsars, we distinguish two cases:
     \begin{enumerate}
      \item 30000 pulsar positions $\left(l_i, b_i\right)$, with the respective DM drawn uniformly between 0 and the respective Galactic DM value at $\left(l_i, b_i\right)$, henceforth called the \textit{ideal} case.
      \item 3000 pulsar positions $\left(l_i, b_i\right)$, with the respective DM drawn uniformly between 0 and $\mathrm{min}\left(\DM{gal}\left(l_i, b_i\right), 1000\, \mathrm{pc\,cm^{-3}} \right)$, henceforth called the \textit{realistic} case.
     \end{enumerate}

     In the first case, we effectively sample the full Milky Way, albeit in an inhomogeneous way.
     The second case is closer to reality, as the number of pulsars is closer to what is available in the ATNF catalog, the high latitude regions are sparsely sampled and the high DM values of the disk are not probed.
     We emphasize the $1000\, \mathrm{pc\,cm^{-3}}$ limit is just a proxy for the real selection effect, as in reality pulsars observations are limited by luminosity and not DM.
    \item We draw error bars uniformly between 1 and 10 \% for the RM and the full sky EM data, and between 0.1 and 1 \% for the pulsar DM.
    This reflects the very low observational errors of pulsar DMs, as discussed in Sect. \ref{subsubsec:DM_data}.
  \end{enumerate}
  \begin{figure*}
    \centering

      \begin{subfigure}{0.45\textwidth}
          \includegraphics[width=\textwidth, draft=False]{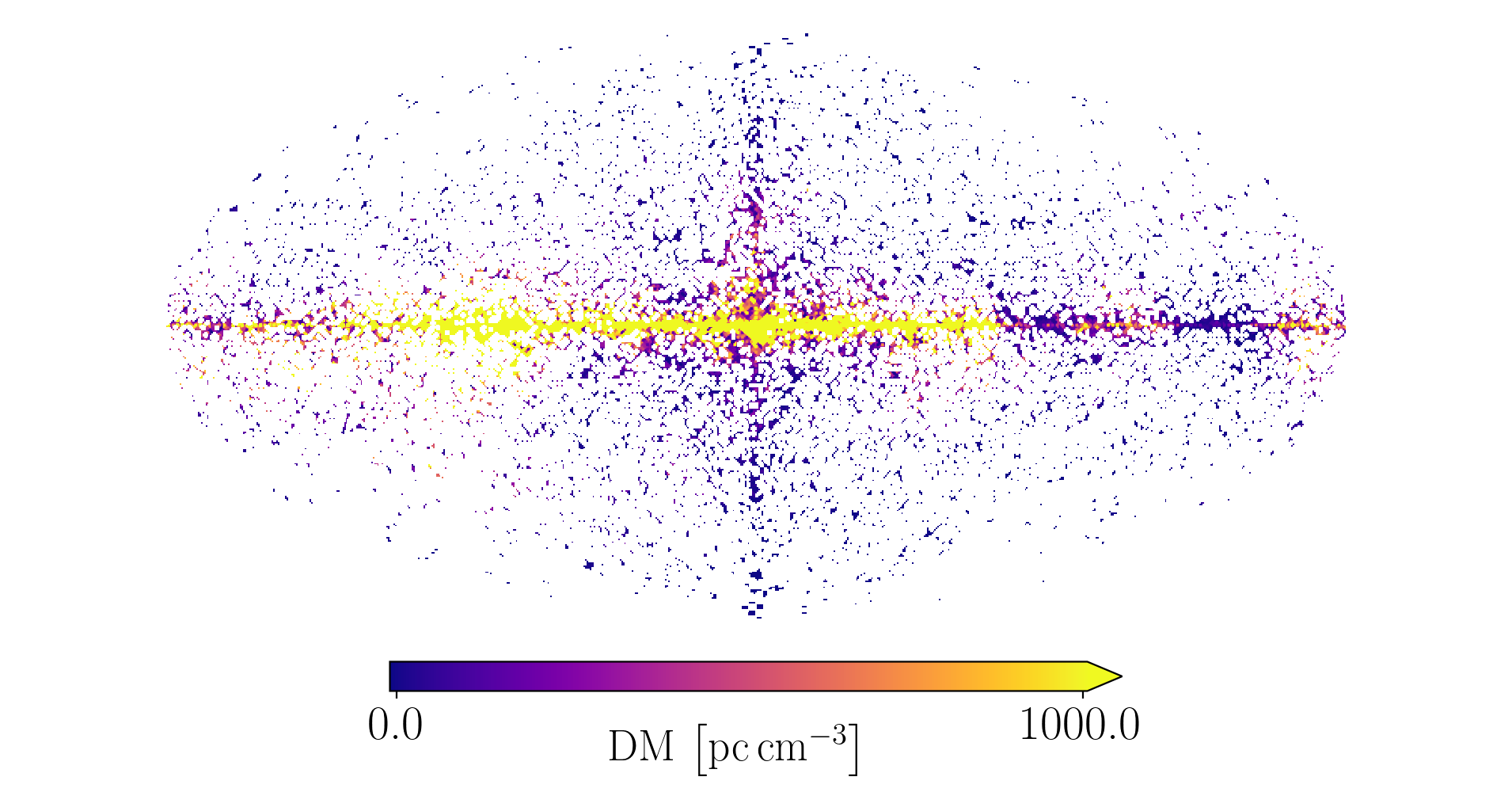}
          \caption{\label{fig:mockdmhigh}}
      \end{subfigure}
      \begin{subfigure}{0.45\textwidth}
        \includegraphics[width=\textwidth, draft=False]{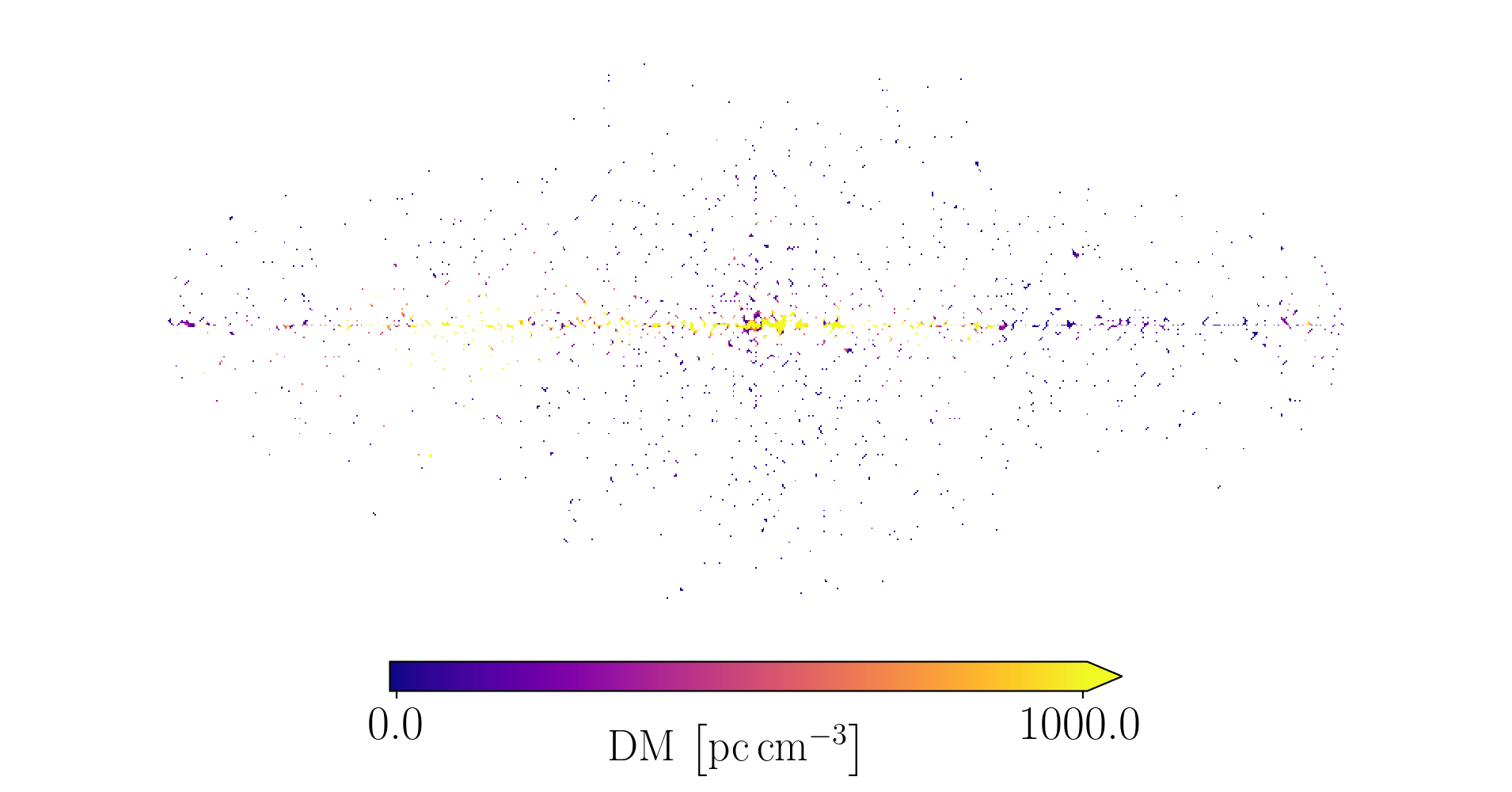}
        \caption{\label{fig:mockdmlow}}
    \end{subfigure}

    \begin{subfigure}{0.45\textwidth}\
      \includegraphics[width=\textwidth, draft=False]{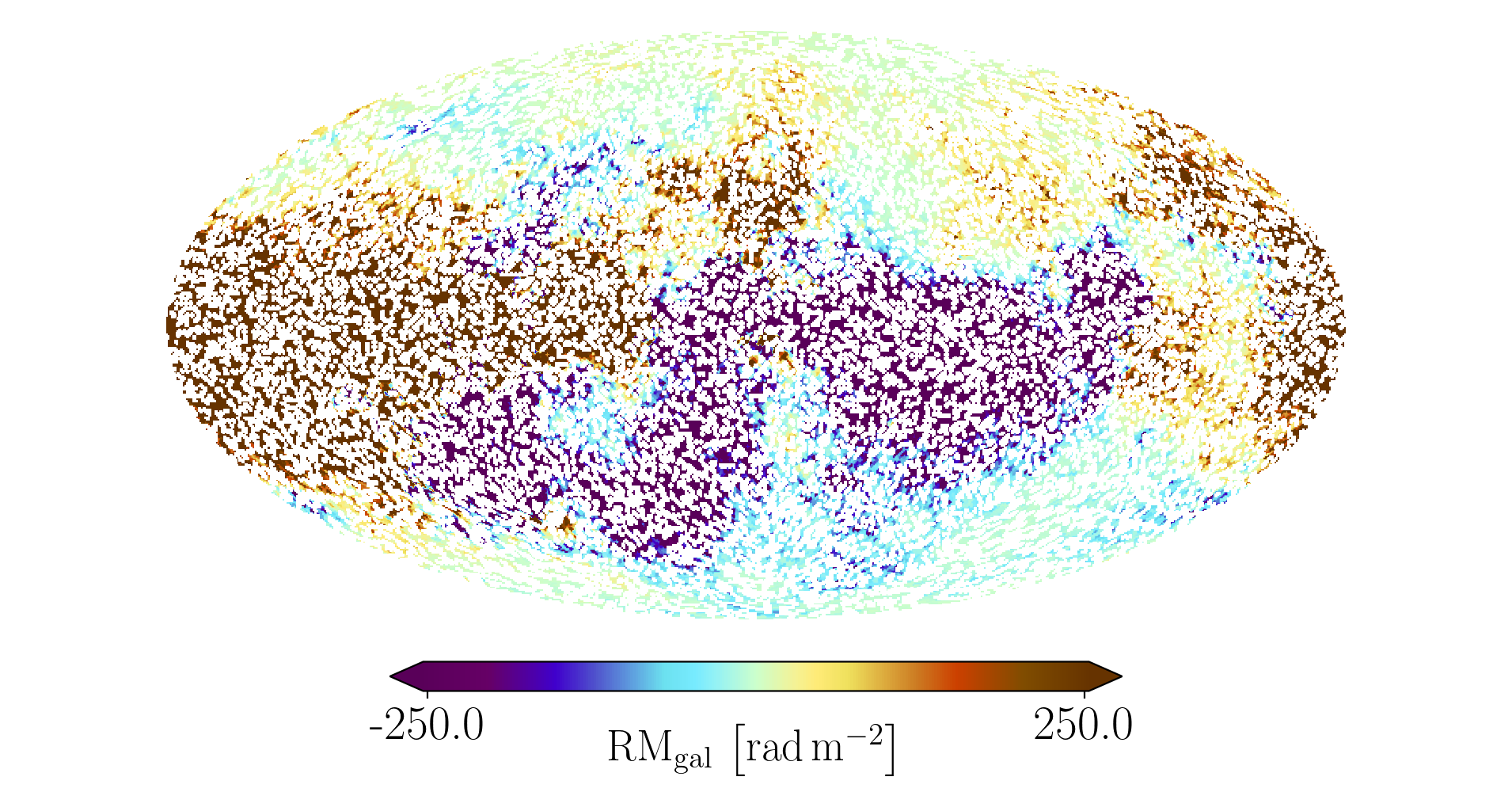}
      \caption{\label{fig:mockrm}}
  \end{subfigure}
      \begin{subfigure}{0.45\textwidth}
        \includegraphics[width=\textwidth, draft=False]{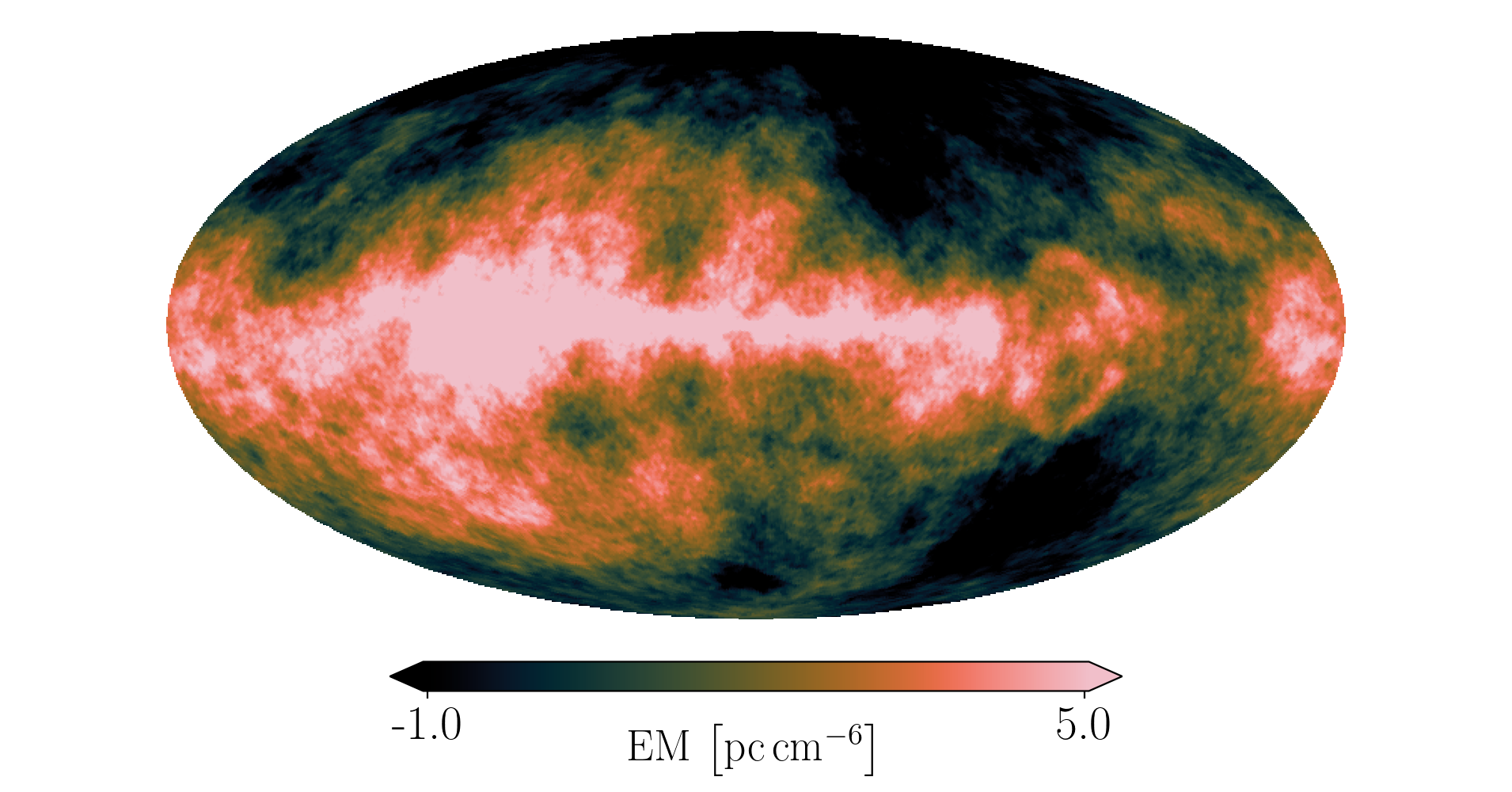}
        \caption{\label{fig:mockem}}
    \end{subfigure}
    \caption{\label{fig:mock_data} The synthetic data sets produced according to the procedure outlined in Sect. \ref{app:synthetic_tests}.
    The plots show projections on the sky for the \textit{ideal} (Fig. a) and \textit{realistic} (Fig. b) cases of pulsar DM, as well as the RM (Fig. c) and EM data (Fig. d)  sets. }
  \end{figure*}

\begin{sidewaysfigure*} %[htbp]
      %\centering
      \begin{subfigure}{0.33\textheight}
          \includegraphics[width=\linewidth, draft=False]{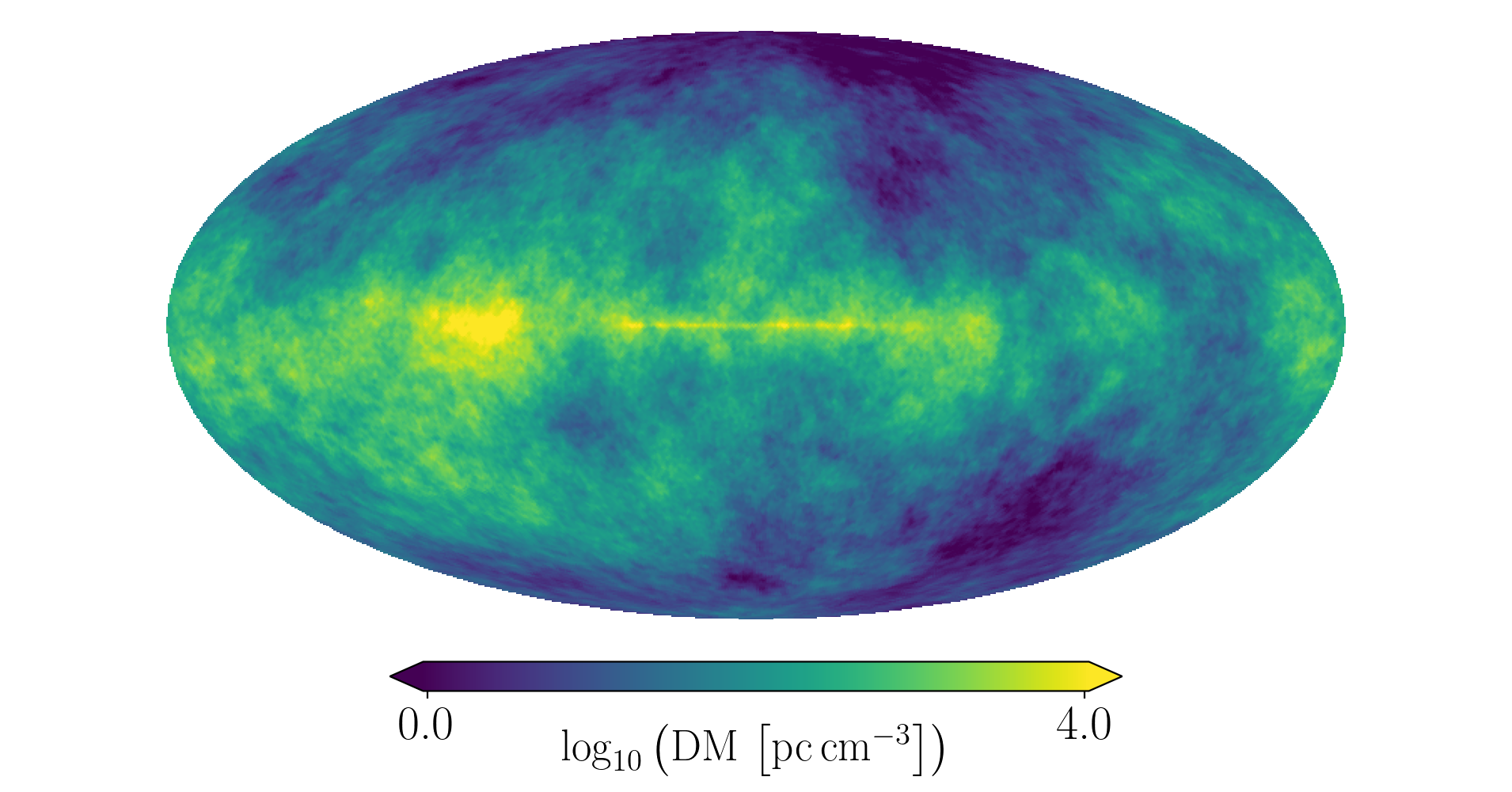}
          \caption{\label{fig:mocklog10dmtruth}}
      \end{subfigure}
      \begin{subfigure}{0.33\textheight}
          \includegraphics[width=\linewidth, draft=False]{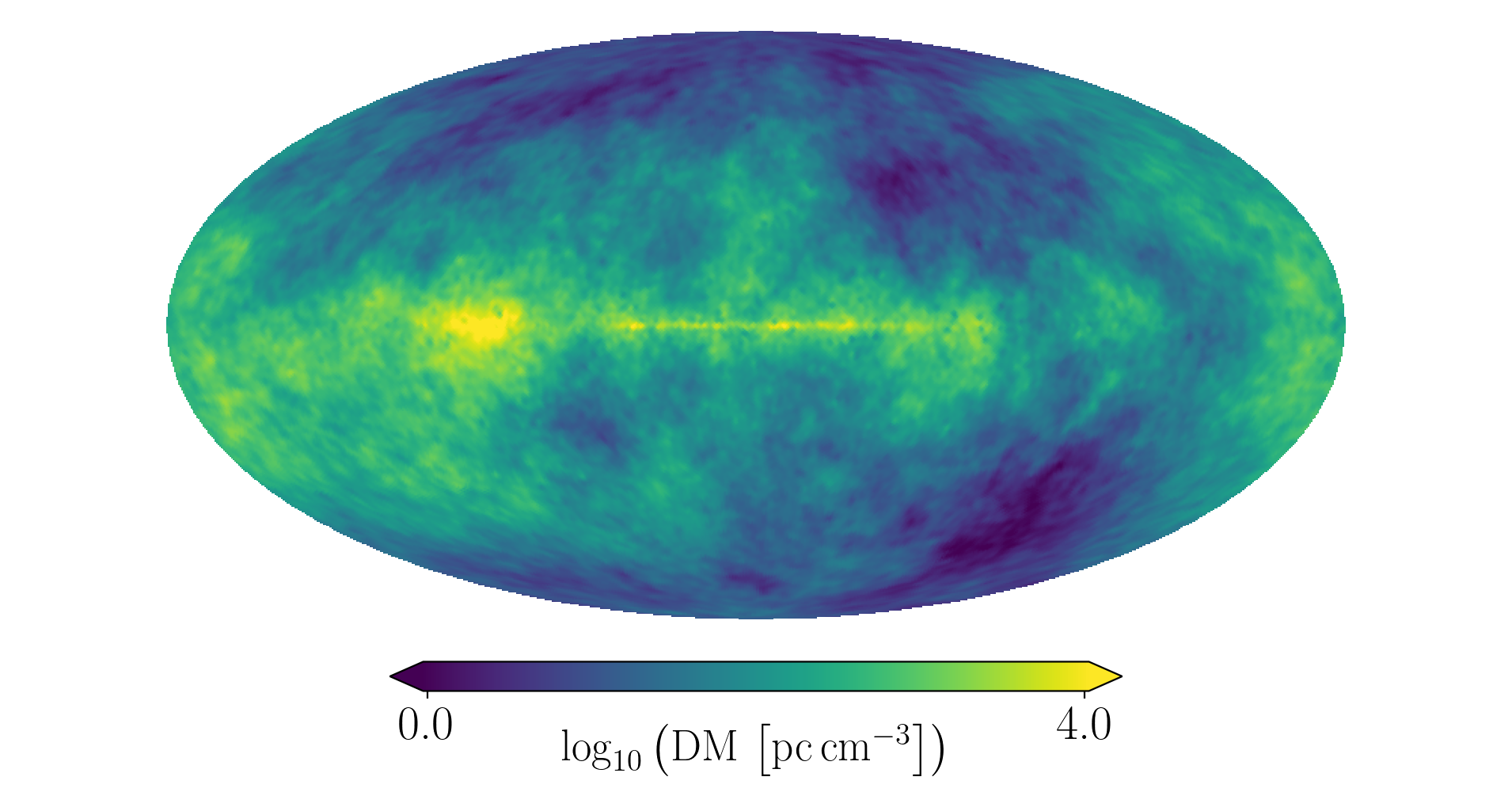}
          \caption{\label{fig:mocklog10dmhigh}}
      \end{subfigure}
      \begin{subfigure}{0.33\textheight}
        \includegraphics[width=\linewidth, draft=False]{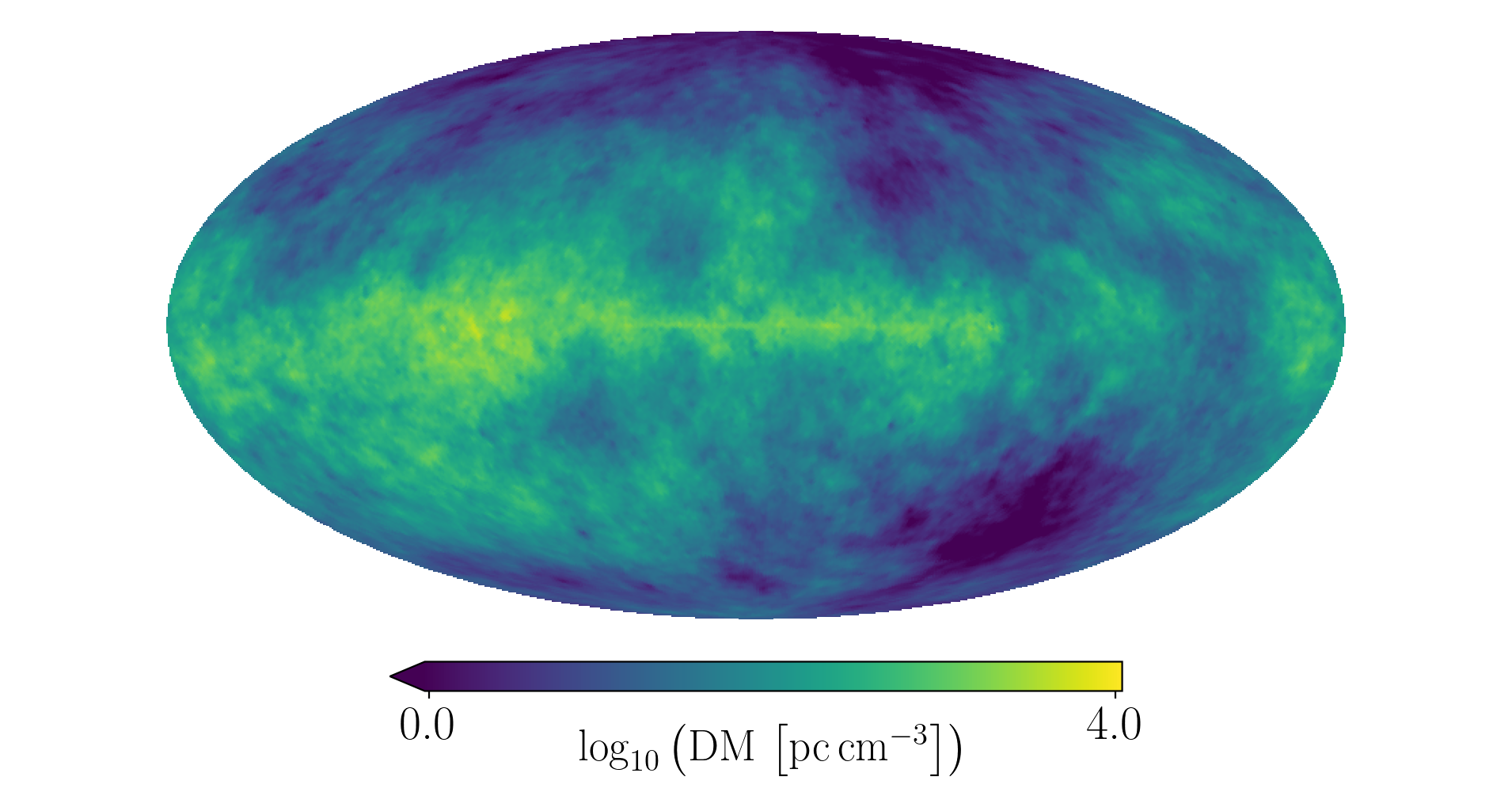}
        \caption{\label{fig:mocklog10dmlow}}
      \end{subfigure}

      \begin{subfigure}{0.33\textheight}
          \includegraphics[width=\linewidth, draft=False]{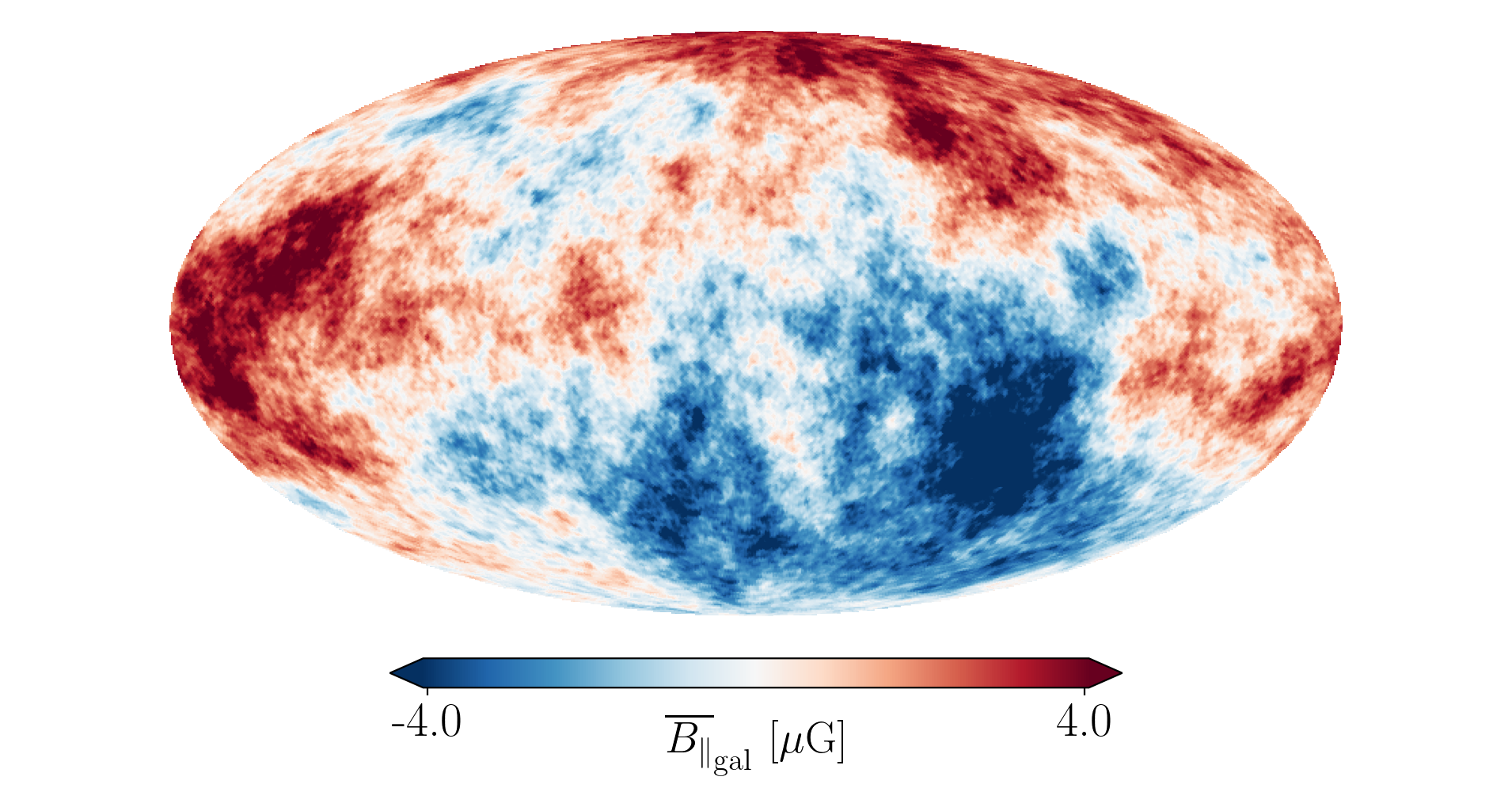}
          \caption{\label{fig:mockbpartruth}}
      \end{subfigure}%
      \begin{subfigure}{0.33\textheight}
          \includegraphics[width=\linewidth, draft=False]{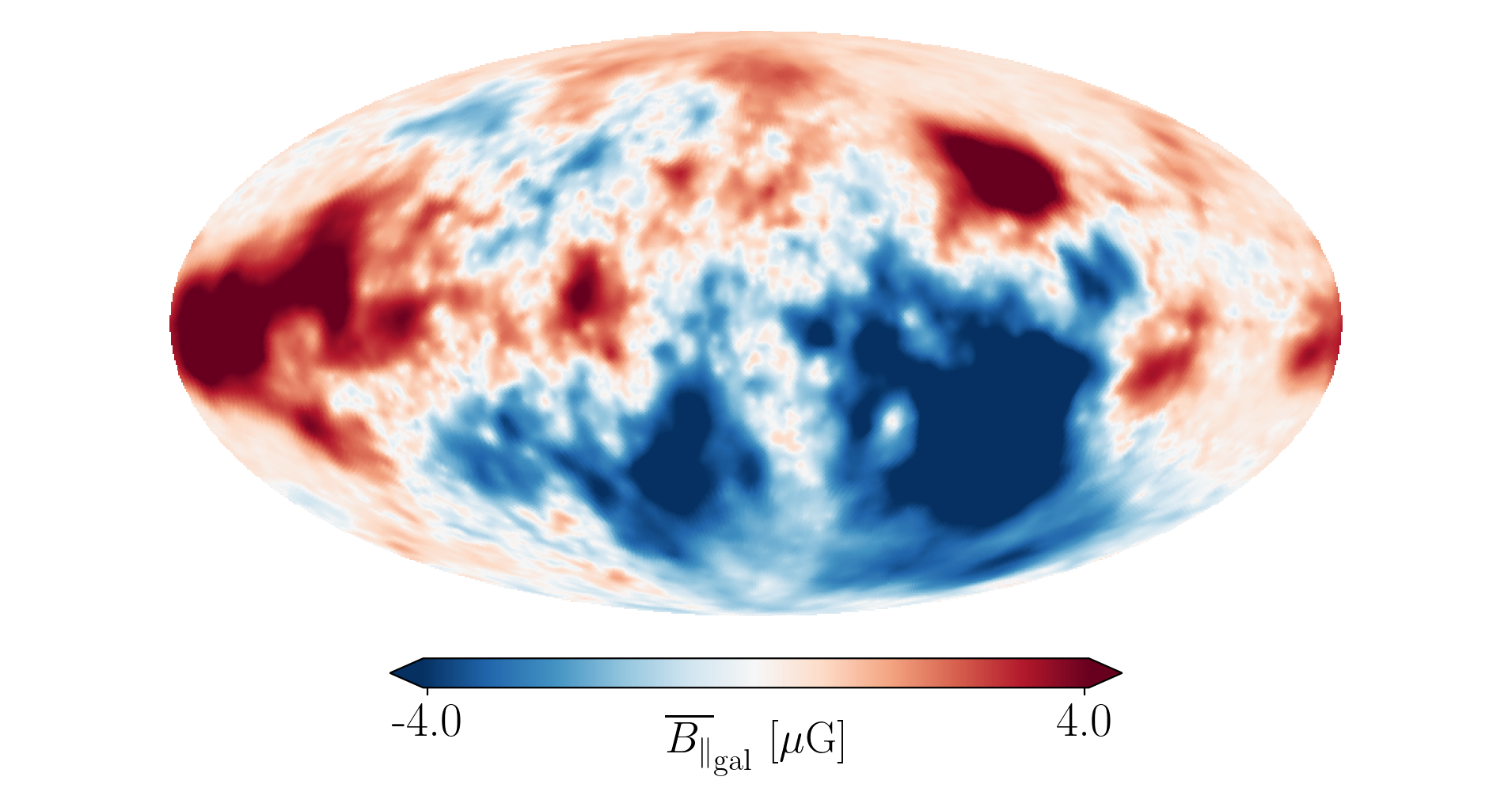}
          \caption{\label{fig:mockbparhigh}}
      \end{subfigure}
      \begin{subfigure}{0.33\textheight}
        \includegraphics[width=\linewidth, draft=False]{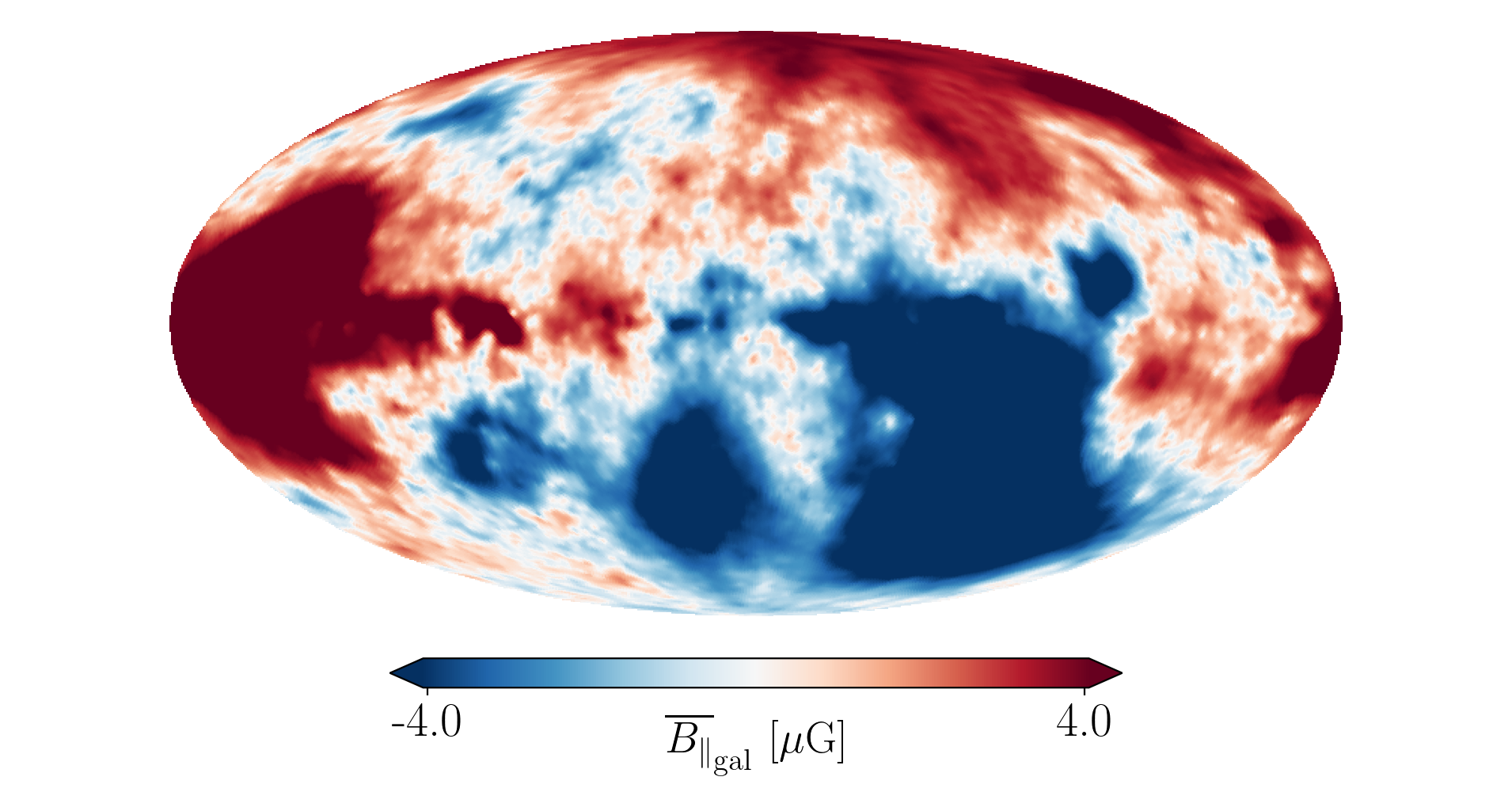}
        \caption{\label{fig:mockbparlow}}
      \end{subfigure}

      \begin{subfigure}{0.33\textheight}
          \includegraphics[width=\linewidth, draft=False]{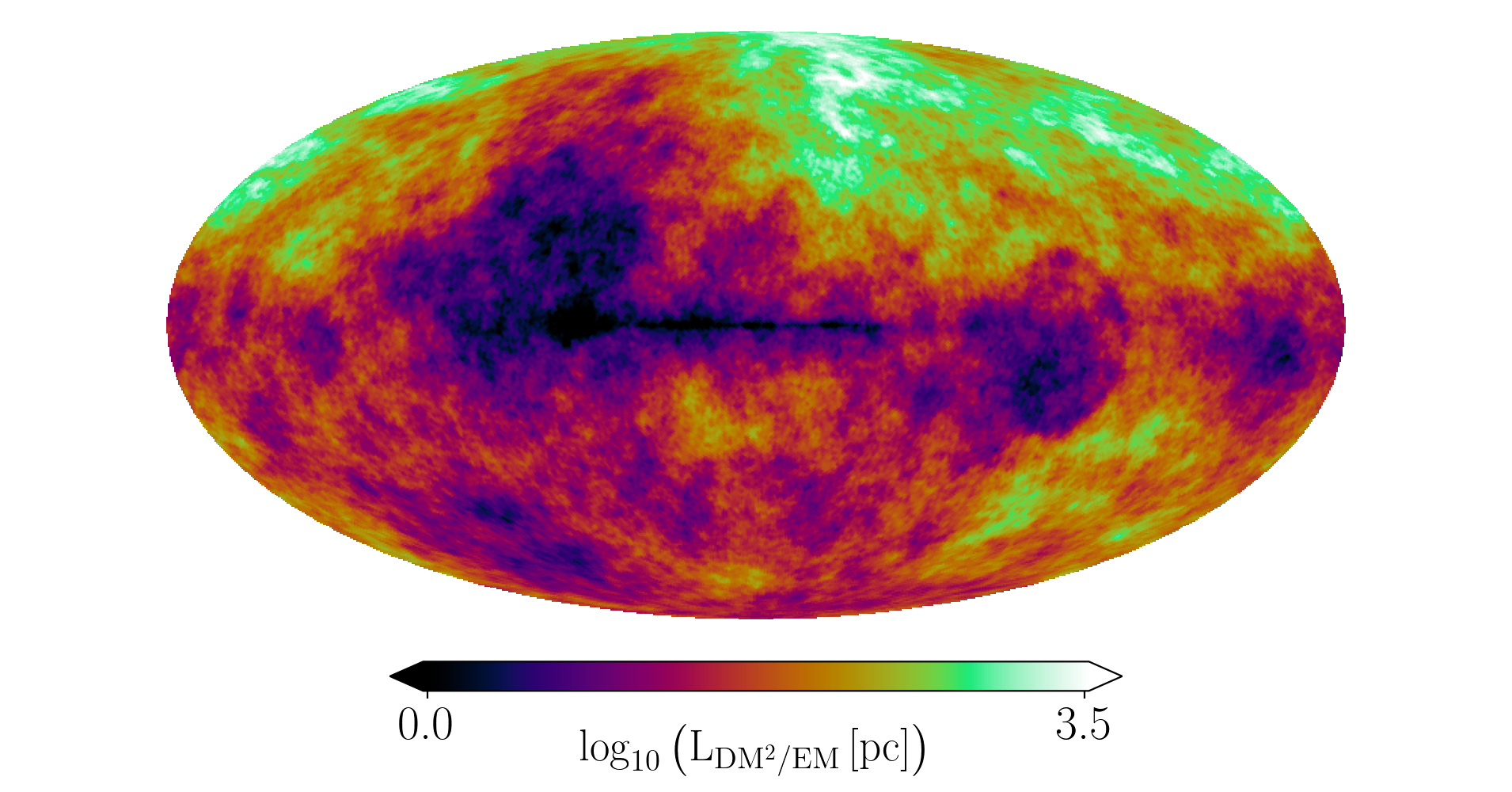}
          \caption{\label{fig:mockdisttruth}}
      \end{subfigure}%
      \begin{subfigure}{0.33\textheight}
          \includegraphics[width=\linewidth, draft=False]{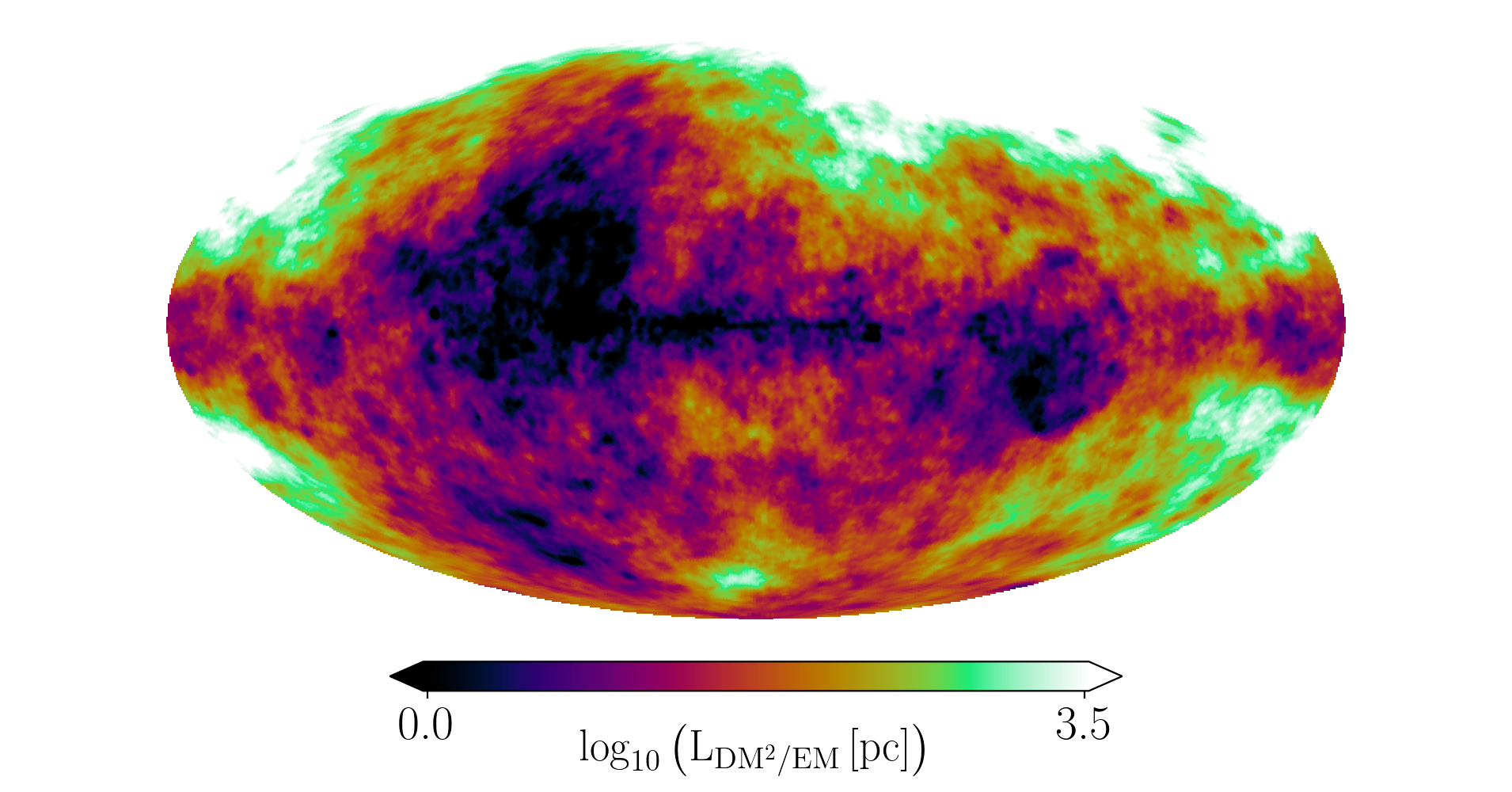}
          \caption{\label{fig:mockdisthigh}}
      \end{subfigure}
      \begin{subfigure}{0.33\textheight}
          \includegraphics[width=\linewidth, draft=False]{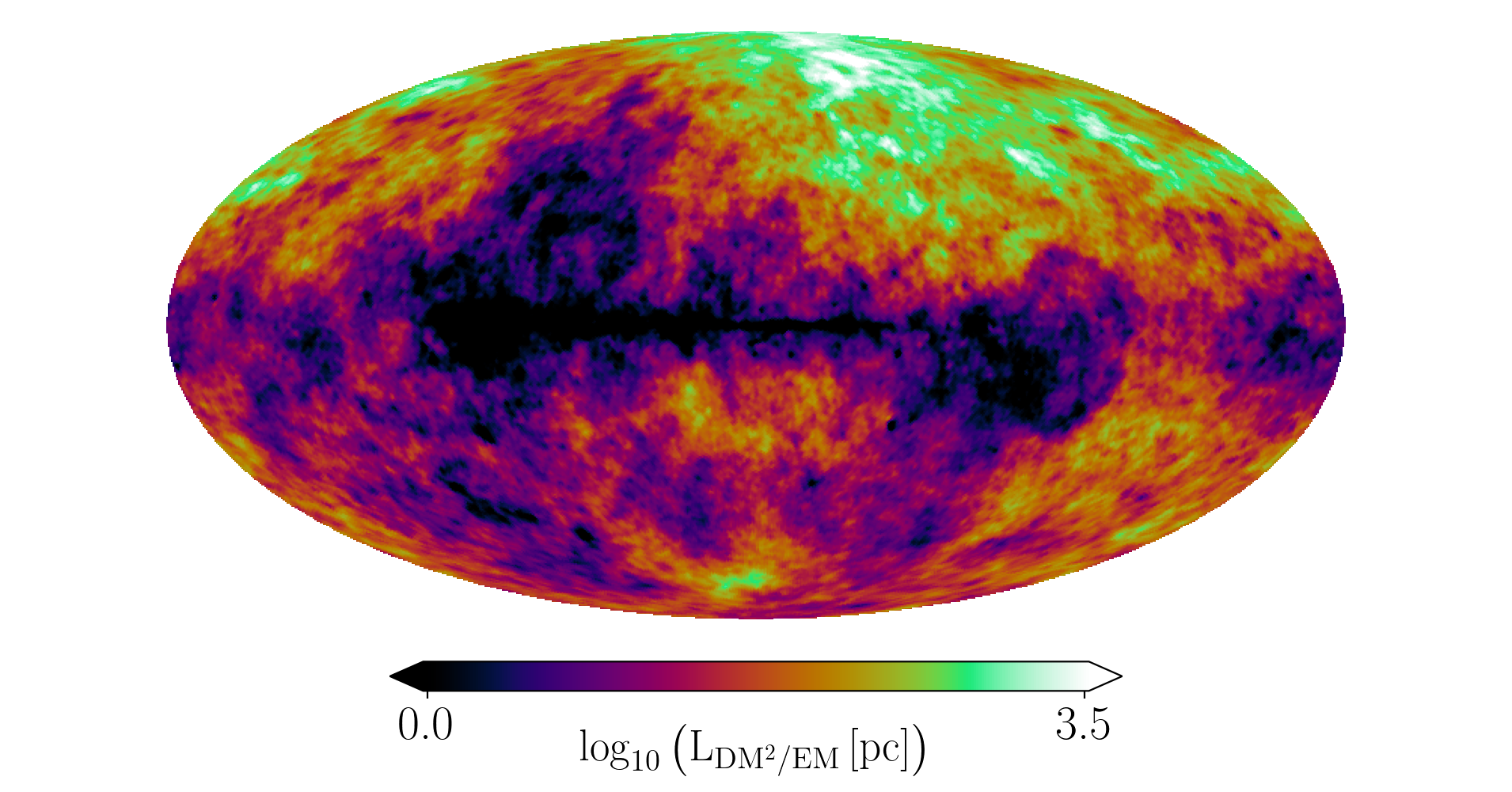}
          \caption{\label{fig:mockdistlow}}
      \end{subfigure}
      \caption{\label{fig:mock_inferences} Ground truth and inference results for the synthetic data.
      The left column shows the ground truth for the $\Bp{gal}$, log-$\DM{gal}$ and log-$\Len{DM^2/EM}$ skies.
      The middle and right column show the corresponding results for the \textit{ideal} and \textit{realistic} cases of pulsar DMs, respectively.}
  \end{sidewaysfigure*}

  The resulting synthetic data sets are shown in Fig. \ref{fig:mock_data}.
  We then run the inference procedure of Sect. \ref{sec:models} using precisely the same set up, i.e. the \textit{main} model with the same set of hyperparameters.
  The inferred sky maps and the synthetic truth are shown in Fig. \ref{fig:mock_inferences}.
  While we recover the general morphology of the maps very well, especially in case of the \textit{ideal} pulsar data set.
  However, the results constrained using the \textit{realistic} pulsar DM data set clearly underestimate the Galactic DM  amplitude in the disk, and this effect is translated in the $\Bp{gal}$ and log-$\Len{DM^2/EM}$ maps.
  The high latitude DM is severely underestimated in certain spots, while other areas appear to be fine.
  We conclude that the disk DM in our case is likely strongly underestimated, and that this effect can also exist at higher latitudes, albeit much less pronounced.

  \section{Faraday and EM sky maps}
  \label{app:results_rm_em}

  For completeness, we show the results of the Faraday rotation sky and logarithmic EM sky in Fig. \ref{fig:em_and_phi}.
  The general agreement with the results obtained by \citet{2022Hutschenreuter} and \citetalias{2019Hutschenreuter} is good, which gives an important validity check for our method.

  \begin{figure*}
  \centering
  \begin{subfigure}{0.99\textwidth}
  \includegraphics[width=\textwidth, draft=False]{\skypath{\rmdmem} RM_mean.png}
  \caption{\label{fig:phi}}
  \end{subfigure}
  \begin{subfigure}{0.99\textwidth}
  \includegraphics[width=\textwidth, draft=False]{\skypath{\rmdmem} log10_EM_mean.png}
  \caption{\label{fig:em}}
  \end{subfigure}
  \caption{\label{fig:em_and_phi}
  Inference results for the Galactic Faraday and EM skies.
  Fig. (a) shows the posterior mean for the Faraday sky, while Fig. (b) shows the posterior mean for the logarithmic EM sky.}
  \end{figure*}

  \section{Secondary models}
  \label{app:results_sky_secondary}

  \begin{figure*}
  \centering
  \begin{subfigure}{0.95\textwidth}
  \includegraphics[width=\textwidth, draft=False]{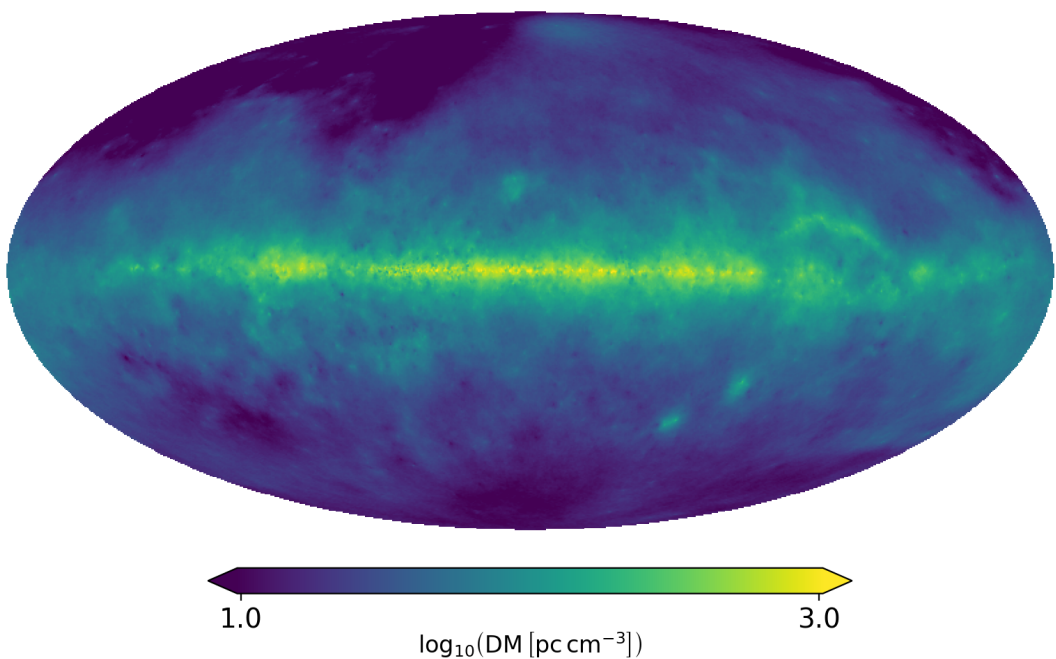}
  \caption{\label{fig:log_dm_phidm}}
  \end{subfigure}
  \begin{subfigure}{0.95\textwidth}
  \includegraphics[width=\textwidth, draft=False]{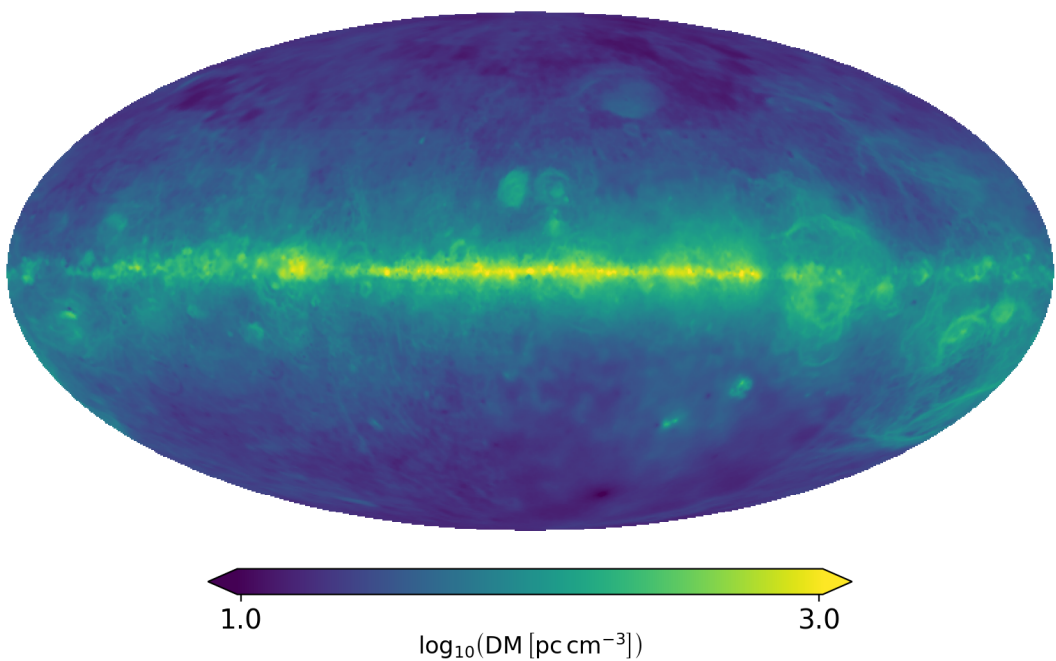}
  \caption{\label{fig:log_dm_dmem}}
  \end{subfigure}
  \caption{\label{fig:dm_compare}
  Inference results for the logarithmic Galactic DM sky of the secondary models introduced in Sect. \ref{subsec:model_summary}.
  Fig. (a) shows the posterior mean of the logarithmic $\DM{gal}$ sky constrained only by RM and DM data, Fig. (b) the corresponding result constrained only by DM and EM data.}
  \end{figure*}

  \begin{figure*}
  \centering
  \begin{subfigure}{0.95\textwidth}
  \includegraphics[width=\textwidth, draft=False]{\skypath{\rmdm}{Bpar_mean.png}}
  \caption{\label{fig:bpar_phidm}}
  \end{subfigure}
  \begin{subfigure}{0.95\textwidth}
  \includegraphics[width=\textwidth, draft=False]{\skypath{\dmem}{log10_distance_mean.png}}
  \caption{\label{fig:log_filling_dmem}}
  \end{subfigure}
  \caption{\label{fig:bpar_filling_secondary}
  Inference results of the secondary models introduced in Sect. \ref{subsec:model_summary}.
  Fig. (a) shows the posterior mean of the $\Bpw{gal}$ sky constrained only by RM and DM data, Fig. (b) the filling factor sky constrained only by DM and EM data.}
  \end{figure*}

  In order to test our results on the dependence, we have run two secondary models, where we have turned off either the RM or EM branch (see. Sect. \ref{sec:models}).
  We show the respective logarithmic $\DM{gal}$ skies in Fig. \ref{fig:dm_compare}.
  These sky maps and the correspondent result for the full model (see. Fig. \ref{fig:dm_log}) reveal that, while the overall scales are similar, several structures and can be clearly associated with either the RM or EM data sets.
  Most filamentary structures can be attributed to the H-$\alpha$ data set, as they are mostly missing in the RM-DM run.
  Conversely, the RM-DM map appears much more granular structure.
  We attribute this to the fact that this run was only constrained by point measurements (i.e. extragalactic RMs and pulsar DMs) and not by the (almost) full sky EM data set, which lead to a much worse constraint on small scales.
  Towards the poles, the RM data seems to favor much lower DM values as the EM-DM map.
  The $\LoS{\Bpw{}}{gal}$ map reveals coherent structures which seem to have absorbed some of the DM amplitude, see the discussion in Sect. \ref{subsec:results_sky_dm} for more details.

\end{appendix}

\clearpage

\end{document}